\documentclass[aps,amsmath,preprint,superscriptaddress,nofootinbib]{revtex4-1}
\usepackage{graphicx}
\usepackage{bm}
\usepackage{color}
\usepackage{amsmath,amssymb}	
\usepackage{hyperref}
\usepackage{setspace}

\usepackage[english]{babel}
\usepackage[utf8x]{inputenc}
\usepackage[colorinlistoftodos]{todonotes}
\usepackage{slashed}

\newcommand{\hc}{\mathrm{h.c.}}
\usepackage{physics}

\begin{document}


\def\a{\alpha}
\def\b{\beta}
\def\c{\varepsilon}
\def\d{\delta}
\def\e{\epsilonsilon}
\def\f{\phi}
\def\g{\gamma}
\def\h{\theta}
\def\k{\kappa}
\def\l{\lambda}
\def\m{\mu}
\def\n{\nu}
\def\p{\psi}
\def\q{\partial}
\def\r{\rho}
\def\s{\sigma}
\def\t{\tau}
\def\u{\upsilon}
\def\v{\varphi}
\def\w{\omega}
\def\x{\xi}
\def\y{\eta}
\def\z{\zeta}
\def\D{\Delta}
\def\G{\Gamma}
\def\H{\Theta}
\def\L{\Lambdabda}
\def\F{\Phi}
\def\P{\Psi}
\def\S{\Sigma}

\def\o{\over}
\def\beq{\begin{eqnarray}}
\def\eeq{\end{eqnarray}}
\newcommand{\gsim}{ \mathop{}_{\textstyle \sim}^{\textstyle >} }
\newcommand{\lsim}{ \mathop{}_{\textstyle \sim}^{\textstyle <} }
\newcommand{\EV}{ {\rm eV} }
\newcommand{\KEV}{ {\rm keV} }
\newcommand{\MEV}{ {\rm MeV} }
\newcommand{\GEV}{ {\rm GeV} }
\newcommand{\TEV}{ {\rm TeV} }
\newcommand{\1}{\mbox{1}\hspace{-0.25em}\mbox{l}}
\newcommand{\headline}[1]{\noindent{\bf #1}}
\def\diag{\mathop{\rm diag}\nolimits}
\def\Spin{\mathop{\rm Spin}}
\def\SO{\mathop{\rm SO}}
\def\O{\mathop{\rm O}}
\def\SU{\mathop{\rm SU}}
\def\U{\mathop{\rm U}}
\def\Sp{\mathop{\rm Sp}}
\def\SL{\mathop{\rm SL}}
\def\tr{\mathop{\rm tr}}
\def\mpl{M_{PL}}

\def\IJMP{Int.~J.~Mod.~Phys. }
\def\MPL{Mod.~Phys.~Lett. }
\def\NP{Nucl.~Phys. }
\def\PL{Phys.~Lett. }
\def\PR{Phys.~Rev. }
\def\PRL{Phys.~Rev.~Lett. }
\def\PTP{Prog.~Theor.~Phys. }
\def\ZP{Z.~Phys. }

\def\dd{\mathrm{d}}
\def\ff{\mathrm{f}}
\def\BH{{\rm BH}}
\def\inf{{\rm inf}}
\def\ev{{\rm evap}}
\def\eq{{\rm eq}}
\def\SM{{\rm sm}}
\def\Mpl{M_{\rm PL}}
\def\GeV{{\rm GeV}}
\newcommand{\Red}[1]{\textcolor{red}{#1}}

%


\title{Oscillating Composite Asymmetric Dark Matter}
\author{Masahiro~Ibe}
\email[e-mail: ]{ibe@icrr.u-tokyo.ac.jp}
\affiliation{Kavli IPMU (WPI), UTIAS, University of Tokyo, Kashiwa, Chiba 277-8583, Japan}
\affiliation{ICRR, University of Tokyo, Kashiwa, Chiba 277-8582, Japan}
\author{Shin Kobayashi}
\email[e-mail: ]{shinkoba@icrr.u-tokyo.ac.jp}
\affiliation{ICRR, University of Tokyo, Kashiwa, Chiba 277-8582, Japan}
\author{Ryo~Nagai}
\email[e-mail: ]{rnagai@icrr.u-tokyo.ac.jp}
\affiliation{ICRR, University of Tokyo, Kashiwa, Chiba 277-8582, Japan}
\author{Wakutaka~Nakano}
\email[e-mail: ]{m156077@icrr.u-tokyo.ac.jp}
\affiliation{ICRR, University of Tokyo, Kashiwa, Chiba 277-8582, Japan}

\begin{abstract}
The asymmetric dark matter (ADM) scenario can solve the coincidence problem between
the baryon and the dark matter (DM) abundance when the DM mass is of ${\cal O}(1)$\,GeV.
In the ADM scenarios, composite dark matter is particularly motivated,
as it can naturally provide the DM mass in the ${\cal O}(1)$\,GeV  range 
and a large annihilation cross section simultaneously.
In this paper, we discuss the indirect detection constraints on the composite ADM model.
The portal operators connecting the $B-L$ asymmetries in the dark and the Standard Model(SM) sectors
are assumed to be generated in association with the seesaw mechanism.
In this model, composite dark matter inevitably obtains a tiny Majorana mass 
which induces a pair-annihilation of ADM at late times.
We show that the model can be efficiently tested by the searches for the $\gamma$-ray 
from the dwarf spheroidal galaxies and the interstellar electron/positron flux.
\end{abstract}

\date{\today}
\maketitle
\preprint{IPMU 19-0103}
%
\section{Introduction}
Asymmetric dark matter (ADM) scenario sheds light on 
the coincidence problem between the observed baryon and dark matter (DM) abundances in the universe
~\cite{Nussinov:1985xr,Barr:1990ca,Barr:1991qn,Dodelson:1991iv,Kaplan:1991ah,Kuzmin:1996he,Foot:2003jt,Foot:2004pq,Kitano:2004sv,Gudnason:2006ug,Kaplan:2009ag} 
(see also~\cite{Davoudiasl:2012uw,Petraki:2013wwa,Zurek:2013wia} for reviews).
If the DM abundance is provided by a mechanism which is unrelated to the baryogenesis, it is quite puzzling why those 
abundances are close with each other despite the fact that the baryon abundance is dominated by 
the contribution from the matter-antimatter asymmetry. 
In the ADM scenario the coincidence problem can be explained when the DM mass is of ${\cal O}(1)$\,GeV,
where the matter-antimatter asymmetry is thermally distributed between the dark and the Standard Model (visible) sectors.

Among various ADM scenarios, composite baryonic DM in QCD-like dynamics is particularly motivated
since it can naturally provide a large annihilation cross section and  the DM mass in the GeV range 
simultaneously~\cite{Foot:2003jt,Foot:2004pq,Berezhiani:2005ek,Alves:2009nf,An:2009vq,Alves:2010dd,Gu:2012fg,
Buckley:2012ky,Detmold:2014qqa,Gu:2014nga,Lonsdale:2018xwd}.
Recently, a minimal composite ADM model and its ultraviolet (UV) completion~\cite{Ibe:2018juk,Ibe:2018tex,Ibe:2019ena} have been proposed where the asymmetry 
generated by the thermal leptogenesis~\cite{Fukugita:1986hr} 
(see also \cite{Giudice:2003jh,Buchmuller:2005eh,Davidson:2008bu} for review) 
is thermally distributed between the two sectors through 
a portal operator associated with the seesaw 
mechanism~\cite{Minkowski:1977sc, Yanagida:1979as, GellMann:1980vs, Glashow:1979nm, Mohapatra:1979ia}.
The dark sector of the model consists of QCD-like dynamics and QED-like interaction, 
which are called as dark QCD and dark QED, respectively.
The lightest baryons of dark QCD play the role of ADM.
The dark QED photon (dark photon) obtains a mass of $\order{10\mbox{-}100}$ MeV,
which plays a crucial role to transfer the excessive entropy of the dark sector
into the visible sector before neutrino decoupling~\cite{Ibe:2018juk,Blennow:2012de}.

In this paper, we discuss the indirect detection of the composite ADM model in \cite{Ibe:2018juk,Ibe:2018tex,Ibe:2019ena}.
The portal operator in this model is generated in association with the seesaw mechanism.
In this model, the dark-neutron, one of the lightest dark baryons, inevitably obtains a tiny Majorana mass. 
Such a tiny Majorana mass induces the oscillation between DM particle and the antiparticle,
which induces a pair-annihilation of ADM at late times~\cite{Cai:2009ia,Buckley:2011ye,Cirelli:2011ac,Tulin:2012re,Okada:2012rm,Hardy:2014dea,Chen:2015yuz}.
A pair of DM particle and the antiparticle annihilates into multiple dark pions, 
and the (neutral) dark pion subsequently decays into a pair of the dark photons.
The dark photon eventually decays into an electron-positron pair. 
Thus, the late time annihilation of ADM results in multiple 
soft electrons/positrons.
In addition, soft photons are also emitted as final state radiation.
As we will see, the model can be efficiently tested 
by the searches for the $\gamma$-ray from the dwarf spheroidal galaxies (dSphs) by the Fermi-LAT.
We also discuss the constraints from 
the observations of the interstellar electron/positron flux by the Voyager-1.

The organization of the paper is as follows.
In section~\ref{sec:model}, we review the composite ADM model in \cite{Ibe:2018juk,Ibe:2018tex,Ibe:2019ena} and show 
how the tiny Majorana mass of the dark neutron appears associated with the seesaw mechanism.
In section~\ref{sec:signal}, we derive the expected $\gamma$-ray flux from the dSphs and discuss the constraints 
on the model by comparing the flux with the Fermi-LAT results. 
We also estimate the interstellar electron/positron flux in cosmic ray from the late time annihilation and compare it with the Voyger-1 result.
The final section is devoted to the conclusions. 

\section{DM anti-DM oscillation in the composite ADM model}\label{sec:model}
\subsection{A Model of Composite ADM}
In this subsection, we briefly review the composite ADM model in~\cite{Ibe:2018juk,Ibe:2018tex,Ibe:2019ena}.
The model is based on $N_g$-generation dark quarks with $SU(3)_D \times U(1)_D$ gauge symmetry.
$SU(3)_D$ provides the dark QCD dynamics and $U(1)_D$ the dark QED interaction.
The dark quarks are the fundamental representations of $SU(3)_D$.
They are charged under the dark QED and the $B-L$ in analogy to
the up-type and the down-type quarks in the visible sector (see Tab.~\ref{tab:Dmatter}).
They have tiny masses, 
\begin{align}
\label{eq:mass}
{\cal L}_\mathrm{mass} = m_{U}' \overline{U}' U' + m_{D}'\overline{D}' D' + \hc\ ,
\end{align}
with $m_{U}'$ and $m_{D}'$ being the mass parameters.
Hereafter, we put primes on the parameters and the fields in the dark sector
when there  are counterparts in the visible sector.

The dark QCD exhibits confinement below the dynamical scale of $SU(3)_D$, 
$\Lambda_{\rm QCD}'$, which leads to the emergence of the dark baryons and the dark mesons.
Throughout this paper, we assume that only one generation of the dark quarks have masses 
smaller than $\Lambda_{\rm QCD}'$.%
\footnote{For $N_g >1$, we assume the heavier dark quarks decay into the lighter 
ones by emitting the dark Higgs boson which has the dark QED charge of $1$.
It should be noted that the dark quark masses are not generated by the vacuum expectation 
value of the dark Higgs boson~\cite{Ibe:2018juk,Ibe:2018tex,Ibe:2019ena},
and hence, the dark Higgs couplings generically violate the flavor symmetry in the dark sector.}
The lightest dark baryons, i.e. the dark nucleons, 
\begin{align}
p' \propto U'U'D'\ , \quad
\bar{p}' \propto \bar{U}'\bar{U}'\bar{D}'\ , \quad
n' \propto U'D'D'\ , \quad
\bar{n}' \propto \bar{U}'\bar{D}'\bar{D}'\ , 
\end{align}
are stable in the decoupling limit from the visible sector due to their $B-L$ charges.
Once the $B-L$ asymmetry is shared between the visible and the dark sector, 
the dark nucleon abundance is dominated by the asymmetric component due to their large annihilation cross section.
Therefore, the dark nucleon with a mass in the GeV range is a good candidate for ADM.

\begin{table}
	\centering
	\caption{The charge assignment of dark quarks. We assume $N_g$ generations of the dark quarks,
although only one generation has a mass smaller than $\Lambda_{\rm QCD}'$.
The $U(1)_{B-L}$ symmetry is the global symmetry which is shared with the visible sector.
	}
	\label{tab:Dmatter}
	\begin{tabular}{|c|c|c||c|}
		\hline
		& $SU(3)_D$ & $U(1)_D$ & $U(1)_{B-L}$ \\ \hline
		$U'$ & $\mathbf{3}$ & $2/3$ & $1/3$ \\
		$\overline U'$ & $\overline{\mathbf{3}}$ & $-2/3$ & $- 1/3$ \\
		$D'$ & $\mathbf{3}$ & $-1/3$ & $1/3$ \\
		$\overline D'$ & $\overline{\mathbf{3}}$ & $1/3$ & $- 1/3$ \\ \hline
	\end{tabular}
\end{table}

When the $B-L$ asymmetry is thermally distributed between the visible and the dark sectors,
the ratio of the $B-L$ asymmetry stored in each sector is given by $A_{\rm DM}/A_{\rm SM} = 44 N_g/237$
for the $B-L$ charges given in Tab.~\ref{tab:Dmatter}~\cite{Fukuda:2014xqa}.%
\footnote{In the presence of additional $B-L$ charged fields in the dark sector, such as dark leptons,
the ratio can be modified.
Besides, the neutrality condition of $U(1)_D$ and the contributions from the dark Higgs sector
also change the ratio by some tens percent for a given $N_g$.}
Thus, the observed ratio of the DM and the baryon abundance can be reproduced when the dark nucleon mass is
\begin{align}
m_{N}' \simeq \frac{\Omega_{\rm DM}}{\Omega_{\rm B}} \frac{A_{\rm B}}{A_{\rm SM}}\frac{A_{\rm SM}}{A_{\rm DM}}
\times m_N
\simeq \frac{8.5\,\rm GeV}{N_g}\ .
\end{align}
Here, we have used the ratio of the baryon asymmetry to the $B-L$ asymmetry in the visible sector, $A_{\rm B}/A_{\rm SM} = 30/97$~\cite{Harvey:1990qw}.
The dark nucleon mass in this range can be naturally realized when $\Lambda_{\rm QCD}'$ is in the GeV range.

The lightest dark mesons,
\begin{eqnarray}
\pi'^0\propto U' \bar U' - D' \bar D' \ , \quad \pi'^+ \propto U'\bar D'
\ , \quad \pi'^- \propto D'\bar U'\ ,
\end{eqnarray}
annihilate or decay into the dark photons.
As a result, they do not contribute to the effective number of neutrino degrees of freedom
nor to the dark matter abundance significantly even if they are stable.
In the following analysis, we assume that the dark charged pions are stable for simplicity.%
\footnote{If $U(1)_D$ is broken by the vacuum expectation value of  a dark Higgs with the dark QED charge of $2$, 
a ${\mathbb Z}_2$ symmetry remains unbroken which makes the dark charged pion stable.
If $U(1)_D$ is broken by the dark Higgs with the charge $1$, the neutral and the charged pions can mix each other, and hence, the charged pions decay.}
The decay of the dark neutral pion into a pair of dark photons, on the other hand, is inevitable due to the chiral anomaly.
As we will see, the decay of the neutral pion plays a central role for the indirect detection of ADM.

The dark photon obtains its mass by the dark Higgs mechanism, 
and it decays into the visible fermions thought the kinetic mixing with the visible QED photon, 
\begin{align}
{\cal L}_{\gamma^\prime} = \frac\epsilon2 F_{\mu\nu} F^{\prime\mu\nu} + \frac{m_{\gamma^\prime}^2}{2} A_\mu^\prime A^{\prime \mu} \,.
\end{align}
Here, $F_{\mu\nu}$ and $F^\prime_{\mu\nu}$ denote the field strengths of 
the visible and the dark QED with $A^\prime_\mu$ being the dark photon gauge field. 
In the following, we assume the kinetic mixing parameters of $\epsilon = 10^{-10}$--$10^{-8}$
and the dark photon mass in $\order{10\mbox{-}100}$ MeV range which 
satisfies all the constraints~\cite{Ibe:2018juk} (see also \cite{Bauer:2018onh,Chang:2016ntp, Chang:2018rso}).%
\footnote{See~\cite{Ibe:2018juk} for discussion on the origin of the tiny kinetic mixing parameters.}
In this parameter range, the dark photon decays when the cosmic temperature is above ${\cal O}(1)$\,MeV.

Finally, let us comment on the ratio between the abundances of the dark protons and the dark neutrons.
In the present model, there is no dark leptons nor dark weak gauge bosons.
Besides, it is expected that 
the mass difference 
between the dark neutron and the dark proton is 
smaller than the mass of the dark pion
when the dark quark masses are smaller than the dynamical
scale of $SU(3)_D$.
Thus, the dark neutron is stable in the limit of the 
vanishing $B-L$ portal interactions (see below).
The ratio between the dark proton abundance 
and the dark neutron abundance is given by~\cite{Ibe:2018tex},
\begin{align}
n_{n}'/n_{p}' \sim e^{-(m_{n}'-m_p')/T_F}\ .
\end{align}
Here, $n_{n,p}'$ and $m_{n,p}'$ are the number densities 
and the masses of the dark neutron and the dark proton, respectively.
$T_F$ denotes the freeze-out temperature of the dark pion annihilation,
$T_F \simeq m_\pi'/{\cal O}(10)$.
Thus, for $m_n'-m_p' \ll m_\pi'$, the dark neutron abundance is comparable 
to that of the dark proton.
In the following, we take $n_n' = n_p'$.

\subsection{The \texorpdfstring{$B-L$}{} portal operator}
The $B-L$ asymmetry generated by thermal leptogenesis is thermally distributed between the visible
and the dark sectors.
For this purpose, there need to be portal interactions which connect the $B-L$ symmetry in the two sectors.
In the model in \cite{Ibe:2018juk} (see also \cite{Ibe:2011hq,Fukuda:2014xqa}), the following operators are assumed 
as the portal operators,
\begin{align}
{\cal L}_{\text{portal}} \sim
\frac{1}{M_*^3} (\overline U' \overline D' \overline D') (L H)
+ \frac{1}{M_*^3} (U'^\dag D'^\dag \overline D') (L H) + \hc \,,
\label{eq:portal}
 \end{align}
where $L$ and $H$ are the lepton and the Higgs doublets in the visible sector, and $M_*$ is a dimensional parameter.%
\footnote{The portal operators require the gauge invariant operators which are charged under the $B-L$ symmetry. 
This is the reason why we need both the up-type and the down-type quarks in the dark sector.}
Here, we omit the $\order{1}$ coefficients.
The effects of the above operators decouple at the cosmic temperature below 
$T_{*} \sim M_{*} (M_{*} / \Mpl)^{1/5}$.
Here, $\Mpl = 2.4 \times 10^{18}~\mathrm{GeV}$ denotes the reduced Planck mass.
For successful ADM with thermal leptogenesis, the decoupling 
temperature, $T_*$, is required to be lower than the temperature, $T_{B-L}$, at which leptogenesis completes.
In the following, we consider the so-called strong washout regime of thermal leptogenesis, 
where the leptogenesis completes at the temperature about $T_{B-L} \simeq M_{R}/z_{B-L}$ with
$z_{B-L} \simeq 10$~\cite{Buchmuller:2005eh}.

In \cite{Ibe:2018juk,Ibe:2018tex}, the UV model has been proposed 
in which the portal operators in Eq.\,\eqref{eq:portal} 
are generated by integrating out the right-handed neutrinos, $\bar{N}$, and the dark colored Higgs boson, $H_C'$.
The gauge charges of $H_C'$ are identical to those of $D'$, while $H_C'$
has the $B-L$ charge $-2/3$.
The right-handed neutrinos couple to both sectors via,
\begin{align}
{\cal L} = \frac{M_R}{2} 
\bar N \bar N + y_N L H \bar N +\frac{1}{2}M_C^2 |H_C'|^2-Y_N H_C'\overline D'\bar{N} 
-Y_C H_C' U' D' - Y_{\bar{C}} H_C'^\dagger \overline U' \overline D'
+ \hc 
\label{eq:UVportal}
\end{align}
Here, $M_C$ denotes the dark colored Higgs mass, $M_R$ the mass of the right-handed neutrinos, 
and $y_N$ and $Y$'s are the Yukawa coupling constants.
The flavor and the gauge indices are suppressed.
It should be noted that the mass terms of the right-handed neutrino 
break $B-L$ symmetry explicitly.
The first two terms are relevant for the seesaw mechanism.

By integrating out $\bar{N}$ and $H_C'$ from Eq.\,\eqref{eq:UVportal}, the portal operators in Eq.\,\eqref{eq:portal} are obtained 
where $M_*$ corresponds to 
\begin{align}
    \frac{1}{M_*^3} = \frac{y_NY_NY_{\bar C}}{2M_C^2 M_R}\ , \quad 
    \frac{1}{M_*^3} = \frac{y_NY_NY_{C}^*}{2M_C^2 M_R}\ ,
    \label{eq:mstar}
\end{align}
for each term of Eq.\,\eqref{eq:portal}, respectively.
From the condition of $T_* < T_{B-L}$, the mass of the dark colored Higgs should satisfy,%
\footnote{Hereafter, we take $M_R >0$ and neglect the complex phases of $Y_N$, $Y_C$ and $Y_{\bar C}$.
We also assume $Y_C = Y_{\bar C}$ for simplicity.}
\begin{align}
\label{eq: upper bound on M_C}
    \frac{M_R}{z_{B-L}}\lesssim M_C\lesssim \frac{10}{z_{B-L}^{5/4}}\left(\frac{\hat{m}_\nu}{0.1\rm eV}\right)^{1/4}\sqrt{Y_N Y_C} M_R\ .
\end{align}
The first inequality comes from a consistency condition of the decoupling limit of 
the dark colored Higgs
at the temperature $T_{B-L}$.
In the right hand side, we have reparameterized the neutrino Yukawa coupling by using a tiny neutrino mass parameter, 
$\hat{m}_\nu$,
\begin{align}
|y_{N}^{2}| \sim 10^{-5} \left( \frac{\hat{m}_{\nu}}{0.1 \, \mathrm{eV}} \right) \left( \frac{M_{R}}{10^{9} \, \mathrm{GeV}} \right) \ .
\label{eq:seesaw}
\end{align}

Incidentally, 
the dark nucleon can decay into the dark pion and the anti-neutrino in the visible sector through the $B-L$ portal operator in Eq.\,\eqref{eq:portal}~\cite{Fukuda:2014xqa}.
The lifetime is roughly given by,
\begin{align}
\label{eq:lifetime}
    \tau_N' \simeq 10^{33}\,{\rm sec} 
     \left(\frac{2\,{\rm GeV}}{\Lambda_{\rm QCD}'}\right)^4
      \left(\frac{0.1\,{\rm eV}}{\hat{m}_\nu}\right)
      \left(\frac{M_R}{10^9\,{\rm GeV}}\right)
    \left(\frac{\tilde{M}_C}{3\times 10^9\,{\rm GeV}}\right)^4
        \left(\frac{10\,{\rm GeV}}{m_{\rm DM}}\right)\ ,   
          \end{align}
where $\tilde{M}_C = M_C/\sqrt{Y_NY_C}$.
Thus, the lifetime of the dark nucleons is 
much longer than the age of the universe 
for $M_R \sim M_C \sim 10^9$\,GeV.

\subsection{The Majorana mass of the dark neutron}
The portal operators in Eq.\,\eqref{eq:portal} are generated in association with the seesaw mechanism.
As a notable feature of the UV completion model in Eq.\,\eqref{eq:UVportal}, it 
also leads to the Majorana mass term of the dark neutron.
This can be observed by integrating out $H_C'$ and $\bar{N}$ one by one.
In the case of $M_C>M_R$, we first integrate out $H_C'$ from Eq.\,\eqref{eq:UVportal}, which reads
	\begin{align}
		\label{eq: integrate out Hc'}
		{\cal L} =\,& \bar{N}^\dagger i\sigma^\mu\partial_\mu \bar{N} + 
		\frac{M_R}{2} \bar{N} \bar{N} + y_N L H \bar{N}\notag \\
		&-\frac{\abs{Y_N}^2}{2M_C^2} \bar{D}'\bar{N}\qty(\bar{D}'\bar{N})^\dagger -\frac{Y_N}{2M_C^2} \bar{D}'\bar{N}\qty[ \qty(Y_CU'D')^\dagger + Y_{\bar{C}}\bar{U}'\bar{D}']
		+ \hc\notag\\
		&+(\text{quartic in dark quark fields}).
	\end{align}
Here, we show the kinetic term of $\bar{N}$ explicitly which were implicit in Eq.\,\eqref{eq:UVportal}.
This formula is of the form
	\begin{align}
	\label{eq: simplyfied N integrate out}
		{\cal L}=\qty(A\bar{N}\bar{N} + B\bar{N}+\hc)-C\bar{N}\bar{N}^\dagger,
	\end{align}
where%
\footnote{Here, $\chi^\dagger\sigma^\mu\overset{\leftrightarrow}{\partial}_\mu\eta 
=\chi^\dagger\sigma^\mu\partial_\mu\eta
-\partial_\mu\chi^\dagger\sigma^\mu\eta$
for the Weyl fermions, $\chi$  and $\eta$.
}
	\begin{align}
	\label{eq: formula for ABC}
		A=\frac{M_R}{2},\quad B=y_NLH-\frac{Y_N}{2M_C^2} \bar{D}'\qty[ Y_{\bar{C}}\bar{U}'\bar{D}' + \qty(Y_CU'D')^\dagger],\quad C=-\frac{i}{2}\sigma^\mu\overset{\leftrightarrow}{\partial}_\mu+\frac{\abs{Y_N}^2}{2M_C^2} \bar{D}'\bar{D}'^\dagger.
	\end{align}
To make $\bar N$ integrated out, it is convenient to complete the square of Eq.\,(\ref{eq: formula for ABC}) with respect to $\bar{N}$.
For this purpose, we shift $\bar{N}$ by $\bar{N}\to\bar{N}+\bar{\psi}$, with which we can eliminate the linear term 
in Eq.\,\eqref{eq: simplyfied N integrate out}.
The condition $\bar{\psi}$ must satisfy is $2A\bar{\psi}+B-C\bar{\psi}^\dagger=0$, which reads
	\begin{align}
		\bar{\psi}=-\frac{2A^*B+CB^\dagger}{4\abs{A}^2-C^2}\simeq -\frac{1}{2\abs{M_R}^2}\qty(1+\frac{C^2}{\abs{M_R}^2})\qty(M_R^*B+CB^\dagger )
	\end{align}
After the shift, we integrate out $\bar{N}$ to obtain
	\begin{align}
	\label{eq: fully integrated out portal}
		{\cal L}&=\qty(A\bar{\psi}\bar{\psi} + B\bar{\psi}+\hc)-C\bar{\psi}\bar{\psi}^\dagger\notag \\
					&=\frac{1}{2}B\bar{\psi}+\hc\notag \\
					&=-\qty(1+\frac{C^2}{\abs{M_R}^2})\qty(\frac{1}{2M_R}BB+\frac{1}{2\abs{M_R}^2}CB^\dagger B)+\hc
	\end{align}
From Eq.\,\eqref{eq: formula for ABC}, we find that $BB$ term includes the Mojorana mass term of the dark neutron
	\begin{align}
	\label{eq: Majorana mass term}
		\frac{1}{2M_R} BB \supset \frac{Y_N^2Y_{\bar{C}}^2}{8M_R M_C^4}\qty(\bar{U}'\bar{D}'\bar{D}')^2\sim
		\frac{Y_N^2Y_{\bar{C}}^2\Lambda_{\rm QCD}'^6}{8M_R M_C^4}\bar{n}'\bar{n}'.
	\end{align}
In this way, Eq.\,\eqref{eq:UVportal} leads to the Majorana mass,
\begin{align}
\label{eq:MajoranaMass}
    m_M = \frac{Y_N^2Y_{\bar{C}}^2\Lambda_{\rm QCD}'^6}{4M_R M_C^4}= \frac{\Lambda_{\rm QCD}'^6}{4M_R \tilde{M}_C^4}\ ,
\end{align}
in addition to the $B-L$ portal  operators in Eq.\,\eqref{eq:portal}.

Once the dark neutron obtains the Majorana mass, the dark neutron and the anti-dark neutron
oscillate with a time scale of $t_{\rm osc} = m_M^{-1}$~\cite{Cai:2009ia,Buckley:2011ye,Cirelli:2011ac,Tulin:2012re,Okada:2012rm,Hardy:2014dea}.
The probability to find an anti-dark neutron at a time $t$
is given by,
\begin{align}
    P(n'\leftrightarrow \bar{n}') = \sin^2(m_Mt) \ .
\end{align}
Here, we assume that the initial state at $t = 0$ is a pure dark neutron state.
As we will see in the next section, the oscillation 
induces a pair-annihilation of ADM which ends up with multiple 
soft electrons/positrons/photons.

\subsection{Washout Interactions and On-Shell Portal}
Before closing this section, let us discuss the $B-L$ washout interactions which 
are also induced from Eq.\,\eqref{eq:UVportal}.
In fact, the term $CB^\dagger B$  in 
Eq.\,\eqref{eq: fully integrated out portal}
includes
	\begin{align}
	\label{eq:washout}
		\frac{1}{2\abs{M_R}^2} CB^\dagger B\supset \frac{y_NY_N^*}{2M_C^2 \abs{M_R}^2}
																				\qty[Y_{\bar C}^*(\overline U'^\dagger \overline D'^\dagger \overline D'^\dagger)
																			(i\sigma^\mu\partial_\mu)(L H)
																				+Y_C(U' D' \overline D'^\dagger)(i\sigma^\mu\partial_\mu)(L H)]\ .
	\end{align}
	In these interaction terms, and those in Eq.\,\eqref{eq:portal}, $L$ couples to the dark sector operators which have the opposite $B-L$ charges with each other.
	Thus, if these operators are also in equilibrium
	at $T_{B-L}$, the $B-L$ asymmetry generated by leptogenesis is washed out.
	To avoid such problems, it is required that
	\begin{align}
	\label{eq: lower bound on M_C}
	 \frac{10}{z_{B-L}^{7/4}}\left(\frac{\hat{m}_\nu}{0.1\rm eV}\right)^{1/4} M_R	\lesssim \tilde{M}_C\ .
	\end{align}
	By comparing Eqs.\,\eqref{eq: upper bound on M_C} and \eqref{eq: lower bound on M_C}, we find that the allowed parameter region for the ADM scenario is highly restricted due to the washout interaction when the portal operators are generated from the UV model in Eq.\,\eqref{eq:UVportal}.
	
	This constraint can be easily relaxed by introducing additional $B-L$ portals.
	For example, we may introduce a pair of gauge singlet fermions, $(X,\bar{X})$ 
	with new scalar fields, $H_p$, and $H_{Cp}'$, whose gauge and $B-L$ charges are 
    the same with those of the Higgs doublet of the SM and the dark colored Higgs, respectively.
	In this case, there can be additional operators,
	\begin{align}
	\label{eq:onshellportal}
	    {\cal L}= 
	    M_X X \bar X +\frac{1}{2}M_{H}^2 |H_{p}|^2+\frac{1}{2}M_{Cp}^2 |H_{Cp}'|^2 
	    +y_X L H_p \bar X -Y_X H_{Cp}'\overline D'\bar X \ .
	\end{align}
	Here, $M_X$, $M_{H}$ and $M_{Cp}$ are the mass parameters of $(X,\bar{X})$, $H_p$ and $H_{Cp}'$, respectively, and $y_X$ and $Y_X$ are Yukawa coupling constants.%
	\footnote{$\bar{N}$ and $\bar{X}$ can be distinguished by an approximate discrete symmetry under which $(X,\bar{X})$, $H_p$ and $H_{Cp}'$ are charged. With the discrete symmetry, we can avoid unnecessarily mixing between $\bar{N}$ and $\bar{X}$. 
	}
	As the mass of $\bar{X}$ is the Dirac type, the interaction terms in Eq.\,\eqref{eq:onshellportal} do not violate the $B-L$ symmetry.
	Thus, these interactions do not washout the asymmetry generated by leptogenesis 
	but thermally distribute the asymmetry between the visible and the dark sector for $M_{X,H,Cp} < T_{B-L}$.

	In the following analysis, we divide the parameter region into two. 
	\begin{itemize}
	    \item Off-shell $B-L$ portal scenario:
	    \begin{align}
	 \frac{10}{z_{B-L}^{7/4}}\left(\frac{\hat{m}_\nu}{0.1\rm eV}\right)^{1/4} M_R	\lesssim \tilde M_C\lesssim \frac{10}{z_{B-L}^{5/4}}\left(\frac{\hat{m}_\nu}{0.1\rm eV}\right)^{1/4}M_R\ .
	 \end{align}
	 \item On-Shell $B-L$ portal scenario:
	 	    \begin{align}
	 	    \label{eq:scenario2}
    \frac{10}{z_{B-L}^{5/4}}\left(\frac{\hat{m}_\nu}{0.1\rm eV}\right)^{1/4} M_R\lesssim \tilde M_C \ .
	 \end{align}
	\end{itemize}
In the on-shell portal scenario, we assume that there are lighter particles than $T_{B-L}$ which mediate the $B-L$ asymmetry between two sectors as in Eq.\,\eqref{eq:onshellportal}.%
\footnote{In the on-shell scenario, we may take $Y_N = 0$, and hence, the Majorana dark neutron mass is not inevitable.}
It should be emphasized that the $B-L$ asymmetries in the two sectors are thermally distributed in both the scenarios.%
\footnote{In the absence of the on-shell portal, 
the region with $M_C < T_{B-L}$ results 
in a dark sector asymmetry which depends on the branching ratio
of $\bar{N}$ for small $Y_N$'s~\cite{Falkowski:2011xh}.
If $Y_N$'s are large for $M_C < T_{B-L}$, on the other hand, the $B-L$ asymmetry 
is washed out very strongly and results in too small asymmetry. }

\section{gamma-ray and electron/positron fluxes }\label{sec:signal}

\begin{figure}
	\centering
	\includegraphics[width=10cm,clip]{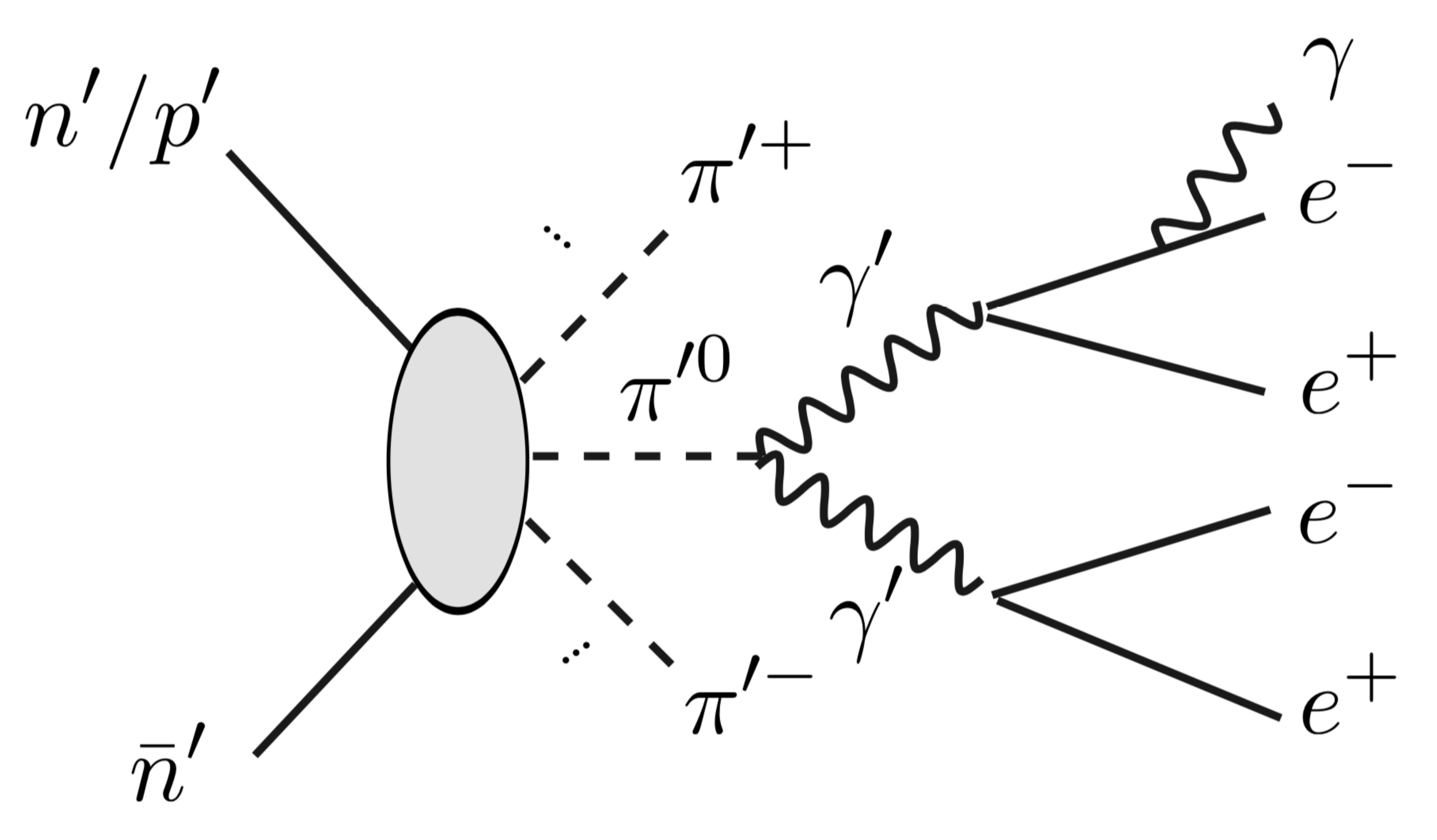}
	\caption{
	ADM annihilation which happens at  late time:
	$\bar{n}'$ can be generated from the ADM oscillation.
	Once the $\bar{n}'$ is generated, dark nucleons ($n'/p'$)
	and $\bar{n}'$ annihilate into dark pions ($\pi'^{\pm}$ and $\pi'^0$). 
	$\pi'^0$ subsequently decays into a pair of dark photons ($\gamma'$). 
	$\gamma'$ eventually decays into $e^++e^-$, and emits $\gamma$ through the FSR process.
	}
	\label{fig:cascade}
\end{figure}

As we have seen in the previous section, the dark neutron obtains a Majorana mass 
when the portal operator is generated in association with the seesaw mechanism.  
Due to the Majorana mass of the dark neutron, the dark neutron can oscillate into the anti-dark neutron. 
The typical time scale of the oscillation, $t_{\rm{osc}}= m_M^{-1}$, is estimated as
\begin{align}
t_{\rm{osc}}
&\simeq
{3.3\times 10^{21}}\,\mbox{sec}\,
\left(\frac{\Lambda'_{\rm{QCD}}}{2\,\mbox{GeV}}\right)^{-6}
\left(\frac{\tilde{M}_C}{3\times 10^9\,\mbox{GeV}}\right)^{4}
\left(\frac{M_R}{10^9\,\mbox{GeV}}\right)\ .
\end{align}
We now see that some fraction of $n'$ can convert into $\bar{n}'$ at  late time,
and then $n'/p'$ and $\bar{n}'$ annihilate into the dark pions. The neutral dark pions decay into the dark photons, and the dark photons finally decay into $e^+e^-$ pairs. $\gamma$ can be also emitted by the final state radiation (FSR) process as depicted in figure \ref{fig:cascade}. 
In this section, we discuss the constraints 
on the late-time annihilation from the 
observations of the $\gamma$-ray from the dSphs and the interstellar $e^++e^-$ flux.

\subsection{Gamma-ray flux from the Dwarf Spheroidal Galaxies}
The $\gamma$-ray signal is one of the most promising channels
to search for dark matter annihilation (e.g., \cite{Gunn:1978gr,Bergstrom:2012fi} for review). 
In particular, dSphs in our galaxy
are the ideal targets to search for the $\gamma$-ray signal,
since they have high dynamical mass-to-light ratios, ($M/L \sim 10-1000$), 
while they lack contaminating astrophysical $\gamma$-ray sources~\cite{Gilmore:2007fy,McConnachie:2012vd}.
In this subsection, we estimate the $\gamma$-ray 
fluxes from the dSphs and compare them with the upper limits 
on the fluxes put by the Fermi-LAT.

First, we calculate the $\gamma$-ray spectrum at production by the $n'\bar{n}'$ annihilation processes:
\begin{align}
n'\bar{n'}
\to
m\pi'^0
+
l\pi'^+ 
+ 
l\pi'^-\ 
,~~~
(m,l = 0,1,2,\cdots)\ .
\end{align}
The cascade spectrum can be calculated by using the technique developed in~\cite{Mardon:2009rc,Elor:2015tva,Elor:2015bho}.

We start to calculate the $\gamma$-ray spectrum at the rest frame of $\gamma'$. 
For $m_{\gamma'} \gg m_e$, the spectrum is given by the Altarelli-Parisi approximation 
formula~\cite{Mardon:2009rc},%
\footnote{In the appendix \ref{sec:FSR}, we compare the direct calculation of
the FSR with the Altarelli-Parisi approximation formula, and confirm the validity of 
the approximation in the parameter region we are interested in.}
\begin{align}
\frac{d\tilde{N}_\gamma}{dx_0}
=
\frac{\alpha_{\rm{EM}}}{\pi}
\frac{1+(1-x_0)^2}{x_0}
\left[-1+\ln\left(\frac{4(1-x_0)}{\epsilon^2_0}\right)\right]\ ,
\label{eq:dNdx0}
\end{align}
where $\epsilon_0={2m_e}/{m_{\gamma'}}$ and
$x_0={2E_0}/{m_{\gamma'}}$ with $E_0$ being the energy of $\gamma$ at the rest frame of $\gamma'$. $\alpha_{\rm{EM}}$ denotes the fine structure constant of SM QED.

The next step is to translate the spectrum in the rest frame of $\gamma'$ to that in the rest frame of $\pi'^0$. 
For the case where $m_{\pi'} \gg m_{\gamma'}$, the spectrum is calculated as
\begin{align}
\frac{d\tilde{N}_{\gamma}}{dx_1}
=
2\int^1_{x_1}
\frac{dx_0}{x_0}
\frac{d\tilde{N}_{\gamma}}{dx_0}
f\left(\frac{2x_1}{x_0}-1\right)
+
\mathcal{O}\left(\frac{m^2_{\gamma'}}{m^2_{\pi'}}\right)\ ,
\label{eq:dNdx1}
\end{align}
where $x_1={2E_1}/{m_{\pi'}}$ with $E_1$ being the energy of $\gamma$ at the rest frame of $\pi'$. 
The function $f$ represents the effect of the anisotropy of the $\gamma'$ decay. 
According to~\cite{Gao:2010qx,Elor:2015tva}, we take
\begin{align}
f(\cos\theta)
=
\frac{3}{8}(1+\cos^2\theta)\ ,
\end{align} 
with $\theta$ being the angle between the $\gamma$ emission line and the boost axis of $\gamma'$.
Note that the angle $\theta$ is kinematically constrained as
\begin{align}
\cos\theta
=
\frac{2x_1}{x_0}-1
+
\mathcal{O}\left(\frac{m^2_{\gamma'}}{m^2_{\pi'}}\right)\ .
\end{align}
This is the reason why we put $f(2x_1/x_0-1)$ in Eq.\,\eqref{eq:dNdx1}.

We next translate the spectrum Eq.\,\eqref{eq:dNdx1} to that in the center of mass (CM) frame for the ADM annihilation.
In order to do that, we need to know how much $\pi'^{0}$ is boosted.
If the total number of the dark pions is two $(m+2l=2)$, 
we can exactly know the energy/boost of the dark pions since they should be emitted back to back in the CM frame.
In this case, the $\gamma$ spectrum is calculated as
\begin{align}
\frac{d\tilde{N}^{(m,l)}_{\gamma}}{dx_2}
=
2\int^1_{x_2}
\frac{dx_1}{x_1}
\frac{d\tilde{N}_{\gamma}}{dx_1}
+
\mathcal{O}\left(\frac{m^2_{\pi'}}{m^2_{\rm{DM}}}\right)\ ,
~~~
\mbox{for}~~m+2l=2\ ,
\label{eq:dNdx2n2}
\end{align}
where $x_2={E_2}/{m_{\rm{DM}}}$ with $E_2$ being the energy of $\gamma$ at the CM frame.
 
On the other hand, in the case of $m+2l\geq3$, 
it becomes highly non-trivial to know how much the $\pi'^0$ can be boosted 
even when we assume that the matrix element of the annihilation is constant as a function 
of the final state momenta.
This is because, in this case, the energy spectrum of the dark pion is given as
\begin{align}
\frac{dN_{\pi'}}{d\xi}
=
\frac{1}{R_n}\frac{dR_n}{d\xi}
 \ ,
\end{align}
where $\xi=E_{\pi'}/m_{\rm{DM}}$ and $R_n$ is the $n=m+2l$ body phase space integration~\cite{Liu:2014cma}.
$E_{\pi'}$ denotes the energy of the dark pion in the CM frame.
In general, it is difficult to perform the phase space integration for $n\geq 3$. 
However, as discussed in \cite{Liu:2014cma,Elor:2015bho}, under the assumption that $m_{\pi'^0}=m_{\pi'^+}\equiv m_{\pi'}\ll m_{\rm{DM}}$, we can perform the phase space integrations analytically as
\begin{align}
\frac{dN_{\pi'}}{d\xi}
=
(n-1)(n-2)(1-\xi)^{n-3}\xi
+
\mathcal{O}\left(\frac{m^2_{\pi'}}{m^2_{\rm{DM}}}\right),
\end{align}
for $n=m+2l\geq3$. Using the results, we finally obtain
\begin{align}
\frac{d\tilde{N}^{(m,l)}_{\gamma}}{dx_2}
=
2(n-1)(n-2)
\int^1_{x_2}d\xi
(1-\xi)^{n-3}
\int^1_{x_2/\xi}
\frac{dx_1}{x_1}
\frac{d\tilde{N}_{\gamma}}{dx_1}
+
\mathcal{O}\left(\frac{m^2_{\pi'}}{m^2_{\rm{DM}}}\right)
 ,
\label{eq:dNdx2}
\end{align}
for $n=m+2l \geq 3$ where we assume $m_{\pi'^0}=m_{\pi'^+}\equiv m_{\pi'}$.

Finally, we sum over the possible intermediate states
and take into account the number of the final states.
It turns out that the total $\gamma$ spectrum from the $n'\bar{n}'$ annihilation is expressed as
\begin{align}
\frac{dN^{(n'\bar{n}')}_\gamma}{dx_2}
=
\sum_{m,l}2m
\left(
\mbox{Br}^{(n'\bar{n}')}(m,l)
\frac{d\tilde{N}^{(m,l)}_{\gamma}}{dx_2}
\right)
\ ,
\label{eq:dNnn}
\end{align}
where $\mbox{Br}^{(n'\bar{n}')}(m,l)$ denotes the branching ratio for the $n'\bar{n'}\to m\pi'^0+l\pi'^+ + l\pi'^-$ annihilation process. The factor $2m$ corresponds to the number of $e^+e^-$ pairs in the annihilation process.

In the same way, we can estimate the $\gamma$ spectrum from the $p'\bar{n}'$ annihilation processes:
\begin{align}
p'\bar{n'}
\to
m\pi'^0
+
l\pi'^+ 
+ 
(l-1)\pi'^-
\ ,~~~
(m = 0,1,2,\cdots, l=1,2,\cdots)\ .
\end{align}
The $\gamma$ spectrum is calculated as
\begin{align}
\frac{dN^{(p'\bar{n}')}_\gamma}{dx_2}
=
\sum_{m,l}2m
\left(
\mbox{Br}^{(p'\bar{n}')}(m,l)
\frac{d\tilde{N}^{(m,l)}_{\gamma}}{dx_2}
\right)\ ,
\label{eq:dNpn}
\end{align}
with replacing $n=m+2l$ by $n=m+2l-1$ in the calculation of ${d\tilde{N}^{(m,l)}_{\gamma}}/{dx_2}$.

In the following analysis, we simply assume that the branching ratio of the dark nucleon annihilation can be estimated as that of nucleon-antinucleon annihilation. According to~\cite{Orfanidis:1973ix}, 
we approximate the branching ratios by the fireball model,%
\footnote{In this approximation, the Parity violating mode, $(m,l)=(2,0)$, is allowed, 
although it is not significant numerically.} 
\begin{align}
\mbox{Br}^{(n'\bar{n}')}(m,l)
&=
\frac{2\alpha^{2l}{} }{(1+\alpha)^n+(1-\alpha)^n}\,
_n C _{2l}\,
P_n\ ,~~\mbox{with}~~
n={m+2l}\ ,\\
\mbox{Br}^{(p'\bar{n}')}(m,l)
&=
\frac{2\alpha^{2l-1}{} }{(1+\alpha)^n+(1-\alpha)^n}\,
_n C _{2l-1}\,
P_n\ ,~~\mbox{with}~~
n={m+2l-1}\ ,
\end{align}
where 
\begin{align}
P_n
=
\frac{1}{\sqrt{2\pi}\sigma}\exp
\left(-\frac{(n-\langle{n}\rangle)^2}{2\sigma^2}\right)\ ,
\end{align}
with $a=1/4$, $\langle{n}\rangle=5.05$, $\sigma^2=a\langle{n}\rangle$ and
\begin{align}
\alpha
 =
   \left\{
     \begin{array}{l}
       \sqrt{2}~~~\mbox{for}~~n=2\ ,\\
       1.5~~~\mbox{for}~~n\neq2 \ .
     \end{array}
   \right.
\end{align}

\begin{figure}
	\centering
	\includegraphics[width=6.3cm,clip]{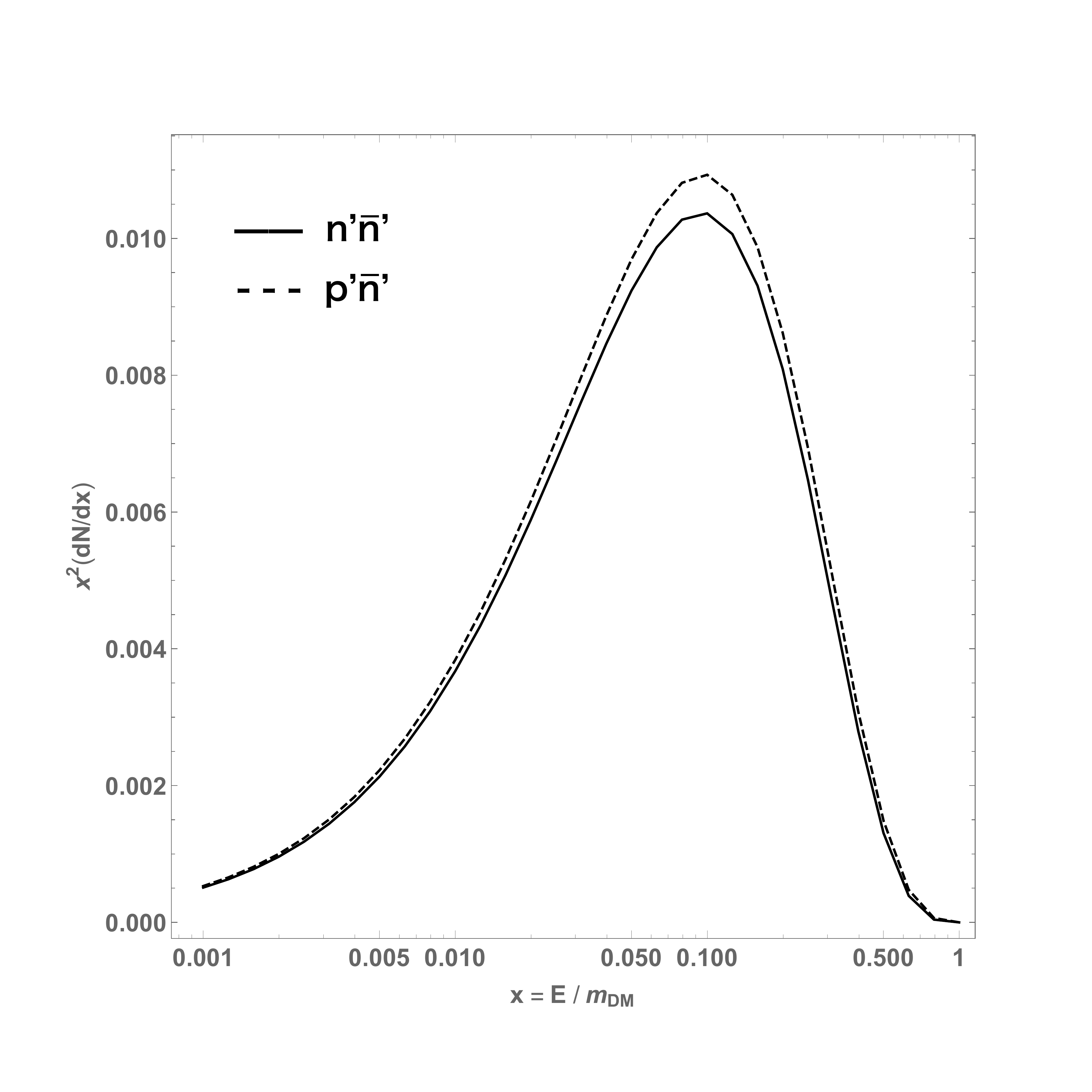}
	\caption{
	The $\gamma$-ray spectrum at production from the $n'\bar{n}'$ (Solid line) and $p'\bar{n}'$ (Dashed line) annihilation. 
	We take $m_{\pi'}=1\,\mbox{GeV}$ and $m_{\gamma'}=40\,\mbox{MeV}$.
	}
	\label{fig:Spectra}
\end{figure}

We are now ready to estimate the $\gamma$-ray spectrum emitted from the ADM annihilation. 
Figure \ref{fig:Spectra} shows the value of the $\gamma$-ray spectrum. 
Here, we take $m_{\mathrm{DM}}=10$\,GeV, $m_{\pi'}=1\,\mbox{GeV}$ and $m_{\gamma'}=40\,\mbox{MeV}$.
The black solid and the dashed lines correspond 
to the spectra predicted from the $n'\bar{n}'$ and $p'\bar{n}'$ annihilation, respectively.
In the analysis, we ignore the contributions from the annihilation with large $(m,l)$ since the branching ratios of them are much suppressed. 
We stop taking the sum over $(m,l)$ if the size of contribution 
is less than $1\%$ of the total amount.

The figure shows that the ADM annihilation predict the continuous $\gamma$-ray spectrum peaked at the energy 
of ${\cal O}(m_{\rm{DM}}/10)$.
This is expected as the typical number of the dark pions for an annihilation is five, and the neutral dark pion
decays into two pairs of $e^+e^-$.

It should be reminded that the $\gamma$-ray emission from the ADM annihilation can happen at the present universe since the ADM
oscillation effectively happens at the late time scale.
The ADM signals can  therefore be tested by $\gamma$-ray telescope experiments from nearby sources, 
while evading the constraints from the observations of the cosmic microwave observations (see e.g. \cite{Elor:2015bho}).

The $\gamma$-ray flux from the dSphs for an energy bin from $E_{\rm{min}}$ to  $E_{\rm{max}}$ is calculated as
\begin{align}
\Phi
=
\int^{E_{\rm{max}}}_{E_{\rm{min}}}dEE
\int_{\Delta\Omega}\frac{d\Omega}{4\pi}
\int_{\rm{l.o.s.}} dl
\left(
n_{n'}n_{\bar{n}'}\langle{\sigma v}\rangle_{n'\bar{n}'}\frac{dN^{(n'\bar{n}')}_\gamma}{dE}
+
n_{p'}n_{\bar{n}'}\langle{\sigma v}\rangle_{p'\bar{n}'}\frac{dN^{(p'\bar{n}')}_\gamma}{dE}
\right)\ ,
\end{align}
where we perform the integrations over
a solid angle, $\Delta\Omega$, and the line-of-sight (l.o.s.).
Here $n_i$ and $\langle{\sigma v}\rangle_{ij}$ denote the number density of a particle $i$ at the dSphs
and the kinematically averaged cross section for $ij$ annihilation, respectively. $N^{(n'\bar{n}')}_\gamma$ and $N^{(p'\bar{n}')}_\gamma$ are the photon spectra from $n'\bar{n}'$ and $p'\bar{n}'$ annihilation which can be calculated from Eqs.\,\eqref{eq:dNnn} and \eqref{eq:dNpn}, respectively.

It should be noted that 
the total amount of the $\gamma$-ray flux can be large enough to be tested by the $\gamma$-ray searches on the dSphs
although the flux is suppressed by the factor,
\begin{align}
\frac{n_{\bar{n}'}}{n_{n'}}
&\simeq
\left(\frac{t_{0}}{t_{\rm{osc}}}\right)^2\nonumber\\
&\simeq
1.6\times 10^{-8}
\left(\frac{\Lambda'_{\rm{QCD}}}{2\,\mbox{GeV}}\right)^{12}
\left(\frac{\tilde{M}_C}{3\times 10^9\,\mbox{GeV}}\right)^{-8}
\left(\frac{M_R}{10^9\,\mbox{GeV}}\right)^{-2}\ .
\end{align}
where $t_0\simeq 4.3\times 10^{17}\,\rm{sec}$ is the age of the universe. 
This is because the thermally-averaged cross section can be large due to the strong interaction. 
In the following analysis, we take the annihilation cross sections to be
\begin{align}
\langle{\sigma v}\rangle_{n'\bar{n}'}
=
\langle{\sigma v}\rangle_{p'\bar{n}'}
=
\frac{4\pi}{m^2_{\rm{DM}}}\ ,
\label{eq:xsec}
\end{align}
to give rough estimation.
Such a large annihilation cross section multiplied by the relative velocity is supported 
by the cross section measurements of the non-relativistic nucleon and anti-nucleon 
annihilation~\cite{Armstrong:1987nu,Bertin:1997gn} 
(see also \cite{Huo:2015nwa,Lee:2015hma}).%
\footnote{The cross section multiplied by the relative velocity in Eq.\,\eqref{eq:xsec} 
is much smaller than the unitarity limit.

In the Appendix~\ref{sec:sommerfeld}, we discuss
the Sommerfeld enhancement effects by the exchange of the dark pions.
There, we find that the enhancement effects are not 
significant in the present setup.
}

\begin{figure}
	\centering
	\includegraphics[width=13cm,clip]{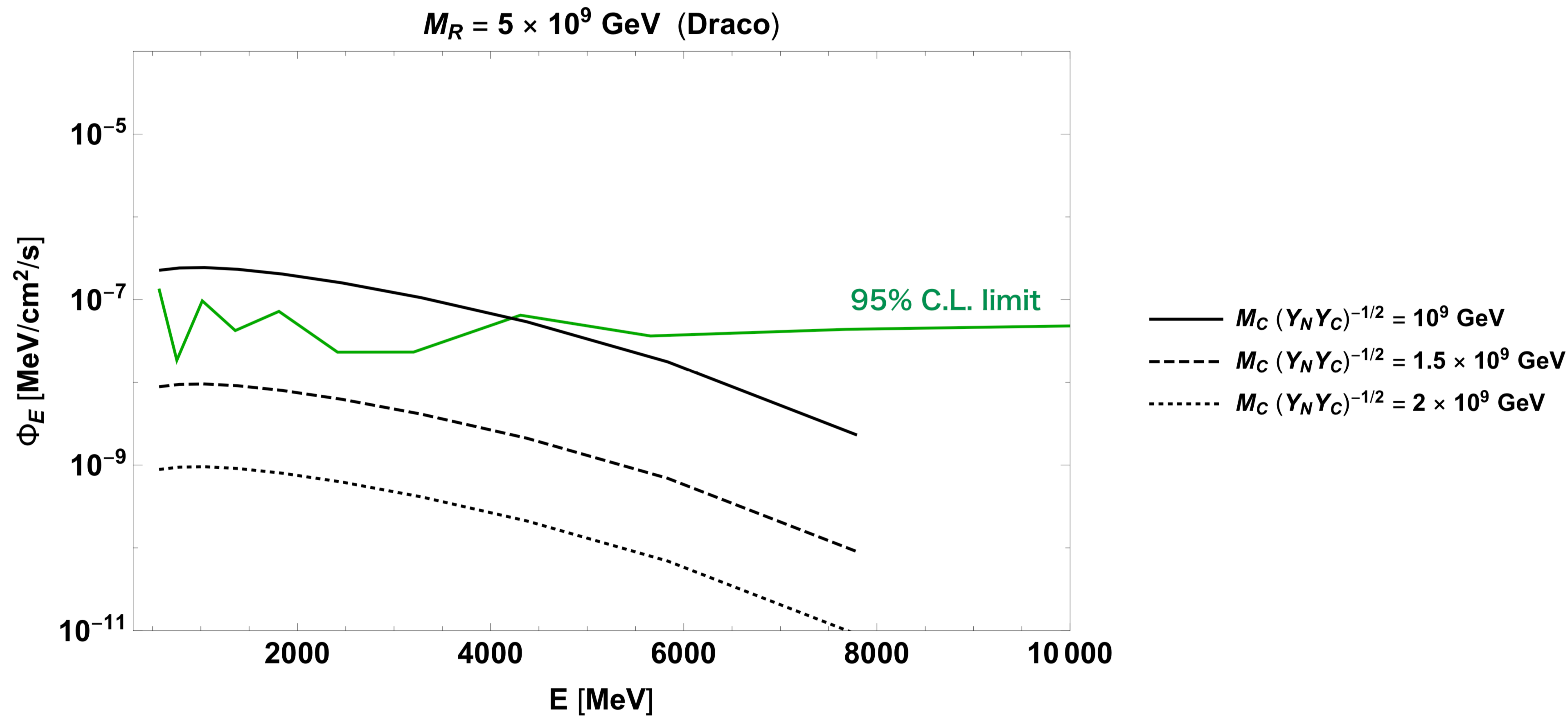}
	\caption{
	The predicted $\gamma$-ray flux from the Draco dSph and the $95\%$ C.L. upper bound obtained by the Fermi-LAT (green line). The black solid, dashed, and dotted lines correspond to the $\gamma$-ray flux when we take $\tilde{M}_C=10^9\,\mbox{GeV}$, $1.5\times 10^9\,\mbox{GeV}$, and $2\times 10^9\,\mbox{GeV}$, respectively. Here, we assume $m_{n'}=m_{p'}=m_{\rm{DM}}$ and fix $m_{\pi'}=1\,\mbox{GeV}$, $M_{R}=5\times 10^9\,\mbox{GeV}$ and $m_{\gamma'}=40\,\mbox{MeV}$. 
	}
	\label{fig:draco}
\end{figure}

 In Figure~\ref{fig:draco}, we show the predicted $\gamma$-ray flux from the Draco dSph. 
 The black solid, dashed, and dotted lines correspond to the $\gamma$-ray flux when we take $\tilde{M}_C=10^9\,\mbox{GeV}$, $1.5\times 10^9\,\mbox{GeV}$, and $2\times 10^9\,\mbox{GeV}$, respectively. Here, we assume $m_{n'}=m_{p'}=m_{\rm{DM}}=10\,\mbox{GeV}$ and fix $m_{\pi'}=1\,\mbox{GeV}$ and $m_{\gamma'}=40\,\mbox{MeV}$. 
 To obtain the predicted $\gamma$-ray spectrum, we use the $J$-factors estimated in \cite{Hayashi:2016kcy}
which takes into account the effects of the non-sphericity of the dSphs.%
\footnote{As for the $J$-factor of the Ursa Minor classical dSphs, we use the value given in
\cite{Geringer-Sameth:2014yza} as it is not analyzed in \cite{Hayashi:2016kcy}.
}
The green line corresponds to the upper bound ($95\%$ C.L.) on the $\gamma$-ray flux based on the  
6 years of Pass 8 data by the Fermi-LAT collaboration~\cite{Ackermann:2015zua}.
The figure shows that the $\gamma$-ray flux from the 
late-time annihilation becomes comparable to the 
upper limit on the observed flux for $\tilde{M}_C = \order{10^9}$\,GeV and $M_R = \order{10^{10}}$\,GeV, which corresponds to the oscillation time scale of 
$t_{\mathrm{osc}}=\order{10^{21}}\,$sec.
We discuss the constraints on the model parameters 
by the Fermi-LAT in subsection~\ref{sec:constraint}.

\subsection{Interstellar Electron/Positron Flux}
The Fermi-LAT observation does not constrain 
the late-time annihilation for $m_{\mathrm{DM}}\lesssim 3$\,GeV, 
since the Fermi-LAT is sensitive to the $\gamma$-ray with 
energy higher than $500$\,MeV.
For such a rather light ADM, 
the most stringent constraint is put by 
the observation of the interstellar $e^++e^-$ flux by the Voyager-1~\cite{Fisk:1976aw,Stone2013} (see also \cite{Boudaud:2016mos}).
In this subsection, we estimate the $e^++e^-$ flux from the late-time annihilation in the Milky Way.

\begin{figure}
	\centering
	\includegraphics[width=6.3cm,clip]{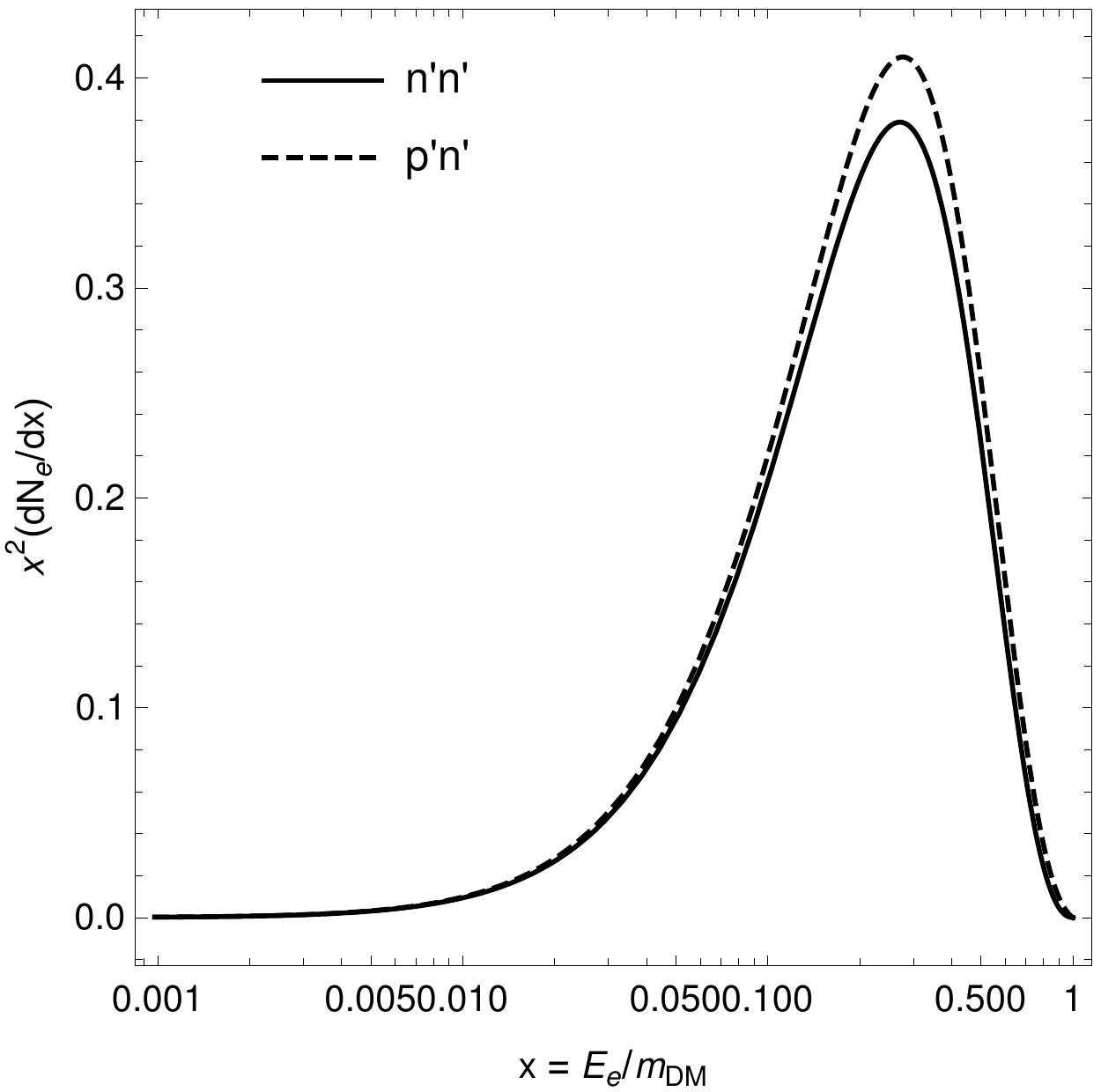}
	\caption{
	The $e^-/e^+$ spectrum predicted from the $n'\bar{n}'$ (Solid line) and $p'\bar{n}'$ (Dashed line) annihilation for 
	$m_e\ll m_{\gamma'}\ll m_{\pi'}\ll m_{\mathrm{DM}}$. 
	}
	\label{fig:SpectraEP}
\end{figure}

The energy spectrum of $e^++e^-$ at production 
by the late-time ADM 
annihilation is obtained by replacing $d\tilde{N}_\gamma/dx_0$ in Eq.\,\eqref{eq:dNdx0} with
the $e^+/e^-$ spectrum in the dark photon rest frame,
\begin{align}
    \frac{d\tilde{N}_e}{dx_0} = \delta(x_0 - 1)\ .
\end{align}
Here, $x_0 = 2E_0/m_{\gamma'}$ with $E_0$ being 
the energy of either $e^-$ or $e^+$.
By repeating the same analysis in the previous section,
we can convert this spectrum to the one in the rest frame 
of the ADM annihilation.
In Figure~\ref{fig:SpectraEP}, we show the $e^+/e^-$ spectrum at production for 	$m_e\ll m_{\gamma'}\ll m_{\pi'}\ll m_{\mathrm{DM}}$. 

For a given $e^+/e^-$ spectra at production, the interstellar $e^++e^-$ flux at around the location of the Earth is 
given by \cite{Cirelli:2010xx,Buch:2015iya},%
\footnote{A typical propagation time of the 
cosmic ray to travel of $\order{1}$\,kpc is 
much shorter than the 
age of the universe.}
\begin{align}
\frac{d\Phi}{dE_e} = \frac{1}{4\pi b(E)}
\left(\frac{\rho_{\mathrm{DM}}}{m_{\mathrm{DM}}}\right)^2
\left(\frac{t_0}{t_{\mathrm{osc}}}\right)^2
\sum_{i=n'\bar{n}',p'\bar{n}'}\langle\sigma v \rangle_{i} \int_E^{E_{\rm DM}}dE_s\,
I(E,E_s)\,\frac{dN_{e\,i}}{dE}(E_s)\ .
\end{align}
Here, $\rho_{\mathrm{DM}}$ denotes a local dark matter 
density at around the location of the Earth, 
$I(E,E_s)$ is a Green function which encodes the propagation of 
$e^\pm$ from a source with a given energy $E_s$ to 
any energy $E$, and $b(E)$ is the $e^\pm$ energy loss function.%
\footnote{The Green function is dimensionless while  $b(E)$ has a unit of GeV/sec
which is typically 
$b(E)\simeq 10^{-(16-15)}$\,GeV/sec 
for $E=\order{10}$\,MeV to $\order{1}$\,GeV~\cite{Cirelli:2010xx,Buch:2015iya}.
}
\begin{figure}
	\centering
	\includegraphics[width=8cm,clip]{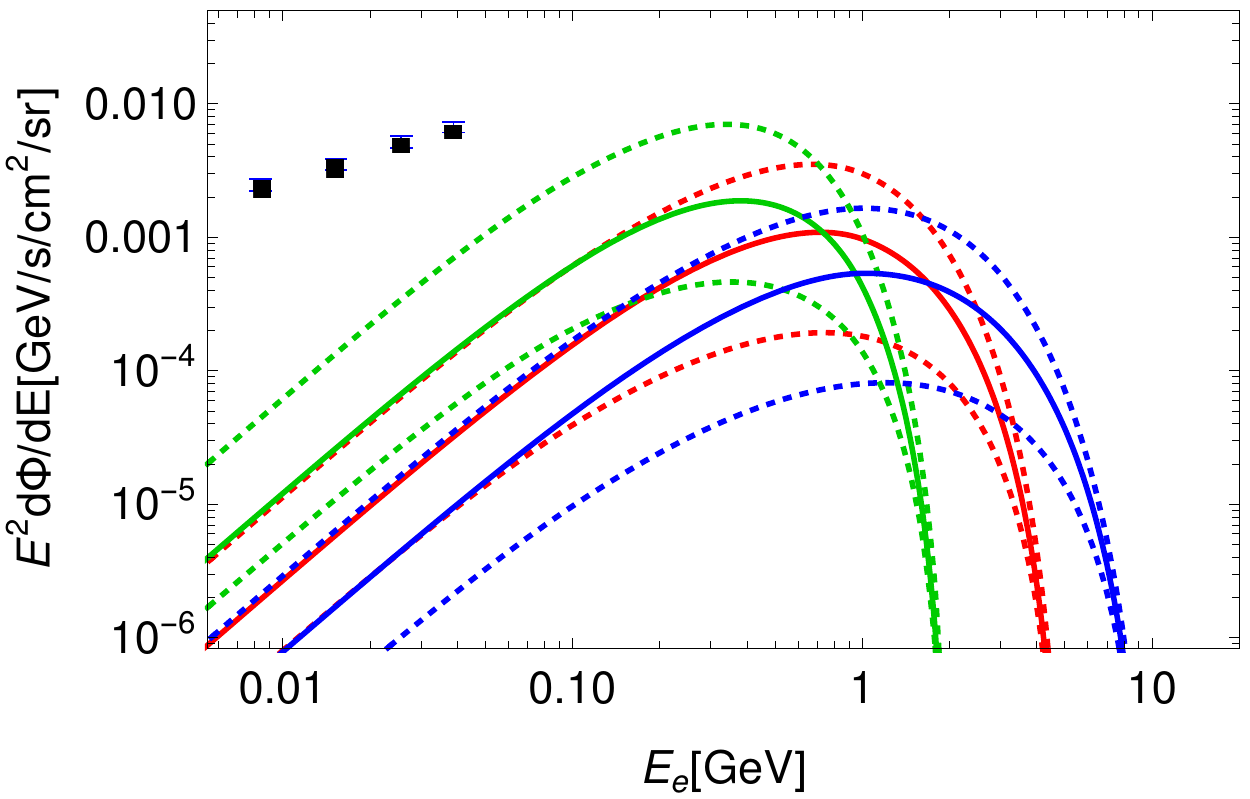}
	\caption{The interstellar $e^++e^-$ flux in cosmic ray at around the location of the Earth from the late-time ADM annihilation. 
	The annihilation cross section and the osicllation time scale is taken to be 
	$(t_0/t_{\mathrm{osc}})^2\times\langle\sigma v\rangle = 1$\,pb ($t_{\mathrm{osc}}\gg t_0$).
	The green, red, and blue lines show the spectrum for $m_{\mathrm{DM}} = 3$\,MeV, $5$\,MeV, and $10$\,MeV, respectively. 
	The solid lines assume the MED propagation model,
	while the upper and the lower dotted lines assume 
	the MAX and the MIN propagation models, respectively.
	The NFW halo profile is used.
	}
	\label{fig:CR}
\end{figure}

In Figure~\ref{fig:CR}, we show the 
interstellar $e^++e^-$ flux at around the location of the Earth from the late-time ADM annihilation. 
Here, the annihilation cross section and the oscillation time scale is set to be $(t_0/t_{\mathrm{osc}})^2\times\langle\sigma v\rangle = 1$\,pb.
The Green function, $I(E,E_s)$, and the energy loss rate, $b(E)$, are those provided by \cite{Cirelli:2010xx,Buch:2015iya}.
In the figure, the solid lines assume the MED propagation model,
while the upper and the lower dotted lines assume 
the MAX and the MIN propagation models, respectively (see \cite{Donato:2003xg}).
The dark matter profile is assumed to be the NFW profile~\cite{Navarro:1996gj},%
\footnote{We numerically checked that the
spectra are not significantly changed even 
for a cored Burkert profile \cite{Burkert:1995yz},
though they are slightly suppressed.}
with the local dark matter density at around the Earth
to be $\rho_{\mathrm{DM}} = 0.3$\,GeV/cm$^3$.

In the figure, we also show the interstellar $e^++e^-$ spectrum observed by the Voyager-1~\cite{Fisk:1976aw,Stone2013}, where the data is taken from \cite{Maurin:2013lwa}. 
The figure shows that the $e^++e^-$ flux 
from the late-time ADM annihilation is much smaller
than the observed flux for $(t_{0}/t_{\mathrm{osc}})^2\times \langle \sigma v\rangle = \order{1}$\,pb.
We will summarize the constraints from the Voyger-1 
in the next subsection.

\subsection{Constraints on Parameter Space}
\label{sec:constraint}
\begin{figure}
	\centering
	\includegraphics[width=8cm,clip]{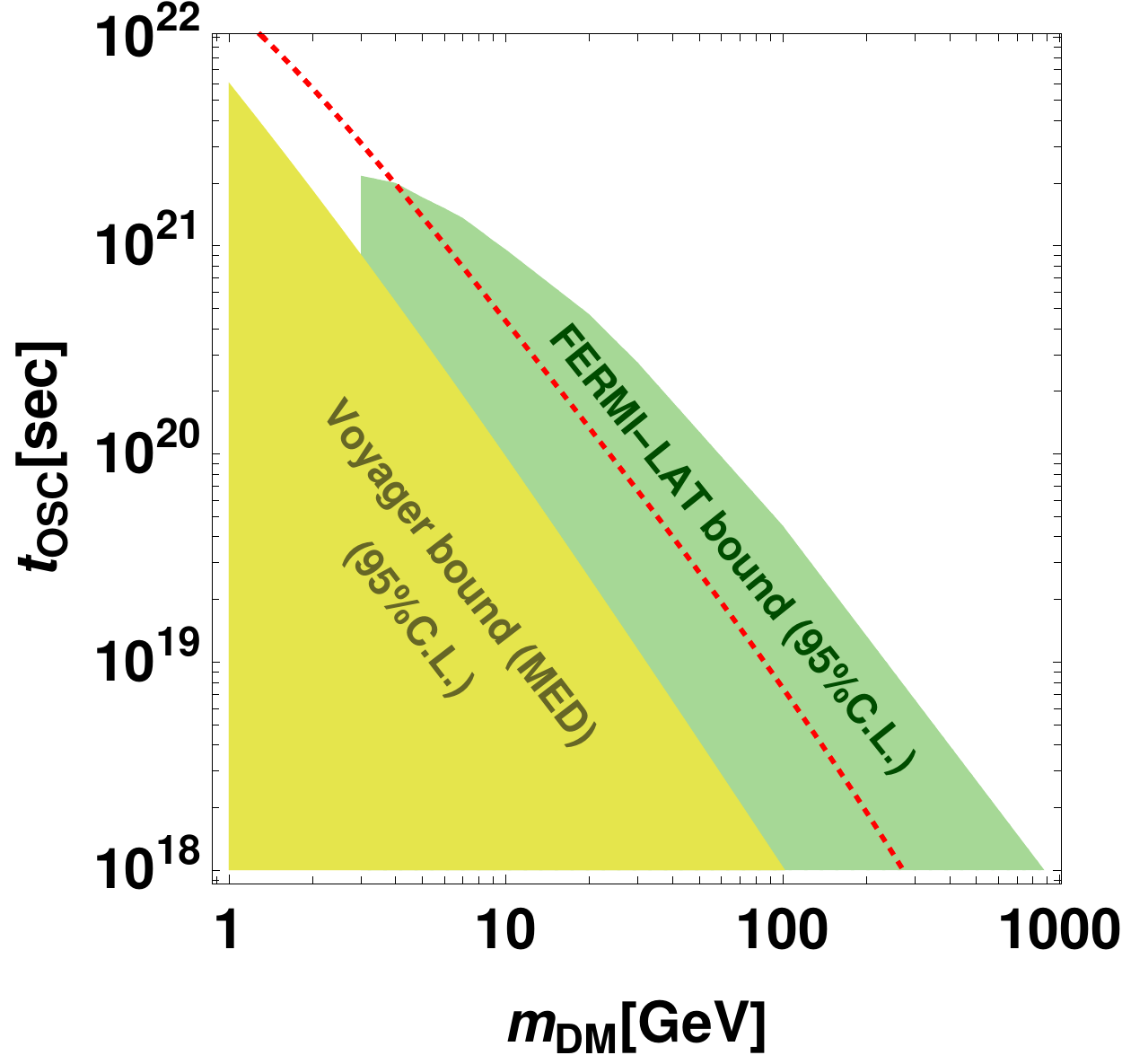}
	\caption{
	Constraints on the oscillation time scale. 
	The green and the yellow shaded regions are excluded by the $\gamma$-ray constraint (Fermi-LAT),
	and by the constraint on the $e^++e^-$ flux (Voyager-1), respectively. 
	We here assume $m_{n'}=m_{p'}=m_{\rm{DM}}$ and take $n_{n'}=n_{p'}$.
	We also assume $m_{\pi'}\ll m_{\mathrm{DM}}$ while fixing $m_{\gamma'}=40\,\mbox{MeV}$. 
	The red dotted line is a prospected lower limit by the $\gamma$-ray search 
	from the dSphs by e-ASTROGAM in one year of effective exposure.
	}
	\label{fig:tosc}
\end{figure}

As we have seen in the previous subsections, 
we can probe the time scale of the matter-antimatter oscillation by the $\gamma$-ray observation up to $t_{\mathrm{osc}}=\order{10^{21}}$\,sec for $m_{\mathrm{DM}}\simeq 10$\,GeV. 
This oscillation time scale corresponds to the effective annihilation cross section,%
\footnote{The effective cross section into the $\gamma$-ray is further suppressed by Eq.\,\eqref{eq:dNdx0}.}
\begin{align}
    \left(\frac{t_0}{t_{\rm osc}}\right)^2 \langle\sigma v\rangle \sim 10\,{\rm pb} \left(\frac{10\,{\rm GeV}}{m_{\rm DM}}\right)^2
    \left(\frac{10^{21}\rm {sec}}{t_{\rm osc}}\right)^2\ .
    \label{eq:effectiveCS}
    \end{align}
A lighter ADM can be also tested by the observation of the interstellar $e^++e^-$ flux.

In Figure~\ref{fig:tosc}, we show the constraints on the oscillation time scale from 
the observations by the Fermi-LAT and the Voyager-1.
Here, we assume $m_e\ll m_{\gamma'} \ll m_{\pi'}\ll m_{\mathrm{DM}}$ while we fix $m_{\gamma'}=40$\,MeV.%
\footnote{The constraints do not depend on $m_{\gamma'}$ significantly, as long as $m_e\ll m_{\gamma'} \ll m_{\pi'}\ll m_{\mathrm{DM}}$.}
The green region corresponds to the $95\%$ C.L. excluded region from the Fermi-LAT 
observations (see also \cite{Fermi-LAT:2016uux,Elor:2015bho}),
where we take into account the $\gamma$-ray fluxes
from the 8-classical dSphs.
The yellow shaded region corresponds to the $95\%$ C.L. excluded region from the Voyager-1 observation
for the MED propagation model with the NFW dark halo profile.
We see that, for $m_{\rm{DM}}\simeq 5\mbox{--} 10\,\mbox{GeV}$, the more stringent 
constraints are put by the Fermi-LAT observation, where the 
oscillation time scale shorter than $t_{\mathrm{osc}}\sim 10^{21}\,\mbox{sec}$ is excluded.
For a lighter mass region,
the Voyager-1 observation excludes the oscillation time scale shorter than 
$t_{\mathrm{osc}}\sim 10^{21-22}$\,sec.

\begin{figure}[t]
	\begin{minipage}{0.4\hsize}
	\centering
	\includegraphics[width=0.95\hsize,clip]{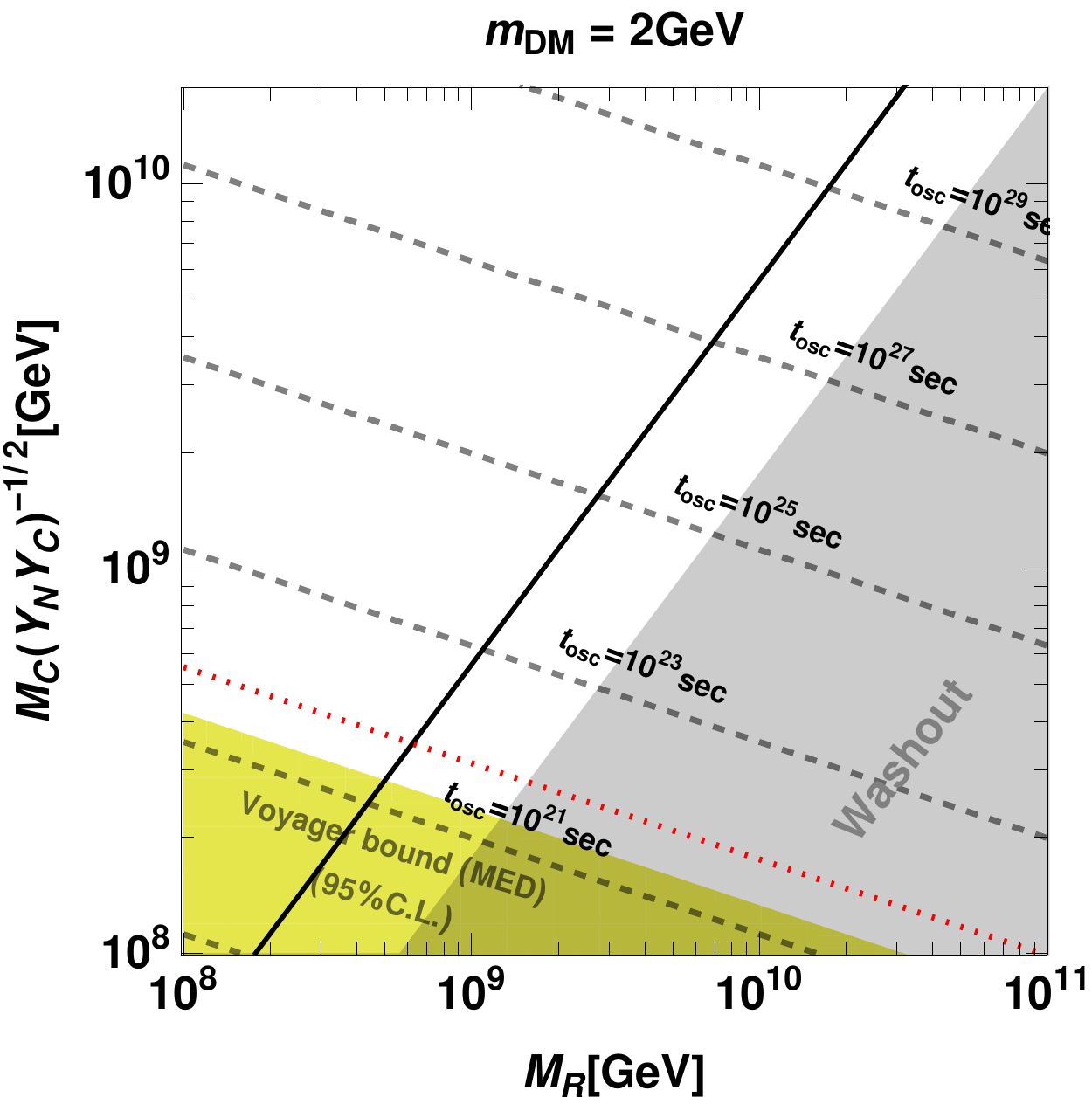}
    \end{minipage}
    \begin{minipage}{0.4\hsize}
	\centering
	\includegraphics[width=0.95\hsize,clip]{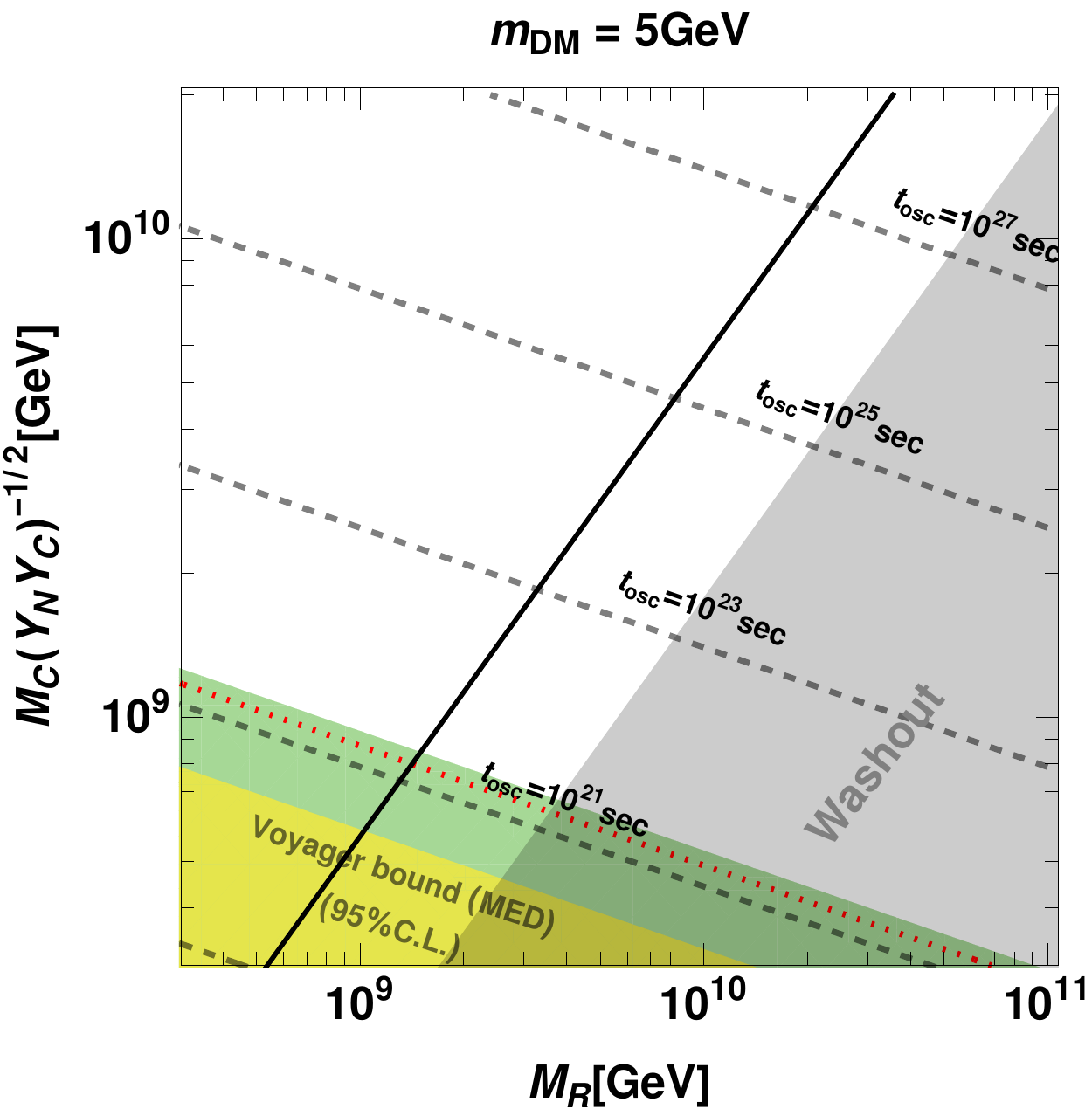}
    \end{minipage}
    \begin{minipage}{0.4\hsize}
	\centering
	\includegraphics[width=0.95\hsize,clip]{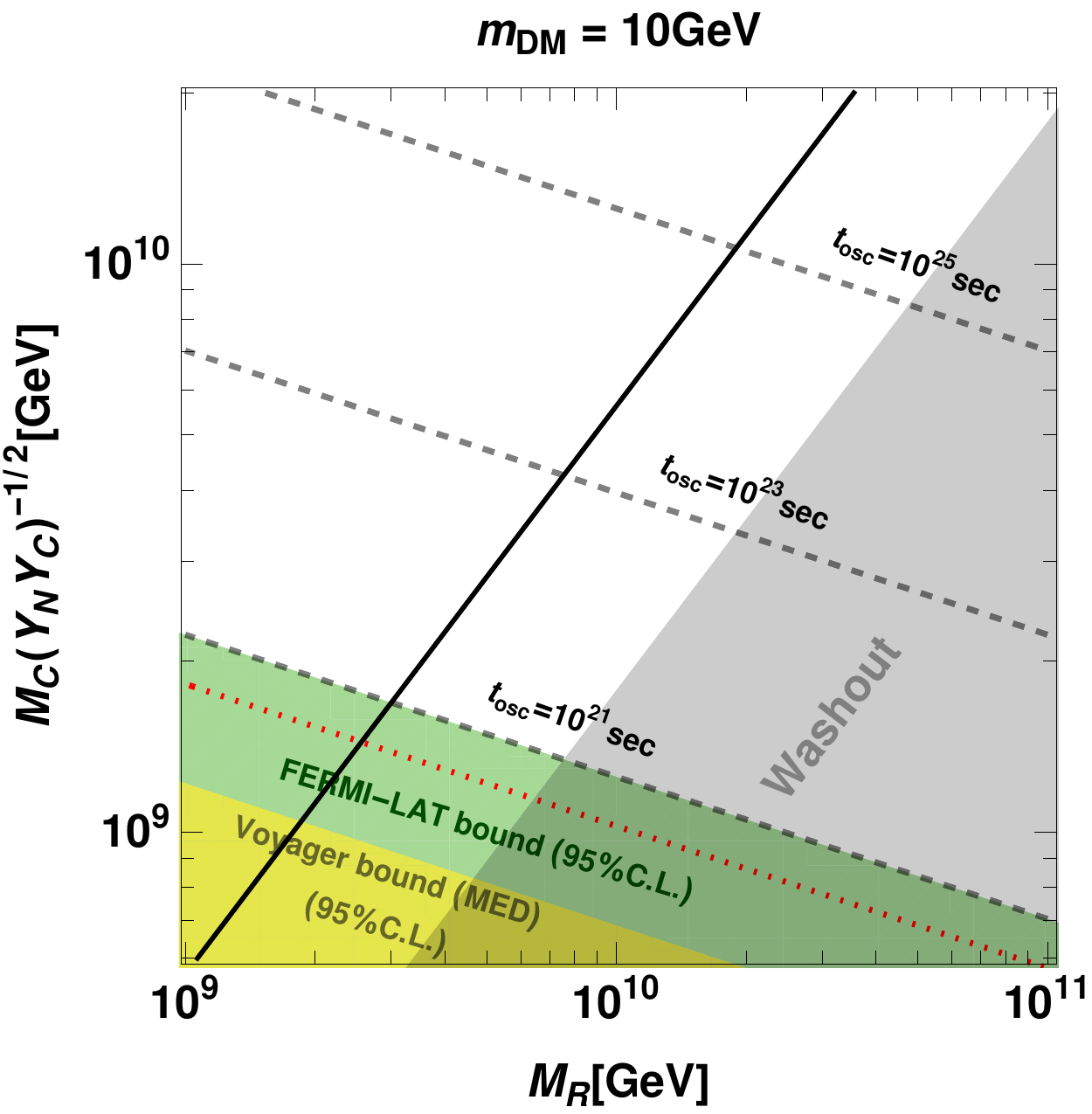}
    \end{minipage}
	\caption{
	Constraint on our ADM scenario for $m_{\mathrm{DM}}=2,5,10$\,GeV.
	The green and yellow shaded regions correspond the $95\%$ C.L. excluded regions by 
	the Fermi-LAT and the Voyager-1, respectively. 
	The lower gray region is excluded 
	in which the $B-L$ asymmetry is washed out (see Eq.\,\eqref{eq: lower bound on M_C}).
	Above the solid line, we require an on-shell $B-L$ portal sector (see Eq.\,\eqref{eq:scenario2}).
	We also show the prospected limits by e-ASTROGAM translated from Figure~\ref{fig:tosc}.}
	\label{fig:limit}
\end{figure}

In Figure~\ref{fig:limit},
we translate the constraints on the oscillation time scale to those on the 
parameters of the present model.
In the figure, we consider $m_{\mathrm{DM}}=2,5,10$\,GeV.
We also take  $\Lambda'_{\rm{QCD}}=2\,\mbox{GeV}\times(m_{\mathrm{DM}}/10\,\mathrm{GeV})$
to mimic QCD for each choice of the dark matter mass.
We also assume  $m_e \ll m_{\gamma'} \ll m_{\pi'}\ll m_{\mathrm{DM}}$.
The green and yellow shaded regions correspond the $95\%$ C.L. excluded regions by 
the Fermi-LAT and the Voyager-1, respectively. 
The lower gray region is excluded 
where the $B-L$ asymmetry is washed out (see Eq.\,\eqref{eq: lower bound on M_C}).
Above the solid line, we require an on-shell $B-L$ portal sector (see Eq.\,\eqref{eq:scenario2}).

We now see that the composite ADM scenario 
with $M_C = {\mathcal O}(10^9)$\,GeV can be tested by the $\gamma$-ray searches from the dSphs by the Fermi-LAT for $m_{\rm DM}\simeq 10$\,GeV.
Even for a lighter ADM scenario, we see that the region with $M_C = \order{10^8}$\,GeV
has been excluded by the Voyager-1 observation.
The resultant constraint is important in view of the fact that the parameter region with $M_C\sim 10^{9}$--$10^{10}$\,GeV
is highly motivated in the UV completion 
model based on $SU(4) \,(\supset SU(3)_{D}\times U(1)_D)$ gauge theory~\cite{Ibe:2018juk,Ibe:2018tex,Ibe:2019ena}.
In this UV completion, the tiny kinetic mixing of $\epsilon = 10^{-10}$--$10^{-9}$ which evades all the phenomenological constraints on the dark photon~\cite{Ibe:2018juk}
is achieved when the $SU(4)$ breaking scale 
is at around $10^{9}$--$10^{10}$\,GeV.
The $SU(4)$ breaking scale also leads 
to the colored dark Higgs mass in a similar range.
The $\gamma$-ray searches are already sensitive to such a well-motivated parameter region 
for $m_{\mathrm{DM}}\simeq 10$\,GeV.

Several comments are in order.
In our discussion, we consider only the $\gamma$-ray emitted by the FSR. 
This should be justified as the $\gamma$-rays made by the Synchrotron radiation and the inverse Compton scattering 
from the sub-GeV $e^+/e^-$ are very soft and below the Fermi-LAT sensitivity~\cite{Cirelli:2010xx}.
It should be also noted that the $\gamma$-ray signal from the galactic center does not lead to more stringent constraints,
despite the signal strength is higher than that from the dSphs.
This is because the $\gamma$-ray background is much higher 
for the galactic center, and hence, it is difficult to 
distinguish the continuous signal spectrum from the background spectrum.

Future $\gamma$-ray searches 
such as 
e-ASTROGAM~\cite{DeAngelis:2017gra,Rando:2019fzq},
SMILE~\cite{smile},
GRAINE~\cite{Aoki:2012nn},
and 
GRAMS~\cite{Aramaki:2019bpi}
projects will be important to test the model further.
It should be emphasized that those experiments 
are sensitive to the MeV $\gamma$-rays, and hence,
they are also able to test the models with $m_{\mathrm{DM}}=$\,a few GeV to which the Fermi-LAT loses 
sensitivity.
In Figure~\ref{fig:tosc}, we show the prospected lower limit on $t_{\mathrm{osc}}$ at 95\%CL 
by the  $\gamma$-ray search 
from the dSphs by e-ASTROGAM in one year of effective exposure.
In our analysis, we used the effective area and the prospected sensitivities for a $\gamma$-ray flux
from a point-like source at a high latitude (in Galactic coordinates) in \cite{DeAngelis:2017gra}.
The testable parameter region can be wider when the $J$-factors of the 
ultra-faint dSphs are determined more precisely by 
future spectroscopic observations such as the Prime Focus Spectrograph~\cite{Ellis:2012rn}.
For example, if the $J$-factor of Triangulum II converges to the central value in \cite{Hayashi:2016kcy},
i.e. $\log_{10} J \simeq 20$, the prospected lower limit on $t_{\mathrm{osc}}$
becomes higher for about a factor of $2^{1/2}$.

\section{Conclusions}
The composite ADM model is particularly motivated as it provides the DM mass of ${\cal O}(1)$\,GeV 
and a large annihilation cross section simultaneously.
In this paper, we discussed the indirect detection of the composite ADM where the portal operators 
of the $B-L$ asymmetry is generated in association with the seesaw mechanism.
In this model, the dark-neutron obtains a tiny Majorana mass,
and hence, ADM can pair-annihilate at later times.

As we have discussed, the late time annihilation of ADM results in multiple soft electrons/positrons and soft photons emitted as the FSR.
As a result, some parameter region of the composite ADM which 
is motivated by thermal leptogenesis and dark UV completion models
has been excluded by the Fermi-LAT and the Voyager-1 observations.
The obtained constraint is tighter than that from  the anti-neutrino flux  made by the 
decay of  ADM via the $B-L$ portal operator~\cite{Fukuda:2014xqa} (see Eq.\,\eqref{eq:lifetime}).
Future experiments which are 
sensitive to sub-GeV $\gamma$-rays such as 
e-ASTROGAM~\cite{DeAngelis:2017gra,Rando:2019fzq},
SMILE~\cite{smile},
GRAINE~\cite{Aoki:2012nn},
and 
GRAMS~\cite{Aramaki:2019bpi}
projects will be important to test the oscillating ADM model
further.

\section*{Acknowledgements}
\vspace{-.3cm}
MI thanks K.~Kohri for useful comments on the possible constraints on the model from the cosmic electron/positron rays.
This work is supported by JSPS KAKENHI Grant Numbers, 19K14701 (R. N.),  No. 15H05889, 
No. 16H03991, NO. 17H02878 
and No. 18H05542 (M. I.) and by World Premier International Research Center Initiative (WPI Initiative), MEXT, Japan. 
This work is also supported by the Advanced Leading Graduate Course for Photon Science (S. K.).
\appendix

\section{Final State Radiation In the Dark Photon Decay}
\label{sec:FSR}

\begin{figure}[htbp]
  \begin{center}
    \includegraphics[width=60mm]{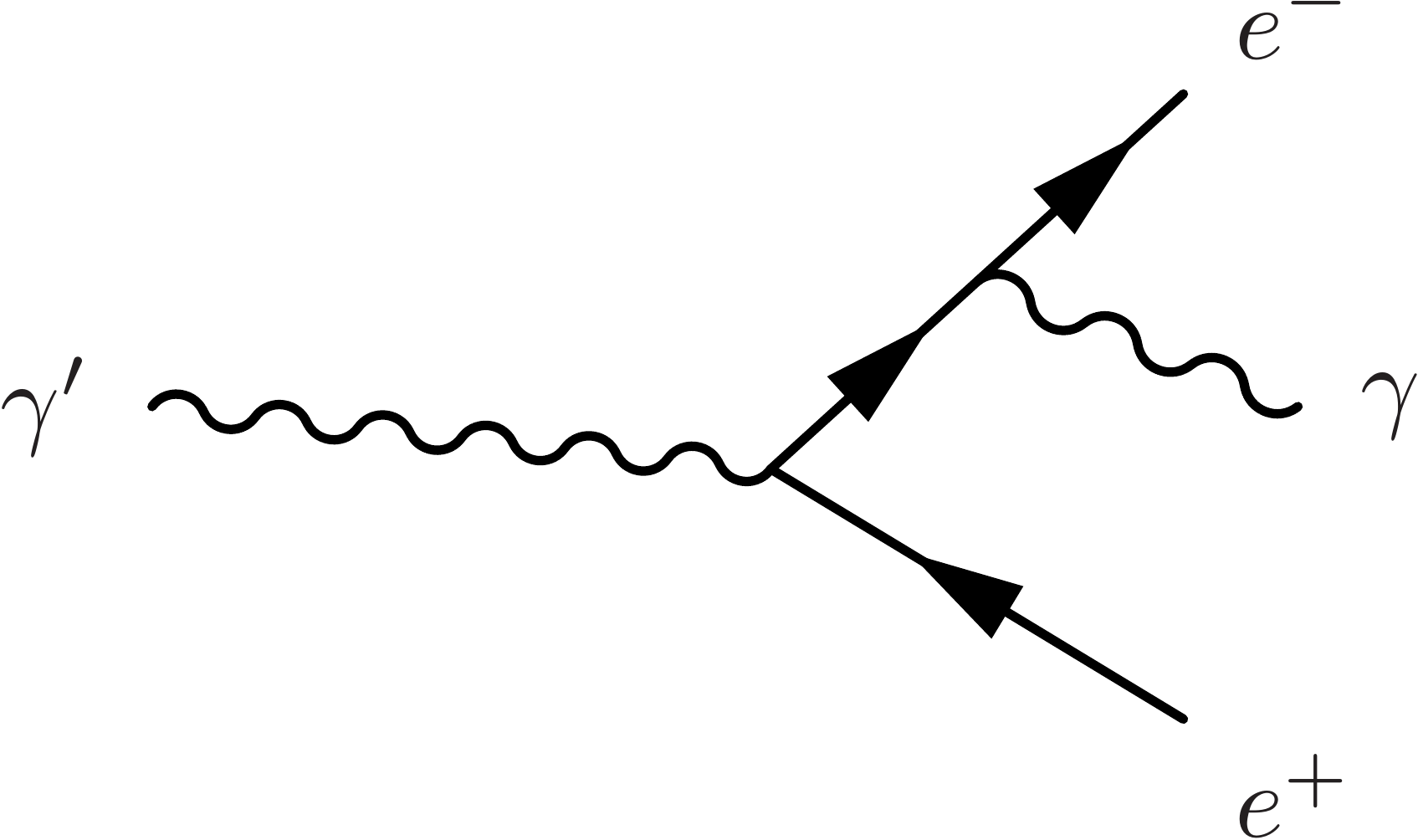}
  \end{center}
   \caption{One of the Feynman diagrams of the final state radiation.}
   \label{fig:g2eeg}
\end{figure}

This appendix is devoted to the photon energy spectrum of the final state radiation in the 
dark photon decay, $\gamma' \rightarrow e^+e^-\gamma$.
One of the diagrams is shown in the figure \,\ref{fig:g2eeg}.

The invariant amplitude for this process is 
\begin{align}
\mathcal{M}=& - 4\pi \epsilon \alpha_{\mathrm{EM}} \bar{u}(p_{1}) \left[ \slashed{\varepsilon}^*(p_3) \frac{\slashed{p}_1+\slashed{p}_3+m_{e}}{(p_1+p_3)^2-m_{e}^2}
\slashed{\varepsilon}(p_0) + \slashed{\varepsilon}(p_0) \frac{-\slashed{p}_2-\slashed{p}_3+m_{e}}{(p_2+p_3)^2-m_{e}^2} \slashed{\varepsilon}^*(p_3) \right] v(p_2) \, ,
\end{align}
where $\epsilon$ represents the strength of kinetic mixing, $\alpha_{\mathrm{EM}}$ the fine structure constant
of QED, $\varepsilon$ the polarization vector, $m_e$ the electron mass, $u$ and $v$ spinors and $p$ momentum vector.
Here the subscripts $(0, 1, 2, 3)$ denote the $(\gamma', e^{-}, e^+, \gamma)$.

Summing over the spins of the final state $e^-,\, e^+$ and averaging over the helicity of initial state $\gamma'$, we obtain
\begin{align}
\frac{1}{3}\sum_{\mathrm{spin}}\abs{\mathcal{M}}^2=& \frac{8 (4\pi \epsilon \alpha_{\mathrm{EM}})^2}{3} \frac{1}{(m_{13}^2-m_e^2)^2 (m_{23}^2-m_e^2)^2} \left[ m_{13}^2 m_{23}^2 \{ 2m_{12}^4 + 2m_{12}^2 (m_{13}^2+m_{23}^2) +m_{13}^4 + m_{23}^4 \}  \right. \notag \\
& - m_e^2 (m_{13}^2+m_{23}^2) \{ 2m_{12}^4+4 m_{12}^2 (m_{13}^2+m_{23}^2) +3 (m_{13}^2+m_{23}^2)^2 \} \notag \\
& + m_e^4 \{ 2m_{12}^4 + 10 m_{12}^2 (m_{13}^2+m_{23}^2) + 11 (m_{13}^2+m_{23}^2)^2 \} \notag \\
& \left. - 4 m_e^6 \{ 2m_{12}^2 + 3 (m_{13}^2+m_{23}^2 ) \} + 2 m_e^8 \right] \ ,
\end{align}
by using the Mandelstam invariants, $m_{ij}^2 =(p_i-p_j)^2$, with the subscripts defined above.
There is a relation between the invariants,
$m_{\gamma'}^2 + 2 m_e^2 = m_{12}^2+m_{13}^2+m_{23}^2$, with $m_{\gamma'}$ being
the dark photon mass.
This expression is symmetric under the exchange between $m_{13}^2$ and $m_{23}^2$ as expected.

Now, let us calculate the decay rate with the final state radiation.
In the following calculation, we use the center of mass frame in which three out-going particles lie in a same plane. 
Thus, we can transform the three-body phase space integral into integration over the energy of two particles and three angles. 
By taking into account of the energy-momentum conservation, the three-body phase space has $9-4=5$ d.o.f.
After fixing the energy of $e^-$, three d.o.f. remain.
Two of them are angles $(\alpha, \beta)$ that specify the direction of $\vec{p}_3$.
The last one is an angle $\delta$ which determines the plane of decay around $\vec{p}_3$.
Thus, $\Gamma_{\gamma'\to e^+e^-\gamma}$ can be written as
\begin{align}
			\Gamma_{\gamma'\to e^+e^-\gamma}&=\int \frac{1}{16m_{\gamma'}}\frac{1}{3}\sum_{\mathrm{spin}}\abs{\mathcal{M}}^2\frac{dE_3 dE_1 d\alpha d(\cos\beta) d\delta }{(2\pi)^5} \ , \\
&=\frac{m_{\gamma'}}{32(2\pi)^3}\int \frac{1}{3}\sum_{\mathrm{spin}}\abs{\mathcal{M}}^2 dx dy  \ , \\
																	&=\frac{m_{\gamma'}}{32(2\pi)^3}\int dx \sum_{n=0}^{2} \epsilon_0^{2n} \left[ f_n(x,y_{\mathrm{max}}(x,\epsilon_0)) - f_n(x,y_{\mathrm{min}}(x,\epsilon_0)) \right] \ .
\end{align}
Here we define $x=2E_3/m_{\gamma'},\, y=2E_1/m_{\gamma'}$ and $\epsilon_0 = 2 m_e/m_{\gamma'}$.
Each $f_n(x,y)$ is defined as the integration of the invariant scattering amplitude over $E_1$, i.e., $y$.
The analytical formula for each $f_n(x,y)$ is as follows:
		\begin{align}
			&f_0(x,y)=\frac{8}{3} (4\pi \epsilon \alpha_{\mathrm{EM}})^2 \qty[2(1-y)-\frac{1+(1-x)^2}{x}\ln\qty(\frac{1-y}{1-x-y})] \ , \\
			&f_1(x,y)=\frac{4}{3} (4\pi \epsilon \alpha_{\mathrm{EM}})^2 \qty[\frac{x+2y-2}{(1-y)(1-x-y)}+2\ln\qty(\frac{1-y}{1-x-y})] \ , \\
			&f_2(x,y)=\frac{2}{3} (4\pi \epsilon \alpha_{\mathrm{EM}})^2 \qty[\frac{x+2y-2}{(1-y)(1-x-y)}+\frac{2}{x}\ln\qty(\frac{1-y}{1-x-y})] \ .
		\end{align}
Here $y_{\mathrm{min}}$ and $y_{\mathrm{max}}$ are the lower and the upper bounds of the integration region of $y$ corresponding to the Dalitz region.
The explicit forms of $y_{\mathrm{min}}$ and $y_{\mathrm{max}}$ are
		\begin{align}
			&y_{\mathrm{min}}=\mathrm{max}\qty[1-\frac{x}{2}-\frac{x}{2}\sqrt{1-\frac{\epsilon_0^2}{1-x}},\epsilon_0] \ , \\
			&y_{\mathrm{max}}=\mathrm{min}\qty[1-\frac{x}{2}+\frac{x}{2}\sqrt{1-\frac{\epsilon_0^2}{1-x}},1] \ .
		\end{align}
		
	\begin{figure}
		\centering
		\includegraphics[width=0.6\linewidth]{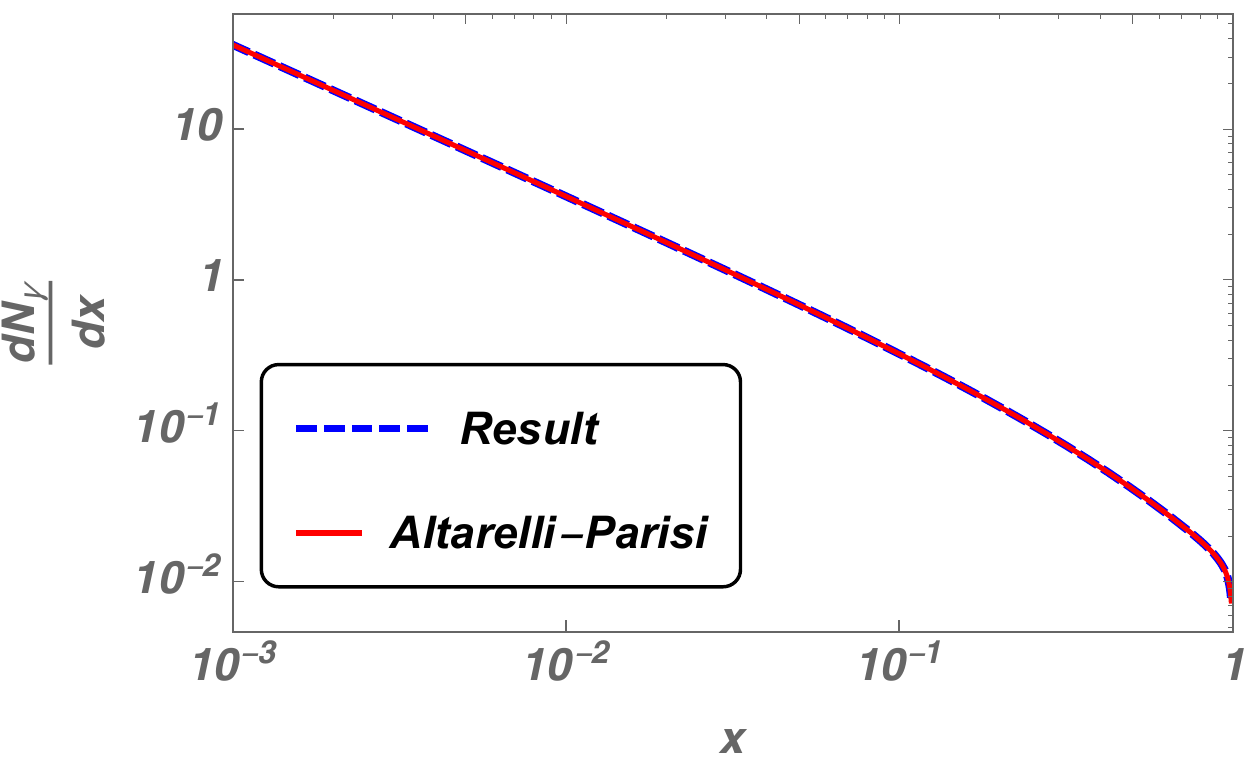}
		\caption{The plot of the analytic formula and the approximation. Here we take $m_{\gamma'} = 40$\,MeV. Two expressions are in good agreement.}
		\label{fig:compare}
	\end{figure}

From above, we obtain the energy spectrum of the final state radiation photon.
The energy spectrum is expressed as \cite{Elor:2015tva}
		\begin{align}
		\label{eq:eq analytic x0 spectrum}
			\frac{1}{N_\gamma}\dv{N_\gamma}{x}& = \frac{1}{\Gamma_{\gamma'\to e^+e^-}}\frac{d \Gamma_{\gamma'\to e^+e^-\gamma}}{dx} \ .
		\end{align}
	Here, $\Gamma_{\gamma'\to e^+e^-} = \frac{1}{3} \epsilon^2 \alpha_{\mathrm{EM}} m_{\gamma'}$ is the decay rate of the process $\gamma'\to e^+e^-$.
We compare the result with twice the Altarelli-Parisi approximation formula \cite{Mardon:2009rc}
		\begin{align}
		\label{eq:eq approx x0 spectrum}
			\frac{1}{\Gamma_{\gamma'\to e^+e^-}}\frac{d \Gamma_{\gamma'\to e^+e^-\gamma}}{dx}= \frac{\alpha_{\mathrm{EM}}}{\pi}\frac{1+(1-x)^2}{x}\ln\qty(\frac{4(1-x)}{\epsilon_0^2}) \ ,
		\end{align}
in the figure\,\ref{fig:compare}.
We take $m_{\gamma'} = 40$\,MeV.
We see that two formulae are in good agreement in a wide range of the photon momentum.
	
\section{Sommerfeld enhancement}
\label{sec:sommerfeld}
The dark pion exchange 
between the dark nucleons 
generates attractive/repulsive forces between 
them depending on their spins and the isospins.%
\footnote{Since the dark quark masses are assumed to be
much smaller than the dark dynamical scale, 
the dark sector possesses the isospin symmetry 
as in the case of the QCD in the SM sector.}
For example, one dark pion exchange results in a
static potential,
\begin{align}
    V({\mathbf r}) = \frac{g_A^{\prime 2}}{16\pi f_{\pi'}^2} (\tau_1\cdot \tau_2)
    (\sigma_1\cdot\mathbf{\partial})
    (\sigma_2\cdot\mathbf{\partial})
    \frac{1}{r}e^{-m_{\pi'}r}\ ,
    \label{eq:ope}
\end{align}
which goes like $1/r^3$ in the region of $r\ll m_{\pi'}^{-1}$.
This potential is obtained from the axial-current interaction,
\begin{align}
{\cal L} =\frac{g_A'}{f_{\pi'}}\, \partial^a
\pi' \bar{N}'\gamma_\mu\gamma_5\left(\frac{\tau^a}{2}\right)N'\ ,
\end{align}
where $f_{\pi'}$ is the decay constant of
the dark pion and $g'_A$ is the form factor of the dark nucleon axial current.%
\footnote{We take the normalization such that 
$f_\pi\simeq 93$\,MeV and $g_A\simeq 1.26$
in the case of the SM.
}
The spin and the isospin indices are implicit,
where ${\sigma}$ and $\tau$ denote the Pauli matrices
applying to the spin and the isospin 
of each nucleon, respectively.
The way of the isospin transition can be read off by noting $\tau_{1\,{ij}}\cdot \tau_{2,{k\ell}} =
2 (\delta_{i\ell}\delta_{jk}-\delta_{ij}\delta_{k\ell}/2)$.

As discussed in \cite{Bedaque:2009ri,Liu:2013vha,Bellazzini:2013foa}, the attractive potential forces mediated by the pseudo-scalar field causes 
the Sommerfeld enhancement of the dark matter annihilation~\cite{Sommerfeld,Hisano:2002fk,Hisano:2003ec,Hisano:2004ds}.
In this appendix, we discuss the Sommerfeld enhancement caused by the dark pion exchange.
In our analysis, we rely on the formalism of the Sommerfeld enhancement in \cite{Blum:2016nrz}, in which the lower cut-off on the relative velocity is taken into account in a self-consistent way.

Following \cite{Bellazzini:2013foa},
we approximate the potential by a spherical one,
\begin{align}
    V(\mathbf{r}) \simeq -\frac{g_A^{\prime 2}}{16\pi f_{\pi'}^2} \frac{1}{r^3}e^{-m_{\pi'}}\ ,
\end{align}
and estimate
the enhancemnt of the $s$-wave annihilation.%
\footnote{Strictly speaking, 
we need to solve a coupled equation 
between the states with angular momenta,
since the potential force in Eq.\,\eqref{eq:ope} changes the nucleon angular momentum by ${\mit\Delta}\ell =\pm2$.
} 
Under this approximation, the Sommerfeld enhancement factor can be obtained by solving the effective Schr\"odinger equation,
\begin{align}
    \left[-\frac{\nabla^2}{2m_{\mathrm{RED}}}+ V(\mathbf r) + u \delta^{(3)}(\mathbf{r})\right]\psi(\mathbf{r}) = \frac{p^2}{2m_{\mathrm{RED}}} \psi(\mathbf{r})\ .
\end{align}
Here, $m_{\mathrm{RED}} = m_{\mathrm{DM}}/2$ is the reduced mass and $p$ denotes relative momentum of the incident dark matter.
The boundary condition of the wave function $\psi(\mathbf{r})$ is taken to be
an incident plane wave with an outgoing spherical wave, i.e. $\psi(\mathbf{r})\to e^{ipz}+f e^{ipr}/r $ at $r\to \infty$.
The complex parameter $u$ encodes the annihilation cross section at a short distance without the Sommerfeld enhancement factor, 
i.e. $u = - i \sigma v_0/2$.%
\footnote{The dark-nucleon self-scattering due to short-range forces can be also encoded in the real part of $u$. In our analysis, 
we assume the self-scattering by short-range forces are subdominant and take $\Re u \simeq 0$. }
 
Since the potential goes to infinity faster than $r^{-2}$ at the origin, it must be regularized at short distances.
In our analysis, we introduce a short distance cutoff $r_0$ 
satisfying  $V(r_0) = m_{\mathrm{DM}}$ and regulate the scalar potential by replacing $V(r)\to V_{\mathrm{reg}}(r) = V(r+r_0)$~\cite{Liu:2013vha,Bellazzini:2013foa}.%
\footnote{Our conclusions do not depend on the choice of the regularization significantly.}
With the regulated potential, the Sommerfeld enhancement factor 
is given by~\cite{Blum:2016nrz},
\begin{align}
   S_{\mathrm{ENF}}(v) = \frac{\sigma v}{\sigma v_0} \simeq \frac{S(v)}
    {\left|1 - i \frac{m_{\mathrm{RED}}^2}{4\pi} \sigma v_0(T(v)+i S(v))v\right|^2} \ .
    \label{eq:SEFF}
\end{align}
Here, $T(v)$ and $S(v)$ are given by,
\begin{align}
    &T(v) = \frac{1}{p}\left(\Re \left.\frac{dg_{p}}{dr}\right|_{r=0}
    - 
   \Re \left.\frac{dg_{p_0}}{dr}\right|_{r=0} \right)\ , \\
    &S(v) = \frac{1}{p} \Im \left.\frac{d g_p}{dr}\right|_{r=0}\ ,
\end{align}
with the function $g_p(r)$ being a solution of 
\begin{align}
    \left[
    -\frac{d^2}{dr^2}
    +2 m_{\mathrm{RED}} V_{\mathrm{reg} }(r) - p^2
    \right] &g_{p}(r) = 0\ ,\\
    &g_{p}(0) = 1\ ,\\
    \lim_{r\to \infty}&g_{p}(r)\propto e^{ipr}\ .
\end{align}
The short distance cross section $\sigma v_0$ is 
fixed at a high momentum $p_0$.

\begin{figure}
\centering
\includegraphics[width=0.4\linewidth]{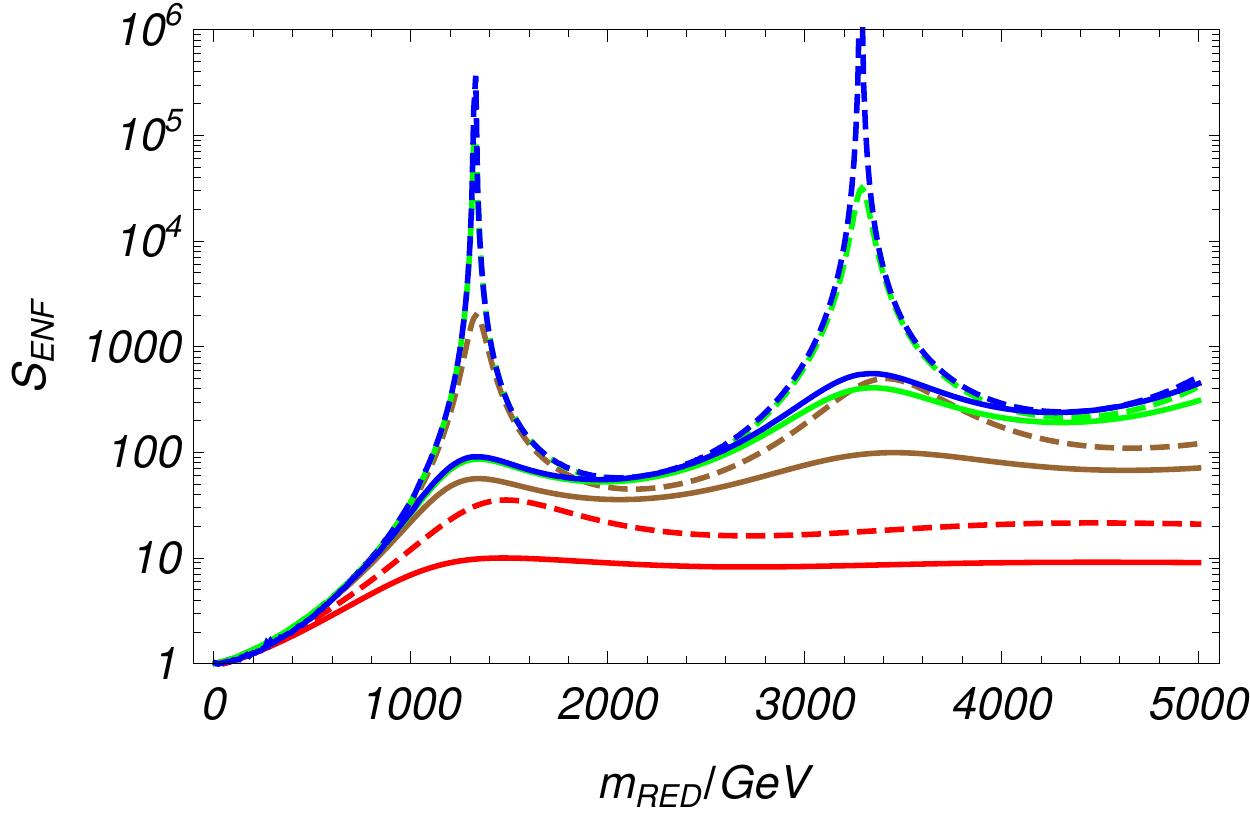}
\caption{The self-consistent Sommerfeld enhancement factor for an $s$-wave annihilation by the $1/r^3$ potential for $v= 10^{-1}$(red), $10^{-2}$(brown),
$10^{-3}$(green) and $10^{-5}$(blue). 
We take the same parameters with \cite{Bellazzini:2013foa} (Figure~3 in the reference) for comparison.
The	short-range annihilation cross section is assumed to be $\sigma v_0 = 4\pi/m_{\mathrm{DM}}^2$.
The solid lines are the enhancement factor in Eq.\,\eqref{eq:SEFF}, and the dashed ones are 
the naive enhancement factor $S(v)$.}
		\label{fig:SEFF1}
	\end{figure}

In Eq.\,\eqref{eq:SEFF}, the factor $S(v)$ corresponds to the naive Sommerfeld enhancement factor.
The denominator, on the other hand, provides an IR cutoff in the limit of $v\to 0$ with which the unitarity violation by the naive Sommerfeld enhancement factor is regulated self-consistently.
The regularization effect is particularly important 
when the short-distance cross section is large as in the case of the ADM scenario.
In Figure~\ref{fig:SEFF1},
we compare the naive enhancement factor shown in \cite{Bellazzini:2013foa} 
and the one in Eq.~\eqref{eq:SEFF} 
by assuming $\sigma v_0 = 4\pi/m_{\mathrm{DM}}^2$.%
\footnote{Due to a slightly different choice of $V_{\mathrm{reg}}(r)$, 
the positions of the resonances appearing in $S(v)$ are shifted from those in \cite{Bellazzini:2015nxw}.}
The figure shows that the enhancement factors 
at the resonances are significantly suppressed 
when the short-distance annihilation cross section is large.

\begin{figure}
\centering
\includegraphics[width=0.4\linewidth]{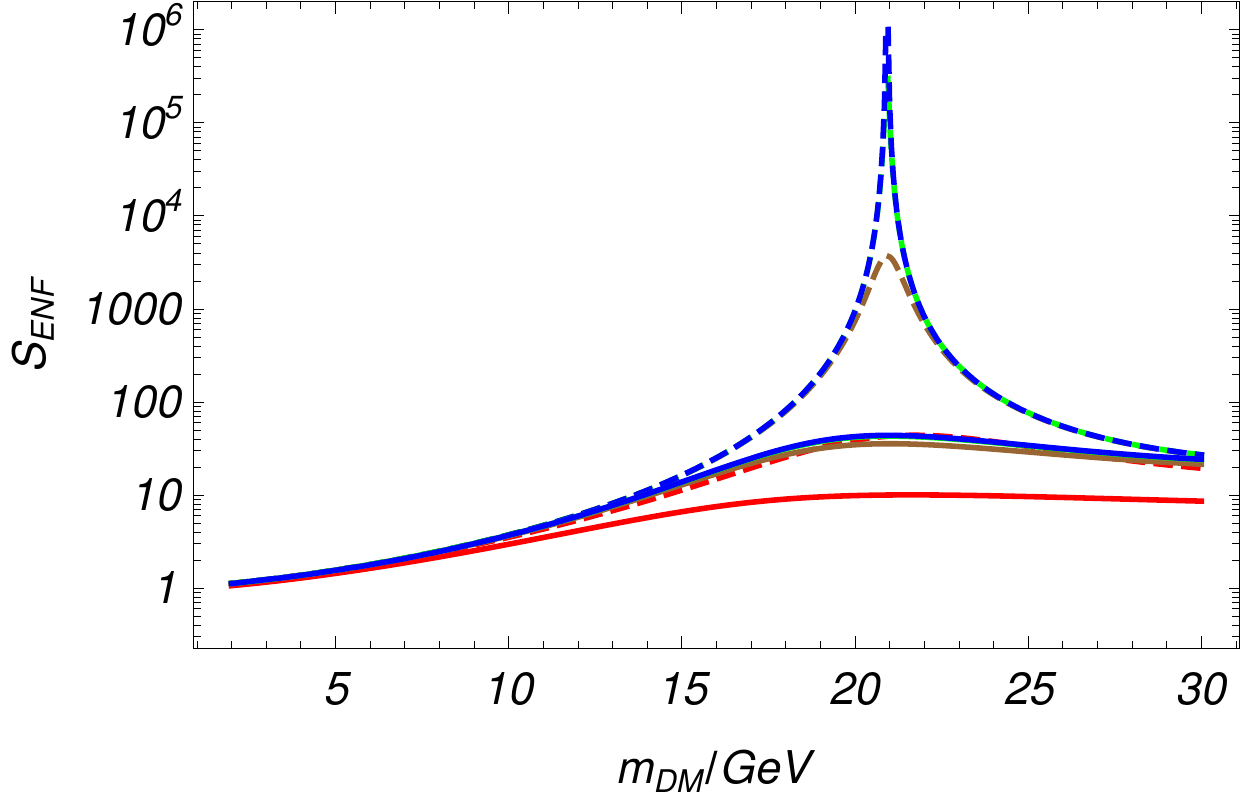}
\caption{The self-consistent Sommerfeld enhancement factor for an $s$-wave dark nucleon annihilation by the $1/r^3$ potential for $v= 10^{-1}$(red), $10^{-2}$(brown), $10^{-3}$(green) and $10^{-4}$(blue). 
The parameters are fixed to be $g_A = 1$, $f_{\pi'} = 1$\,GeV, and $m_{\pi'}=1$\,GeV.
The solid lines are the enhancement factor in Eq.\,\eqref{eq:SEFF}, and the dashed ones are 
the naive enhancement factor $S(v)$.}
		\label{fig:SEFFPI}
	\end{figure}

Now, let us apply Eq.\,\eqref{eq:SEFF} to the 
dark nucleon annihilation.
In Figure~\ref{fig:SEFFPI}, we show the Sommerfeld enhancement factor as a function of $m_{\mathrm{DM}}$ for $g_A = 1$, $f_{\pi'} = 1$\,GeV, and $m_{\pi'}=1$\,GeV.
The figure shows that the regularization effects
are important at around the resonance, $m_{\mathrm{DM}}\simeq 21$\,GeV.
The figure also shows that the Sommerfeld enhancement factor for the mass region of the ADM, 
$m_{\mathrm{DM}} \lesssim 10$\,GeV, is less significant.

As we fix the short-range cross section of the ADM, $\sigma v_0 \simeq 4\pi/m_{\mathrm{DM}}^2$,
to mimic the measured nucleon annihilation cross section at $v = \order{10^{-1}}$~\cite{Armstrong:1987nu,Bertin:1997gn},
the effective Sommerfeld enhancement factor corresponds to $S_{\mathrm{ENF}}(v)/S_{\mathrm{ENF}}(10^{-1})$.
The figure shows that the effective enhancement factor is close to unity for $m_{\mathrm{DM}} \lesssim 10$\,GeV.

\begin{figure}
\centering
\includegraphics[width=0.4\linewidth]{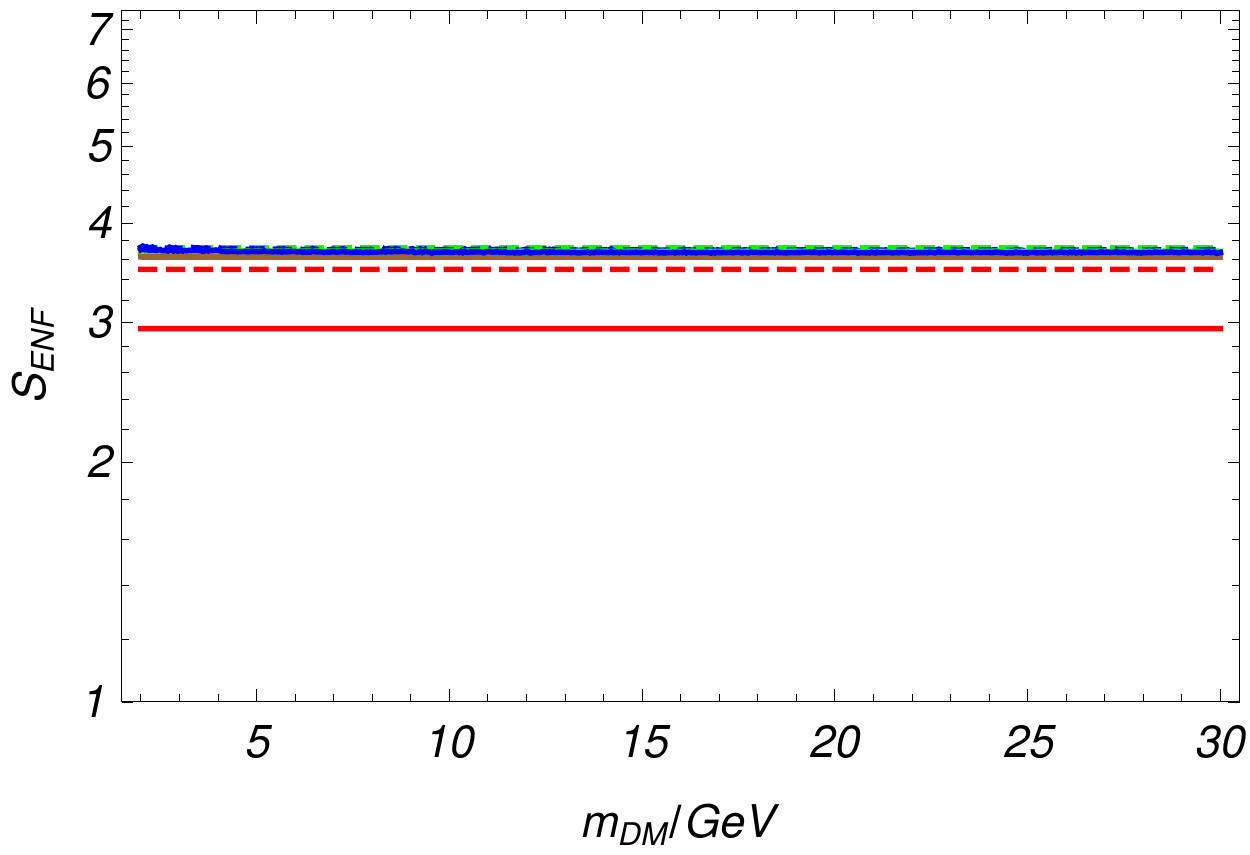}
\caption{The self-consistent Sommerfeld enhancement factor for an $s$-wave dark nucleon annihilation by the $1/r^3$ potential for $v= 10^{-1}$ (red), $10^{-2}$ (brown), $10^{-3}$ (green) and $10^{-4}$ (blue). 
The parameters are chosen to be 
$g_A = 1$, $f_{\pi'} = 0.1\times m_{\mathrm{DM}}$, and $m_{\pi'}=0.1\times m_{\rm DM}$.
The solid lines are the enhancement factor in Eq.\,\eqref{eq:SEFF}, and the dashed ones are 
the naive enhancement factor $S(v)$.}
		\label{fig:SEFFPI2}
	\end{figure}

In Figure~\ref{fig:SEFFPI2}, we also show the 
Sommerfeld enhancement factor for more realistic relations between the parameters,
\begin{align}
    f_{\pi'} = 0.1\times m_{\rm DM}\ , 
    \quad 
    m_{\pi'} = 0.1\times m_{\rm DM}\ ,
\end{align}
which mimic QCD.
The figure shows that no resonance appears 
when the parameters satisfy these relations.
As a result, we find that the effective enhancement factor, $S_{\mathrm{ENF}}(v)/S_{\mathrm{ENF}}(10^{-1})$, is of $\order{1}$.%
\footnote{The Sommerfeld enhancement
for coupled channels between different 
angular momenta requires more careful analysis.
However, as the centrifugal barriers of the higher angular momenta make the attractive potential wells shallower 
and smaller in spatial size, 
the resonances are expected to appear at a 
higher dark nucleon mass than those 
for $\ell = 0$.
Thus, the coupled equations do not lead to resonances in the mass range $m_{\mathrm{DM}}\lesssim 10$\,GeV.}
We also numerically confirmed that the results do not depend on the dark pion mass as long as it is much lighter than the dark nucleon.
Therefore, we conclude that the Sommerfeld enhancement is not significant in the present setup.

\bibliography{papers}

\begin{thebibliography}{92}%
\makeatletter
\providecommand \@ifxundefined [1]{%
 \@ifx{#1\undefined}
}%
\providecommand \@ifnum [1]{%
 \ifnum #1\expandafter \@firstoftwo
 \else \expandafter \@secondoftwo
 \fi
}%
\providecommand \@ifx [1]{%
 \ifx #1\expandafter \@firstoftwo
 \else \expandafter \@secondoftwo
 \fi
}%
\providecommand \natexlab [1]{#1}%
\providecommand \enquote  [1]{``#1''}%
\providecommand \bibnamefont  [1]{#1}%
\providecommand \bibfnamefont [1]{#1}%
\providecommand \citenamefont [1]{#1}%
\providecommand \href@noop [0]{\@secondoftwo}%
\providecommand \href [0]{\begingroup \@sanitize@url \@href}%
\providecommand \@href[1]{\@@startlink{#1}\@@href}%
\providecommand \@@href[1]{\endgroup#1\@@endlink}%
\providecommand \@sanitize@url [0]{\catcode `\\12\catcode `\$12\catcode
  `\&12\catcode `\#12\catcode `\^12\catcode `\_12\catcode `\%12\relax}%
\providecommand \@@startlink[1]{}%
\providecommand \@@endlink[0]{}%
\providecommand \url  [0]{\begingroup\@sanitize@url \@url }%
\providecommand \@url [1]{\endgroup\@href {#1}{\urlprefix }}%
\providecommand \urlprefix  [0]{URL }%
\providecommand \Eprint [0]{\href }%
\providecommand \doibase [0]{http://dx.doi.org/}%
\providecommand \selectlanguage [0]{\@gobble}%
\providecommand \bibinfo  [0]{\@secondoftwo}%
\providecommand \bibfield  [0]{\@secondoftwo}%
\providecommand \translation [1]{[#1]}%
\providecommand \BibitemOpen [0]{}%
\providecommand \bibitemStop [0]{}%
\providecommand \bibitemNoStop [0]{.\EOS\space}%
\providecommand \EOS [0]{\spacefactor3000\relax}%
\providecommand \BibitemShut  [1]{\csname bibitem#1\endcsname}%
\let\auto@bib@innerbib\@empty
\bibitem [{\citenamefont {Nussinov}(1985)}]{Nussinov:1985xr}%
  \BibitemOpen
  \bibfield  {author} {\bibinfo {author} {\bibfnamefont {S.}~\bibnamefont
  {Nussinov}},\ }\href {\doibase 10.1016/0370-2693(85)90689-6} {\bibfield
  {journal} {\bibinfo  {journal} {Phys. Lett.}\ }\textbf {\bibinfo {volume}
  {165B}},\ \bibinfo {pages} {55} (\bibinfo {year} {1985})}\BibitemShut
  {NoStop}%
\bibitem [{\citenamefont {Barr}\ \emph {et~al.}(1990)\citenamefont {Barr},
  \citenamefont {Chivukula},\ and\ \citenamefont {Farhi}}]{Barr:1990ca}%
  \BibitemOpen
  \bibfield  {author} {\bibinfo {author} {\bibfnamefont {S.~M.}\ \bibnamefont
  {Barr}}, \bibinfo {author} {\bibfnamefont {R.~S.}\ \bibnamefont {Chivukula}},
  \ and\ \bibinfo {author} {\bibfnamefont {E.}~\bibnamefont {Farhi}},\ }\href
  {\doibase 10.1016/0370-2693(90)91661-T} {\bibfield  {journal} {\bibinfo
  {journal} {Phys. Lett.}\ }\textbf {\bibinfo {volume} {B241}},\ \bibinfo
  {pages} {387} (\bibinfo {year} {1990})}\BibitemShut {NoStop}%
\bibitem [{\citenamefont {Barr}(1991)}]{Barr:1991qn}%
  \BibitemOpen
  \bibfield  {author} {\bibinfo {author} {\bibfnamefont {S.~M.}\ \bibnamefont
  {Barr}},\ }\href {\doibase 10.1103/PhysRevD.44.3062} {\bibfield  {journal}
  {\bibinfo  {journal} {Phys. Rev.}\ }\textbf {\bibinfo {volume} {D44}},\
  \bibinfo {pages} {3062} (\bibinfo {year} {1991})}\BibitemShut {NoStop}%
\bibitem [{\citenamefont {Dodelson}\ \emph {et~al.}(1992)\citenamefont
  {Dodelson}, \citenamefont {Greene},\ and\ \citenamefont
  {Widrow}}]{Dodelson:1991iv}%
  \BibitemOpen
  \bibfield  {author} {\bibinfo {author} {\bibfnamefont {S.}~\bibnamefont
  {Dodelson}}, \bibinfo {author} {\bibfnamefont {B.~R.}\ \bibnamefont
  {Greene}}, \ and\ \bibinfo {author} {\bibfnamefont {L.~M.}\ \bibnamefont
  {Widrow}},\ }\href {\doibase 10.1016/0550-3213(92)90328-9} {\bibfield
  {journal} {\bibinfo  {journal} {Nucl. Phys.}\ }\textbf {\bibinfo {volume}
  {B372}},\ \bibinfo {pages} {467} (\bibinfo {year} {1992})}\BibitemShut
  {NoStop}%
\bibitem [{\citenamefont {Kaplan}(1992)}]{Kaplan:1991ah}%
  \BibitemOpen
  \bibfield  {author} {\bibinfo {author} {\bibfnamefont {D.~B.}\ \bibnamefont
  {Kaplan}},\ }\href {\doibase 10.1103/PhysRevLett.68.741} {\bibfield
  {journal} {\bibinfo  {journal} {Phys. Rev. Lett.}\ }\textbf {\bibinfo
  {volume} {68}},\ \bibinfo {pages} {741} (\bibinfo {year} {1992})}\BibitemShut
  {NoStop}%
\bibitem [{\citenamefont {Kuzmin}(1998)}]{Kuzmin:1996he}%
  \BibitemOpen
  \bibfield  {author} {\bibinfo {author} {\bibfnamefont {V.~A.}\ \bibnamefont
  {Kuzmin}},\ }\bibfield  {booktitle} {\emph {\bibinfo {booktitle}
  {{Nonaccelerator new physics. Proceedings, 1st International Workshop,
  NANP'97, Dubna, Russia, July 7-11, 1997}}},\ }\href {\doibase
  10.1134/1.953070} {\bibfield  {journal} {\bibinfo  {journal} {Phys. Part.
  Nucl.}\ }\textbf {\bibinfo {volume} {29}},\ \bibinfo {pages} {257} (\bibinfo
  {year} {1998})},\ \bibinfo {note} {[Phys. Atom. Nucl.61,1107(1998)]},\
  \Eprint {http://arxiv.org/abs/hep-ph/9701269} {arXiv:hep-ph/9701269 [hep-ph]}
  \BibitemShut {NoStop}%
\bibitem [{\citenamefont {Foot}\ and\ \citenamefont
  {Volkas}(2003)}]{Foot:2003jt}%
  \BibitemOpen
  \bibfield  {author} {\bibinfo {author} {\bibfnamefont {R.}~\bibnamefont
  {Foot}}\ and\ \bibinfo {author} {\bibfnamefont {R.~R.}\ \bibnamefont
  {Volkas}},\ }\href {\doibase 10.1103/PhysRevD.68.021304} {\bibfield
  {journal} {\bibinfo  {journal} {Phys. Rev.}\ }\textbf {\bibinfo {volume}
  {D68}},\ \bibinfo {pages} {021304} (\bibinfo {year} {2003})},\ \Eprint
  {http://arxiv.org/abs/hep-ph/0304261} {arXiv:hep-ph/0304261 [hep-ph]}
  \BibitemShut {NoStop}%
\bibitem [{\citenamefont {Foot}\ and\ \citenamefont
  {Volkas}(2004)}]{Foot:2004pq}%
  \BibitemOpen
  \bibfield  {author} {\bibinfo {author} {\bibfnamefont {R.}~\bibnamefont
  {Foot}}\ and\ \bibinfo {author} {\bibfnamefont {R.~R.}\ \bibnamefont
  {Volkas}},\ }\href {\doibase 10.1103/PhysRevD.69.123510} {\bibfield
  {journal} {\bibinfo  {journal} {Phys. Rev.}\ }\textbf {\bibinfo {volume}
  {D69}},\ \bibinfo {pages} {123510} (\bibinfo {year} {2004})},\ \Eprint
  {http://arxiv.org/abs/hep-ph/0402267} {arXiv:hep-ph/0402267 [hep-ph]}
  \BibitemShut {NoStop}%
\bibitem [{\citenamefont {Kitano}\ and\ \citenamefont
  {Low}(2005)}]{Kitano:2004sv}%
  \BibitemOpen
  \bibfield  {author} {\bibinfo {author} {\bibfnamefont {R.}~\bibnamefont
  {Kitano}}\ and\ \bibinfo {author} {\bibfnamefont {I.}~\bibnamefont {Low}},\
  }\href {\doibase 10.1103/PhysRevD.71.023510} {\bibfield  {journal} {\bibinfo
  {journal} {Phys. Rev.}\ }\textbf {\bibinfo {volume} {D71}},\ \bibinfo {pages}
  {023510} (\bibinfo {year} {2005})},\ \Eprint
  {http://arxiv.org/abs/hep-ph/0411133} {arXiv:hep-ph/0411133 [hep-ph]}
  \BibitemShut {NoStop}%
\bibitem [{\citenamefont {Gudnason}\ \emph {et~al.}(2006)\citenamefont
  {Gudnason}, \citenamefont {Kouvaris},\ and\ \citenamefont
  {Sannino}}]{Gudnason:2006ug}%
  \BibitemOpen
  \bibfield  {author} {\bibinfo {author} {\bibfnamefont {S.~B.}\ \bibnamefont
  {Gudnason}}, \bibinfo {author} {\bibfnamefont {C.}~\bibnamefont {Kouvaris}},
  \ and\ \bibinfo {author} {\bibfnamefont {F.}~\bibnamefont {Sannino}},\ }\href
  {\doibase 10.1103/PhysRevD.73.115003} {\bibfield  {journal} {\bibinfo
  {journal} {Phys. Rev.}\ }\textbf {\bibinfo {volume} {D73}},\ \bibinfo {pages}
  {115003} (\bibinfo {year} {2006})},\ \Eprint
  {http://arxiv.org/abs/hep-ph/0603014} {arXiv:hep-ph/0603014 [hep-ph]}
  \BibitemShut {NoStop}%
\bibitem [{\citenamefont {Kaplan}\ \emph {et~al.}(2009)\citenamefont {Kaplan},
  \citenamefont {Luty},\ and\ \citenamefont {Zurek}}]{Kaplan:2009ag}%
  \BibitemOpen
  \bibfield  {author} {\bibinfo {author} {\bibfnamefont {D.~E.}\ \bibnamefont
  {Kaplan}}, \bibinfo {author} {\bibfnamefont {M.~A.}\ \bibnamefont {Luty}}, \
  and\ \bibinfo {author} {\bibfnamefont {K.~M.}\ \bibnamefont {Zurek}},\ }\href
  {\doibase 10.1103/PhysRevD.79.115016} {\bibfield  {journal} {\bibinfo
  {journal} {Phys. Rev.}\ }\textbf {\bibinfo {volume} {D79}},\ \bibinfo {pages}
  {115016} (\bibinfo {year} {2009})},\ \Eprint {http://arxiv.org/abs/0901.4117}
  {arXiv:0901.4117 [hep-ph]} \BibitemShut {NoStop}%
\bibitem [{\citenamefont {Davoudiasl}\ and\ \citenamefont
  {Mohapatra}(2012)}]{Davoudiasl:2012uw}%
  \BibitemOpen
  \bibfield  {author} {\bibinfo {author} {\bibfnamefont {H.}~\bibnamefont
  {Davoudiasl}}\ and\ \bibinfo {author} {\bibfnamefont {R.~N.}\ \bibnamefont
  {Mohapatra}},\ }\href {\doibase 10.1088/1367-2630/14/9/095011} {\bibfield
  {journal} {\bibinfo  {journal} {New J. Phys.}\ }\textbf {\bibinfo {volume}
  {14}},\ \bibinfo {pages} {095011} (\bibinfo {year} {2012})},\ \Eprint
  {http://arxiv.org/abs/1203.1247} {arXiv:1203.1247 [hep-ph]} \BibitemShut
  {NoStop}%
\bibitem [{\citenamefont {Petraki}\ and\ \citenamefont
  {Volkas}(2013)}]{Petraki:2013wwa}%
  \BibitemOpen
  \bibfield  {author} {\bibinfo {author} {\bibfnamefont {K.}~\bibnamefont
  {Petraki}}\ and\ \bibinfo {author} {\bibfnamefont {R.~R.}\ \bibnamefont
  {Volkas}},\ }\href {\doibase 10.1142/S0217751X13300287} {\bibfield  {journal}
  {\bibinfo  {journal} {Int. J. Mod. Phys.}\ }\textbf {\bibinfo {volume}
  {A28}},\ \bibinfo {pages} {1330028} (\bibinfo {year} {2013})},\ \Eprint
  {http://arxiv.org/abs/1305.4939} {arXiv:1305.4939 [hep-ph]} \BibitemShut
  {NoStop}%
\bibitem [{\citenamefont {Zurek}(2014)}]{Zurek:2013wia}%
  \BibitemOpen
  \bibfield  {author} {\bibinfo {author} {\bibfnamefont {K.~M.}\ \bibnamefont
  {Zurek}},\ }\href {\doibase 10.1016/j.physrep.2013.12.001} {\bibfield
  {journal} {\bibinfo  {journal} {Phys. Rept.}\ }\textbf {\bibinfo {volume}
  {537}},\ \bibinfo {pages} {91} (\bibinfo {year} {2014})},\ \Eprint
  {http://arxiv.org/abs/1308.0338} {arXiv:1308.0338 [hep-ph]} \BibitemShut
  {NoStop}%
\bibitem [{\citenamefont {Berezhiani}(2005)}]{Berezhiani:2005ek}%
  \BibitemOpen
  \bibfield  {author} {\bibinfo {author} {\bibfnamefont {Z.}~\bibnamefont
  {Berezhiani}},\ }\href {\doibase 10.1142/9789812775344_0055} {\bibfield
  {journal} {\bibinfo  {journal} {From fields to strings}\ ,\ \bibinfo {pages}
  {2147}} (\bibinfo {year} {2005})},\ \Eprint
  {http://arxiv.org/abs/hep-ph/0508233} {arXiv:hep-ph/0508233 [hep-ph]}
  \BibitemShut {NoStop}%
\bibitem [{\citenamefont {Alves}\ \emph {et~al.}(2010)\citenamefont {Alves},
  \citenamefont {Behbahani}, \citenamefont {Schuster},\ and\ \citenamefont
  {Wacker}}]{Alves:2009nf}%
  \BibitemOpen
  \bibfield  {author} {\bibinfo {author} {\bibfnamefont {D.~S.~M.}\
  \bibnamefont {Alves}}, \bibinfo {author} {\bibfnamefont {S.~R.}\ \bibnamefont
  {Behbahani}}, \bibinfo {author} {\bibfnamefont {P.}~\bibnamefont {Schuster}},
  \ and\ \bibinfo {author} {\bibfnamefont {J.~G.}\ \bibnamefont {Wacker}},\
  }\href {\doibase 10.1016/j.physletb.2010.08.006} {\bibfield  {journal}
  {\bibinfo  {journal} {Phys. Lett.}\ }\textbf {\bibinfo {volume} {B692}},\
  \bibinfo {pages} {323} (\bibinfo {year} {2010})},\ \Eprint
  {http://arxiv.org/abs/0903.3945} {arXiv:0903.3945 [hep-ph]} \BibitemShut
  {NoStop}%
\bibitem [{\citenamefont {An}\ \emph {et~al.}(2010)\citenamefont {An},
  \citenamefont {Chen}, \citenamefont {Mohapatra},\ and\ \citenamefont
  {Zhang}}]{An:2009vq}%
  \BibitemOpen
  \bibfield  {author} {\bibinfo {author} {\bibfnamefont {H.}~\bibnamefont
  {An}}, \bibinfo {author} {\bibfnamefont {S.-L.}\ \bibnamefont {Chen}},
  \bibinfo {author} {\bibfnamefont {R.~N.}\ \bibnamefont {Mohapatra}}, \ and\
  \bibinfo {author} {\bibfnamefont {Y.}~\bibnamefont {Zhang}},\ }\href
  {\doibase 10.1007/JHEP03(2010)124} {\bibfield  {journal} {\bibinfo  {journal}
  {JHEP}\ }\textbf {\bibinfo {volume} {03}},\ \bibinfo {pages} {124} (\bibinfo
  {year} {2010})},\ \Eprint {http://arxiv.org/abs/0911.4463} {arXiv:0911.4463
  [hep-ph]} \BibitemShut {NoStop}%
\bibitem [{\citenamefont {Spier Moreira~Alves}\ \emph
  {et~al.}(2010)\citenamefont {Spier Moreira~Alves}, \citenamefont {Behbahani},
  \citenamefont {Schuster},\ and\ \citenamefont {Wacker}}]{Alves:2010dd}%
  \BibitemOpen
  \bibfield  {author} {\bibinfo {author} {\bibfnamefont {D.}~\bibnamefont
  {Spier Moreira~Alves}}, \bibinfo {author} {\bibfnamefont {S.~R.}\
  \bibnamefont {Behbahani}}, \bibinfo {author} {\bibfnamefont {P.}~\bibnamefont
  {Schuster}}, \ and\ \bibinfo {author} {\bibfnamefont {J.~G.}\ \bibnamefont
  {Wacker}},\ }\href {\doibase 10.1007/JHEP06(2010)113} {\bibfield  {journal}
  {\bibinfo  {journal} {JHEP}\ }\textbf {\bibinfo {volume} {06}},\ \bibinfo
  {pages} {113} (\bibinfo {year} {2010})},\ \Eprint
  {http://arxiv.org/abs/1003.4729} {arXiv:1003.4729 [hep-ph]} \BibitemShut
  {NoStop}%
\bibitem [{\citenamefont {Gu}(2013)}]{Gu:2012fg}%
  \BibitemOpen
  \bibfield  {author} {\bibinfo {author} {\bibfnamefont {P.-H.}\ \bibnamefont
  {Gu}},\ }\href {\doibase 10.1016/j.nuclphysb.2013.03.014} {\bibfield
  {journal} {\bibinfo  {journal} {Nucl. Phys.}\ }\textbf {\bibinfo {volume}
  {B872}},\ \bibinfo {pages} {38} (\bibinfo {year} {2013})},\ \Eprint
  {http://arxiv.org/abs/1209.4579} {arXiv:1209.4579 [hep-ph]} \BibitemShut
  {NoStop}%
\bibitem [{\citenamefont {Buckley}\ and\ \citenamefont
  {Neil}(2013)}]{Buckley:2012ky}%
  \BibitemOpen
  \bibfield  {author} {\bibinfo {author} {\bibfnamefont {M.~R.}\ \bibnamefont
  {Buckley}}\ and\ \bibinfo {author} {\bibfnamefont {E.~T.}\ \bibnamefont
  {Neil}},\ }\href {\doibase 10.1103/PhysRevD.87.043510} {\bibfield  {journal}
  {\bibinfo  {journal} {Phys. Rev.}\ }\textbf {\bibinfo {volume} {D87}},\
  \bibinfo {pages} {043510} (\bibinfo {year} {2013})},\ \Eprint
  {http://arxiv.org/abs/1209.6054} {arXiv:1209.6054 [hep-ph]} \BibitemShut
  {NoStop}%
\bibitem [{\citenamefont {Detmold}\ \emph {et~al.}(2014)\citenamefont
  {Detmold}, \citenamefont {McCullough},\ and\ \citenamefont
  {Pochinsky}}]{Detmold:2014qqa}%
  \BibitemOpen
  \bibfield  {author} {\bibinfo {author} {\bibfnamefont {W.}~\bibnamefont
  {Detmold}}, \bibinfo {author} {\bibfnamefont {M.}~\bibnamefont {McCullough}},
  \ and\ \bibinfo {author} {\bibfnamefont {A.}~\bibnamefont {Pochinsky}},\
  }\href {\doibase 10.1103/PhysRevD.90.115013} {\bibfield  {journal} {\bibinfo
  {journal} {Phys. Rev.}\ }\textbf {\bibinfo {volume} {D90}},\ \bibinfo {pages}
  {115013} (\bibinfo {year} {2014})},\ \Eprint {http://arxiv.org/abs/1406.2276}
  {arXiv:1406.2276 [hep-ph]} \BibitemShut {NoStop}%
\bibitem [{\citenamefont {Gu}(2014)}]{Gu:2014nga}%
  \BibitemOpen
  \bibfield  {author} {\bibinfo {author} {\bibfnamefont {P.-H.}\ \bibnamefont
  {Gu}},\ }\href {\doibase 10.1088/1475-7516/2014/12/046} {\bibfield  {journal}
  {\bibinfo  {journal} {JCAP}\ }\textbf {\bibinfo {volume} {1412}},\ \bibinfo
  {pages} {046} (\bibinfo {year} {2014})},\ \Eprint
  {http://arxiv.org/abs/1410.5759} {arXiv:1410.5759 [hep-ph]} \BibitemShut
  {NoStop}%
\bibitem [{\citenamefont {Lonsdale}\ and\ \citenamefont
  {Volkas}(2018)}]{Lonsdale:2018xwd}%
  \BibitemOpen
  \bibfield  {author} {\bibinfo {author} {\bibfnamefont {S.~J.}\ \bibnamefont
  {Lonsdale}}\ and\ \bibinfo {author} {\bibfnamefont {R.~R.}\ \bibnamefont
  {Volkas}},\ }\href {\doibase 10.1103/PhysRevD.97.103510} {\bibfield
  {journal} {\bibinfo  {journal} {Phys. Rev.}\ }\textbf {\bibinfo {volume}
  {D97}},\ \bibinfo {pages} {103510} (\bibinfo {year} {2018})},\ \Eprint
  {http://arxiv.org/abs/1801.05561} {arXiv:1801.05561 [hep-ph]} \BibitemShut
  {NoStop}%
\bibitem [{\citenamefont {Ibe}\ \emph {et~al.}(2018)\citenamefont {Ibe},
  \citenamefont {Kamada}, \citenamefont {Kobayashi},\ and\ \citenamefont
  {Nakano}}]{Ibe:2018juk}%
  \BibitemOpen
  \bibfield  {author} {\bibinfo {author} {\bibfnamefont {M.}~\bibnamefont
  {Ibe}}, \bibinfo {author} {\bibfnamefont {A.}~\bibnamefont {Kamada}},
  \bibinfo {author} {\bibfnamefont {S.}~\bibnamefont {Kobayashi}}, \ and\
  \bibinfo {author} {\bibfnamefont {W.}~\bibnamefont {Nakano}},\ }\href
  {\doibase 10.1007/JHEP11(2018)203} {\bibfield  {journal} {\bibinfo  {journal}
  {JHEP}\ }\textbf {\bibinfo {volume} {11}},\ \bibinfo {pages} {203} (\bibinfo
  {year} {2018})},\ \Eprint {http://arxiv.org/abs/1805.06876} {arXiv:1805.06876
  [hep-ph]} \BibitemShut {NoStop}%
\bibitem [{\citenamefont {Ibe}\ \emph {et~al.}(2019{\natexlab{a}})\citenamefont
  {Ibe}, \citenamefont {Kamada}, \citenamefont {Kobayashi}, \citenamefont
  {Kuwahara},\ and\ \citenamefont {Nakano}}]{Ibe:2018tex}%
  \BibitemOpen
  \bibfield  {author} {\bibinfo {author} {\bibfnamefont {M.}~\bibnamefont
  {Ibe}}, \bibinfo {author} {\bibfnamefont {A.}~\bibnamefont {Kamada}},
  \bibinfo {author} {\bibfnamefont {S.}~\bibnamefont {Kobayashi}}, \bibinfo
  {author} {\bibfnamefont {T.}~\bibnamefont {Kuwahara}}, \ and\ \bibinfo
  {author} {\bibfnamefont {W.}~\bibnamefont {Nakano}},\ }\href {\doibase
  10.1007/JHEP03(2019)173} {\bibfield  {journal} {\bibinfo  {journal} {JHEP}\
  }\textbf {\bibinfo {volume} {03}},\ \bibinfo {pages} {173} (\bibinfo {year}
  {2019}{\natexlab{a}})},\ \Eprint {http://arxiv.org/abs/1811.10232}
  {arXiv:1811.10232 [hep-ph]} \BibitemShut {NoStop}%
\bibitem [{\citenamefont {Ibe}\ \emph {et~al.}(2019{\natexlab{b}})\citenamefont
  {Ibe}, \citenamefont {Kamada}, \citenamefont {Kobayashi}, \citenamefont
  {Kuwahara},\ and\ \citenamefont {Nakano}}]{Ibe:2019ena}%
  \BibitemOpen
  \bibfield  {author} {\bibinfo {author} {\bibfnamefont {M.}~\bibnamefont
  {Ibe}}, \bibinfo {author} {\bibfnamefont {A.}~\bibnamefont {Kamada}},
  \bibinfo {author} {\bibfnamefont {S.}~\bibnamefont {Kobayashi}}, \bibinfo
  {author} {\bibfnamefont {T.}~\bibnamefont {Kuwahara}}, \ and\ \bibinfo
  {author} {\bibfnamefont {W.}~\bibnamefont {Nakano}},\ }\href {\doibase
  10.1103/PhysRevD.100.075022} {\bibfield  {journal} {\bibinfo  {journal}
  {Phys. Rev.}\ }\textbf {\bibinfo {volume} {D100}},\ \bibinfo {pages} {075022}
  (\bibinfo {year} {2019}{\natexlab{b}})},\ \Eprint
  {http://arxiv.org/abs/1907.03404} {arXiv:1907.03404 [hep-ph]} \BibitemShut
  {NoStop}%
\bibitem [{\citenamefont {Fukugita}\ and\ \citenamefont
  {Yanagida}(1986)}]{Fukugita:1986hr}%
  \BibitemOpen
  \bibfield  {author} {\bibinfo {author} {\bibfnamefont {M.}~\bibnamefont
  {Fukugita}}\ and\ \bibinfo {author} {\bibfnamefont {T.}~\bibnamefont
  {Yanagida}},\ }\href {\doibase 10.1016/0370-2693(86)91126-3} {\bibfield
  {journal} {\bibinfo  {journal} {Phys. Lett.}\ }\textbf {\bibinfo {volume}
  {B174}},\ \bibinfo {pages} {45} (\bibinfo {year} {1986})}\BibitemShut
  {NoStop}%
\bibitem [{\citenamefont {Giudice}\ \emph {et~al.}(2004)\citenamefont
  {Giudice}, \citenamefont {Notari}, \citenamefont {Raidal}, \citenamefont
  {Riotto},\ and\ \citenamefont {Strumia}}]{Giudice:2003jh}%
  \BibitemOpen
  \bibfield  {author} {\bibinfo {author} {\bibfnamefont {G.~F.}\ \bibnamefont
  {Giudice}}, \bibinfo {author} {\bibfnamefont {A.}~\bibnamefont {Notari}},
  \bibinfo {author} {\bibfnamefont {M.}~\bibnamefont {Raidal}}, \bibinfo
  {author} {\bibfnamefont {A.}~\bibnamefont {Riotto}}, \ and\ \bibinfo {author}
  {\bibfnamefont {A.}~\bibnamefont {Strumia}},\ }\href {\doibase
  10.1016/j.nuclphysb.2004.02.019} {\bibfield  {journal} {\bibinfo  {journal}
  {Nucl. Phys.}\ }\textbf {\bibinfo {volume} {B685}},\ \bibinfo {pages} {89}
  (\bibinfo {year} {2004})},\ \Eprint {http://arxiv.org/abs/hep-ph/0310123}
  {arXiv:hep-ph/0310123 [hep-ph]} \BibitemShut {NoStop}%
\bibitem [{\citenamefont {Buchmuller}\ \emph {et~al.}(2005)\citenamefont
  {Buchmuller}, \citenamefont {Peccei},\ and\ \citenamefont
  {Yanagida}}]{Buchmuller:2005eh}%
  \BibitemOpen
  \bibfield  {author} {\bibinfo {author} {\bibfnamefont {W.}~\bibnamefont
  {Buchmuller}}, \bibinfo {author} {\bibfnamefont {R.~D.}\ \bibnamefont
  {Peccei}}, \ and\ \bibinfo {author} {\bibfnamefont {T.}~\bibnamefont
  {Yanagida}},\ }\href {\doibase 10.1146/annurev.nucl.55.090704.151558}
  {\bibfield  {journal} {\bibinfo  {journal} {Ann. Rev. Nucl. Part. Sci.}\
  }\textbf {\bibinfo {volume} {55}},\ \bibinfo {pages} {311} (\bibinfo {year}
  {2005})},\ \Eprint {http://arxiv.org/abs/hep-ph/0502169}
  {arXiv:hep-ph/0502169 [hep-ph]} \BibitemShut {NoStop}%
\bibitem [{\citenamefont {Davidson}\ \emph {et~al.}(2008)\citenamefont
  {Davidson}, \citenamefont {Nardi},\ and\ \citenamefont
  {Nir}}]{Davidson:2008bu}%
  \BibitemOpen
  \bibfield  {author} {\bibinfo {author} {\bibfnamefont {S.}~\bibnamefont
  {Davidson}}, \bibinfo {author} {\bibfnamefont {E.}~\bibnamefont {Nardi}}, \
  and\ \bibinfo {author} {\bibfnamefont {Y.}~\bibnamefont {Nir}},\ }\href
  {\doibase 10.1016/j.physrep.2008.06.002} {\bibfield  {journal} {\bibinfo
  {journal} {Phys. Rept.}\ }\textbf {\bibinfo {volume} {466}},\ \bibinfo
  {pages} {105} (\bibinfo {year} {2008})},\ \Eprint
  {http://arxiv.org/abs/0802.2962} {arXiv:0802.2962 [hep-ph]} \BibitemShut
  {NoStop}%
\bibitem [{\citenamefont {Minkowski}(1977)}]{Minkowski:1977sc}%
  \BibitemOpen
  \bibfield  {author} {\bibinfo {author} {\bibfnamefont {P.}~\bibnamefont
  {Minkowski}},\ }\href {\doibase 10.1016/0370-2693(77)90435-X} {\bibfield
  {journal} {\bibinfo  {journal} {Phys. Lett.}\ }\textbf {\bibinfo {volume}
  {B67}},\ \bibinfo {pages} {421} (\bibinfo {year} {1977})}\BibitemShut
  {NoStop}%
\bibitem [{\citenamefont {Yanagida}(1979)}]{Yanagida:1979as}%
  \BibitemOpen
  \bibfield  {author} {\bibinfo {author} {\bibfnamefont {T.}~\bibnamefont
  {Yanagida}},\ }\bibfield  {booktitle} {\emph {\bibinfo {booktitle}
  {{Proceedings: Workshop on the Unified Theories and the Baryon Number in the
  Universe: Tsukuba, Japan, February 13-14, 1979}}},\ }\href@noop {} {\bibfield
   {journal} {\bibinfo  {journal} {Conf. Proc.}\ }\textbf {\bibinfo {volume}
  {C7902131}},\ \bibinfo {pages} {95} (\bibinfo {year} {1979})}\BibitemShut
  {NoStop}%
\bibitem [{\citenamefont {Gell-Mann}\ \emph {et~al.}(1979)\citenamefont
  {Gell-Mann}, \citenamefont {Ramond},\ and\ \citenamefont
  {Slansky}}]{GellMann:1980vs}%
  \BibitemOpen
  \bibfield  {author} {\bibinfo {author} {\bibfnamefont {M.}~\bibnamefont
  {Gell-Mann}}, \bibinfo {author} {\bibfnamefont {P.}~\bibnamefont {Ramond}}, \
  and\ \bibinfo {author} {\bibfnamefont {R.}~\bibnamefont {Slansky}},\
  }\bibfield  {booktitle} {\emph {\bibinfo {booktitle} {{Supergravity Workshop
  Stony Brook, New York, September 27-28, 1979}}},\ }\href@noop {} {\bibfield
  {journal} {\bibinfo  {journal} {Conf. Proc.}\ }\textbf {\bibinfo {volume}
  {C790927}},\ \bibinfo {pages} {315} (\bibinfo {year} {1979})},\ \Eprint
  {http://arxiv.org/abs/1306.4669} {arXiv:1306.4669 [hep-th]} \BibitemShut
  {NoStop}%
\bibitem [{\citenamefont {Glashow}(1980)}]{Glashow:1979nm}%
  \BibitemOpen
  \bibfield  {author} {\bibinfo {author} {\bibfnamefont {S.~L.}\ \bibnamefont
  {Glashow}},\ }\bibfield  {booktitle} {\emph {\bibinfo {booktitle} {{Cargese
  Summer Institute: Quarks and Leptons Cargese, France, July 9-29, 1979}}},\
  }\href {\doibase 10.1007/978-1-4684-7197-7_15} {\bibfield  {journal}
  {\bibinfo  {journal} {NATO Sci. Ser. B}\ }\textbf {\bibinfo {volume} {61}},\
  \bibinfo {pages} {687} (\bibinfo {year} {1980})}\BibitemShut {NoStop}%
\bibitem [{\citenamefont {Mohapatra}\ and\ \citenamefont
  {Senjanovic}(1980)}]{Mohapatra:1979ia}%
  \BibitemOpen
  \bibfield  {author} {\bibinfo {author} {\bibfnamefont {R.~N.}\ \bibnamefont
  {Mohapatra}}\ and\ \bibinfo {author} {\bibfnamefont {G.}~\bibnamefont
  {Senjanovic}},\ }\href {\doibase 10.1103/PhysRevLett.44.912} {\bibfield
  {journal} {\bibinfo  {journal} {Phys. Rev. Lett.}\ }\textbf {\bibinfo
  {volume} {44}},\ \bibinfo {pages} {912} (\bibinfo {year} {1980})},\ \bibinfo
  {note} {[,231(1979)]}\BibitemShut {NoStop}%
\bibitem [{\citenamefont {Blennow}\ \emph {et~al.}(2012)\citenamefont
  {Blennow}, \citenamefont {Fernandez-Martinez}, \citenamefont {Mena},
  \citenamefont {Redondo},\ and\ \citenamefont {Serra}}]{Blennow:2012de}%
  \BibitemOpen
  \bibfield  {author} {\bibinfo {author} {\bibfnamefont {M.}~\bibnamefont
  {Blennow}}, \bibinfo {author} {\bibfnamefont {E.}~\bibnamefont
  {Fernandez-Martinez}}, \bibinfo {author} {\bibfnamefont {O.}~\bibnamefont
  {Mena}}, \bibinfo {author} {\bibfnamefont {J.}~\bibnamefont {Redondo}}, \
  and\ \bibinfo {author} {\bibfnamefont {P.}~\bibnamefont {Serra}},\ }\href
  {\doibase 10.1088/1475-7516/2012/07/022} {\bibfield  {journal} {\bibinfo
  {journal} {JCAP}\ }\textbf {\bibinfo {volume} {1207}},\ \bibinfo {pages}
  {022} (\bibinfo {year} {2012})},\ \Eprint {http://arxiv.org/abs/1203.5803}
  {arXiv:1203.5803 [hep-ph]} \BibitemShut {NoStop}%
\bibitem [{\citenamefont {Cai}\ \emph {et~al.}(2009)\citenamefont {Cai},
  \citenamefont {Luty},\ and\ \citenamefont {Kaplan}}]{Cai:2009ia}%
  \BibitemOpen
  \bibfield  {author} {\bibinfo {author} {\bibfnamefont {Y.}~\bibnamefont
  {Cai}}, \bibinfo {author} {\bibfnamefont {M.~A.}\ \bibnamefont {Luty}}, \
  and\ \bibinfo {author} {\bibfnamefont {D.~E.}\ \bibnamefont {Kaplan}},\
  }\href@noop {} {\  (\bibinfo {year} {2009})},\ \Eprint
  {http://arxiv.org/abs/0909.5499} {arXiv:0909.5499 [hep-ph]} \BibitemShut
  {NoStop}%
\bibitem [{\citenamefont {Buckley}\ and\ \citenamefont
  {Profumo}(2012)}]{Buckley:2011ye}%
  \BibitemOpen
  \bibfield  {author} {\bibinfo {author} {\bibfnamefont {M.~R.}\ \bibnamefont
  {Buckley}}\ and\ \bibinfo {author} {\bibfnamefont {S.}~\bibnamefont
  {Profumo}},\ }\href {\doibase 10.1103/PhysRevLett.108.011301} {\bibfield
  {journal} {\bibinfo  {journal} {Phys. Rev. Lett.}\ }\textbf {\bibinfo
  {volume} {108}},\ \bibinfo {pages} {011301} (\bibinfo {year} {2012})},\
  \Eprint {http://arxiv.org/abs/1109.2164} {arXiv:1109.2164 [hep-ph]}
  \BibitemShut {NoStop}%
\bibitem [{\citenamefont {Cirelli}\ \emph {et~al.}(2012)\citenamefont
  {Cirelli}, \citenamefont {Panci}, \citenamefont {Servant},\ and\
  \citenamefont {Zaharijas}}]{Cirelli:2011ac}%
  \BibitemOpen
  \bibfield  {author} {\bibinfo {author} {\bibfnamefont {M.}~\bibnamefont
  {Cirelli}}, \bibinfo {author} {\bibfnamefont {P.}~\bibnamefont {Panci}},
  \bibinfo {author} {\bibfnamefont {G.}~\bibnamefont {Servant}}, \ and\
  \bibinfo {author} {\bibfnamefont {G.}~\bibnamefont {Zaharijas}},\ }\href
  {\doibase 10.1088/1475-7516/2012/03/015} {\bibfield  {journal} {\bibinfo
  {journal} {JCAP}\ }\textbf {\bibinfo {volume} {1203}},\ \bibinfo {pages}
  {015} (\bibinfo {year} {2012})},\ \Eprint {http://arxiv.org/abs/1110.3809}
  {arXiv:1110.3809 [hep-ph]} \BibitemShut {NoStop}%
\bibitem [{\citenamefont {Tulin}\ \emph {et~al.}(2012)\citenamefont {Tulin},
  \citenamefont {Yu},\ and\ \citenamefont {Zurek}}]{Tulin:2012re}%
  \BibitemOpen
  \bibfield  {author} {\bibinfo {author} {\bibfnamefont {S.}~\bibnamefont
  {Tulin}}, \bibinfo {author} {\bibfnamefont {H.-B.}\ \bibnamefont {Yu}}, \
  and\ \bibinfo {author} {\bibfnamefont {K.~M.}\ \bibnamefont {Zurek}},\ }\href
  {\doibase 10.1088/1475-7516/2012/05/013} {\bibfield  {journal} {\bibinfo
  {journal} {JCAP}\ }\textbf {\bibinfo {volume} {1205}},\ \bibinfo {pages}
  {013} (\bibinfo {year} {2012})},\ \Eprint {http://arxiv.org/abs/1202.0283}
  {arXiv:1202.0283 [hep-ph]} \BibitemShut {NoStop}%
\bibitem [{\citenamefont {Okada}\ and\ \citenamefont
  {Seto}(2012)}]{Okada:2012rm}%
  \BibitemOpen
  \bibfield  {author} {\bibinfo {author} {\bibfnamefont {N.}~\bibnamefont
  {Okada}}\ and\ \bibinfo {author} {\bibfnamefont {O.}~\bibnamefont {Seto}},\
  }\href {\doibase 10.1103/PhysRevD.86.063525} {\bibfield  {journal} {\bibinfo
  {journal} {Phys. Rev.}\ }\textbf {\bibinfo {volume} {D86}},\ \bibinfo {pages}
  {063525} (\bibinfo {year} {2012})},\ \Eprint {http://arxiv.org/abs/1205.2844}
  {arXiv:1205.2844 [hep-ph]} \BibitemShut {NoStop}%
\bibitem [{\citenamefont {Hardy}\ \emph {et~al.}(2014)\citenamefont {Hardy},
  \citenamefont {Lasenby},\ and\ \citenamefont {Unwin}}]{Hardy:2014dea}%
  \BibitemOpen
  \bibfield  {author} {\bibinfo {author} {\bibfnamefont {E.}~\bibnamefont
  {Hardy}}, \bibinfo {author} {\bibfnamefont {R.}~\bibnamefont {Lasenby}}, \
  and\ \bibinfo {author} {\bibfnamefont {J.}~\bibnamefont {Unwin}},\ }\href
  {\doibase 10.1007/JHEP07(2014)049} {\bibfield  {journal} {\bibinfo  {journal}
  {JHEP}\ }\textbf {\bibinfo {volume} {07}},\ \bibinfo {pages} {049} (\bibinfo
  {year} {2014})},\ \Eprint {http://arxiv.org/abs/1402.4500} {arXiv:1402.4500
  [hep-ph]} \BibitemShut {NoStop}%
\bibitem [{\citenamefont {Chen}\ and\ \citenamefont
  {Kang}(2016)}]{Chen:2015yuz}%
  \BibitemOpen
  \bibfield  {author} {\bibinfo {author} {\bibfnamefont {S.-L.}\ \bibnamefont
  {Chen}}\ and\ \bibinfo {author} {\bibfnamefont {Z.}~\bibnamefont {Kang}},\
  }\href {\doibase 10.1016/j.physletb.2016.08.051} {\bibfield  {journal}
  {\bibinfo  {journal} {Phys. Lett.}\ }\textbf {\bibinfo {volume} {B761}},\
  \bibinfo {pages} {296} (\bibinfo {year} {2016})},\ \Eprint
  {http://arxiv.org/abs/1512.08780} {arXiv:1512.08780 [hep-ph]} \BibitemShut
  {NoStop}%
\bibitem [{\citenamefont {Fukuda}\ \emph {et~al.}(2015)\citenamefont {Fukuda},
  \citenamefont {Matsumoto},\ and\ \citenamefont
  {Mukhopadhyay}}]{Fukuda:2014xqa}%
  \BibitemOpen
  \bibfield  {author} {\bibinfo {author} {\bibfnamefont {H.}~\bibnamefont
  {Fukuda}}, \bibinfo {author} {\bibfnamefont {S.}~\bibnamefont {Matsumoto}}, \
  and\ \bibinfo {author} {\bibfnamefont {S.}~\bibnamefont {Mukhopadhyay}},\
  }\href {\doibase 10.1103/PhysRevD.92.013008} {\bibfield  {journal} {\bibinfo
  {journal} {Phys. Rev.}\ }\textbf {\bibinfo {volume} {D92}},\ \bibinfo {pages}
  {013008} (\bibinfo {year} {2015})},\ \Eprint {http://arxiv.org/abs/1411.4014}
  {arXiv:1411.4014 [hep-ph]} \BibitemShut {NoStop}%
\bibitem [{\citenamefont {Harvey}\ and\ \citenamefont
  {Turner}(1990)}]{Harvey:1990qw}%
  \BibitemOpen
  \bibfield  {author} {\bibinfo {author} {\bibfnamefont {J.~A.}\ \bibnamefont
  {Harvey}}\ and\ \bibinfo {author} {\bibfnamefont {M.~S.}\ \bibnamefont
  {Turner}},\ }\href {\doibase 10.1103/PhysRevD.42.3344} {\bibfield  {journal}
  {\bibinfo  {journal} {Phys. Rev.}\ }\textbf {\bibinfo {volume} {D42}},\
  \bibinfo {pages} {3344} (\bibinfo {year} {1990})}\BibitemShut {NoStop}%
\bibitem [{\citenamefont {Bauer}\ \emph {et~al.}(2018)\citenamefont {Bauer},
  \citenamefont {Foldenauer},\ and\ \citenamefont {Jaeckel}}]{Bauer:2018onh}%
  \BibitemOpen
  \bibfield  {author} {\bibinfo {author} {\bibfnamefont {M.}~\bibnamefont
  {Bauer}}, \bibinfo {author} {\bibfnamefont {P.}~\bibnamefont {Foldenauer}}, \
  and\ \bibinfo {author} {\bibfnamefont {J.}~\bibnamefont {Jaeckel}},\ }\href
  {\doibase 10.1007/JHEP07(2018)094} {\bibfield  {journal} {\bibinfo  {journal}
  {JHEP}\ }\textbf {\bibinfo {volume} {07}},\ \bibinfo {pages} {094} (\bibinfo
  {year} {2018})},\ \bibinfo {note} {[JHEP18,094(2020)]},\ \Eprint
  {http://arxiv.org/abs/1803.05466} {arXiv:1803.05466 [hep-ph]} \BibitemShut
  {NoStop}%
\bibitem [{\citenamefont {Chang}\ \emph {et~al.}(2017)\citenamefont {Chang},
  \citenamefont {Essig},\ and\ \citenamefont {McDermott}}]{Chang:2016ntp}%
  \BibitemOpen
  \bibfield  {author} {\bibinfo {author} {\bibfnamefont {J.~H.}\ \bibnamefont
  {Chang}}, \bibinfo {author} {\bibfnamefont {R.}~\bibnamefont {Essig}}, \ and\
  \bibinfo {author} {\bibfnamefont {S.~D.}\ \bibnamefont {McDermott}},\ }\href
  {\doibase 10.1007/JHEP01(2017)107} {\bibfield  {journal} {\bibinfo  {journal}
  {JHEP}\ }\textbf {\bibinfo {volume} {01}},\ \bibinfo {pages} {107} (\bibinfo
  {year} {2017})},\ \Eprint {http://arxiv.org/abs/1611.03864} {arXiv:1611.03864
  [hep-ph]} \BibitemShut {NoStop}%
\bibitem [{\citenamefont {Chang}\ \emph {et~al.}(2018)\citenamefont {Chang},
  \citenamefont {Essig},\ and\ \citenamefont {McDermott}}]{Chang:2018rso}%
  \BibitemOpen
  \bibfield  {author} {\bibinfo {author} {\bibfnamefont {J.~H.}\ \bibnamefont
  {Chang}}, \bibinfo {author} {\bibfnamefont {R.}~\bibnamefont {Essig}}, \ and\
  \bibinfo {author} {\bibfnamefont {S.~D.}\ \bibnamefont {McDermott}},\ }\href
  {\doibase 10.1007/JHEP09(2018)051} {\bibfield  {journal} {\bibinfo  {journal}
  {JHEP}\ }\textbf {\bibinfo {volume} {09}},\ \bibinfo {pages} {051} (\bibinfo
  {year} {2018})},\ \Eprint {http://arxiv.org/abs/1803.00993} {arXiv:1803.00993
  [hep-ph]} \BibitemShut {NoStop}%
\bibitem [{\citenamefont {Ibe}\ \emph {et~al.}(2012)\citenamefont {Ibe},
  \citenamefont {Matsumoto},\ and\ \citenamefont {Yanagida}}]{Ibe:2011hq}%
  \BibitemOpen
  \bibfield  {author} {\bibinfo {author} {\bibfnamefont {M.}~\bibnamefont
  {Ibe}}, \bibinfo {author} {\bibfnamefont {S.}~\bibnamefont {Matsumoto}}, \
  and\ \bibinfo {author} {\bibfnamefont {T.~T.}\ \bibnamefont {Yanagida}},\
  }\href {\doibase 10.1016/j.physletb.2012.01.032} {\bibfield  {journal}
  {\bibinfo  {journal} {Phys. Lett.}\ }\textbf {\bibinfo {volume} {B708}},\
  \bibinfo {pages} {112} (\bibinfo {year} {2012})},\ \Eprint
  {http://arxiv.org/abs/1110.5452} {arXiv:1110.5452 [hep-ph]} \BibitemShut
  {NoStop}%
\bibitem [{\citenamefont {Falkowski}\ \emph {et~al.}(2011)\citenamefont
  {Falkowski}, \citenamefont {Ruderman},\ and\ \citenamefont
  {Volansky}}]{Falkowski:2011xh}%
  \BibitemOpen
  \bibfield  {author} {\bibinfo {author} {\bibfnamefont {A.}~\bibnamefont
  {Falkowski}}, \bibinfo {author} {\bibfnamefont {J.~T.}\ \bibnamefont
  {Ruderman}}, \ and\ \bibinfo {author} {\bibfnamefont {T.}~\bibnamefont
  {Volansky}},\ }\href {\doibase 10.1007/JHEP05(2011)106} {\bibfield  {journal}
  {\bibinfo  {journal} {JHEP}\ }\textbf {\bibinfo {volume} {05}},\ \bibinfo
  {pages} {106} (\bibinfo {year} {2011})},\ \Eprint
  {http://arxiv.org/abs/1101.4936} {arXiv:1101.4936 [hep-ph]} \BibitemShut
  {NoStop}%
\bibitem [{\citenamefont {Gunn}\ \emph {et~al.}(1978)\citenamefont {Gunn},
  \citenamefont {Lee}, \citenamefont {Lerche}, \citenamefont {Schramm},\ and\
  \citenamefont {Steigman}}]{Gunn:1978gr}%
  \BibitemOpen
  \bibfield  {author} {\bibinfo {author} {\bibfnamefont {J.~E.}\ \bibnamefont
  {Gunn}}, \bibinfo {author} {\bibfnamefont {B.~W.}\ \bibnamefont {Lee}},
  \bibinfo {author} {\bibfnamefont {I.}~\bibnamefont {Lerche}}, \bibinfo
  {author} {\bibfnamefont {D.~N.}\ \bibnamefont {Schramm}}, \ and\ \bibinfo
  {author} {\bibfnamefont {G.}~\bibnamefont {Steigman}},\ }\href {\doibase
  10.1086/156335} {\bibfield  {journal} {\bibinfo  {journal} {Astrophys. J.}\
  }\textbf {\bibinfo {volume} {223}},\ \bibinfo {pages} {1015} (\bibinfo {year}
  {1978})},\ \bibinfo {note} {[,190(1978)]}\BibitemShut {NoStop}%
\bibitem [{\citenamefont {Bergstrom}(2012)}]{Bergstrom:2012fi}%
  \BibitemOpen
  \bibfield  {author} {\bibinfo {author} {\bibfnamefont {L.}~\bibnamefont
  {Bergstrom}},\ }\href {\doibase 10.1002/andp.201200116} {\bibfield  {journal}
  {\bibinfo  {journal} {Annalen Phys.}\ }\textbf {\bibinfo {volume} {524}},\
  \bibinfo {pages} {479} (\bibinfo {year} {2012})},\ \Eprint
  {http://arxiv.org/abs/1205.4882} {arXiv:1205.4882 [astro-ph.HE]} \BibitemShut
  {NoStop}%
\bibitem [{\citenamefont {Gilmore}\ \emph {et~al.}(2007)\citenamefont
  {Gilmore}, \citenamefont {Wilkinson}, \citenamefont {Wyse}, \citenamefont
  {Kleyna}, \citenamefont {Koch}, \citenamefont {Evans},\ and\ \citenamefont
  {Grebel}}]{Gilmore:2007fy}%
  \BibitemOpen
  \bibfield  {author} {\bibinfo {author} {\bibfnamefont {G.}~\bibnamefont
  {Gilmore}}, \bibinfo {author} {\bibfnamefont {M.~I.}\ \bibnamefont
  {Wilkinson}}, \bibinfo {author} {\bibfnamefont {R.~F.~G.}\ \bibnamefont
  {Wyse}}, \bibinfo {author} {\bibfnamefont {J.~T.}\ \bibnamefont {Kleyna}},
  \bibinfo {author} {\bibfnamefont {A.}~\bibnamefont {Koch}}, \bibinfo {author}
  {\bibfnamefont {N.~W.}\ \bibnamefont {Evans}}, \ and\ \bibinfo {author}
  {\bibfnamefont {E.~K.}\ \bibnamefont {Grebel}},\ }\href {\doibase
  10.1086/518025} {\bibfield  {journal} {\bibinfo  {journal} {Astrophys. J.}\
  }\textbf {\bibinfo {volume} {663}},\ \bibinfo {pages} {948} (\bibinfo {year}
  {2007})},\ \Eprint {http://arxiv.org/abs/astro-ph/0703308}
  {arXiv:astro-ph/0703308 [ASTRO-PH]} \BibitemShut {NoStop}%
\bibitem [{\citenamefont {McConnachie}(2012)}]{McConnachie:2012vd}%
  \BibitemOpen
  \bibfield  {author} {\bibinfo {author} {\bibfnamefont {A.~W.}\ \bibnamefont
  {McConnachie}},\ }\href {\doibase 10.1088/0004-6256/144/1/4} {\bibfield
  {journal} {\bibinfo  {journal} {Astron. J.}\ }\textbf {\bibinfo {volume}
  {144}},\ \bibinfo {pages} {4} (\bibinfo {year} {2012})},\ \Eprint
  {http://arxiv.org/abs/1204.1562} {arXiv:1204.1562 [astro-ph.CO]} \BibitemShut
  {NoStop}%
\bibitem [{\citenamefont {Mardon}\ \emph {et~al.}(2009)\citenamefont {Mardon},
  \citenamefont {Nomura}, \citenamefont {Stolarski},\ and\ \citenamefont
  {Thaler}}]{Mardon:2009rc}%
  \BibitemOpen
  \bibfield  {author} {\bibinfo {author} {\bibfnamefont {J.}~\bibnamefont
  {Mardon}}, \bibinfo {author} {\bibfnamefont {Y.}~\bibnamefont {Nomura}},
  \bibinfo {author} {\bibfnamefont {D.}~\bibnamefont {Stolarski}}, \ and\
  \bibinfo {author} {\bibfnamefont {J.}~\bibnamefont {Thaler}},\ }\href
  {\doibase 10.1088/1475-7516/2009/05/016} {\bibfield  {journal} {\bibinfo
  {journal} {JCAP}\ }\textbf {\bibinfo {volume} {0905}},\ \bibinfo {pages}
  {016} (\bibinfo {year} {2009})},\ \Eprint {http://arxiv.org/abs/0901.2926}
  {arXiv:0901.2926 [hep-ph]} \BibitemShut {NoStop}%
\bibitem [{\citenamefont {Elor}\ \emph {et~al.}(2015)\citenamefont {Elor},
  \citenamefont {Rodd},\ and\ \citenamefont {Slatyer}}]{Elor:2015tva}%
  \BibitemOpen
  \bibfield  {author} {\bibinfo {author} {\bibfnamefont {G.}~\bibnamefont
  {Elor}}, \bibinfo {author} {\bibfnamefont {N.~L.}\ \bibnamefont {Rodd}}, \
  and\ \bibinfo {author} {\bibfnamefont {T.~R.}\ \bibnamefont {Slatyer}},\
  }\href {\doibase 10.1103/PhysRevD.91.103531} {\bibfield  {journal} {\bibinfo
  {journal} {Phys. Rev.}\ }\textbf {\bibinfo {volume} {D91}},\ \bibinfo {pages}
  {103531} (\bibinfo {year} {2015})},\ \Eprint
  {http://arxiv.org/abs/1503.01773} {arXiv:1503.01773 [hep-ph]} \BibitemShut
  {NoStop}%
\bibitem [{\citenamefont {Elor}\ \emph {et~al.}(2016)\citenamefont {Elor},
  \citenamefont {Rodd}, \citenamefont {Slatyer},\ and\ \citenamefont
  {Xue}}]{Elor:2015bho}%
  \BibitemOpen
  \bibfield  {author} {\bibinfo {author} {\bibfnamefont {G.}~\bibnamefont
  {Elor}}, \bibinfo {author} {\bibfnamefont {N.~L.}\ \bibnamefont {Rodd}},
  \bibinfo {author} {\bibfnamefont {T.~R.}\ \bibnamefont {Slatyer}}, \ and\
  \bibinfo {author} {\bibfnamefont {W.}~\bibnamefont {Xue}},\ }\href {\doibase
  10.1088/1475-7516/2016/06/024} {\bibfield  {journal} {\bibinfo  {journal}
  {JCAP}\ }\textbf {\bibinfo {volume} {1606}},\ \bibinfo {pages} {024}
  (\bibinfo {year} {2016})},\ \Eprint {http://arxiv.org/abs/1511.08787}
  {arXiv:1511.08787 [hep-ph]} \BibitemShut {NoStop}%
\bibitem [{\citenamefont {Gao}\ \emph {et~al.}(2010)\citenamefont {Gao},
  \citenamefont {Gritsan}, \citenamefont {Guo}, \citenamefont {Melnikov},
  \citenamefont {Schulze},\ and\ \citenamefont {Tran}}]{Gao:2010qx}%
  \BibitemOpen
  \bibfield  {author} {\bibinfo {author} {\bibfnamefont {Y.}~\bibnamefont
  {Gao}}, \bibinfo {author} {\bibfnamefont {A.~V.}\ \bibnamefont {Gritsan}},
  \bibinfo {author} {\bibfnamefont {Z.}~\bibnamefont {Guo}}, \bibinfo {author}
  {\bibfnamefont {K.}~\bibnamefont {Melnikov}}, \bibinfo {author}
  {\bibfnamefont {M.}~\bibnamefont {Schulze}}, \ and\ \bibinfo {author}
  {\bibfnamefont {N.~V.}\ \bibnamefont {Tran}},\ }\href {\doibase
  10.1103/PhysRevD.81.075022} {\bibfield  {journal} {\bibinfo  {journal} {Phys.
  Rev.}\ }\textbf {\bibinfo {volume} {D81}},\ \bibinfo {pages} {075022}
  (\bibinfo {year} {2010})},\ \Eprint {http://arxiv.org/abs/1001.3396}
  {arXiv:1001.3396 [hep-ph]} \BibitemShut {NoStop}%
\bibitem [{\citenamefont {Liu}\ \emph {et~al.}(2015)\citenamefont {Liu},
  \citenamefont {Weiner},\ and\ \citenamefont {Xue}}]{Liu:2014cma}%
  \BibitemOpen
  \bibfield  {author} {\bibinfo {author} {\bibfnamefont {J.}~\bibnamefont
  {Liu}}, \bibinfo {author} {\bibfnamefont {N.}~\bibnamefont {Weiner}}, \ and\
  \bibinfo {author} {\bibfnamefont {W.}~\bibnamefont {Xue}},\ }\href {\doibase
  10.1007/JHEP08(2015)050} {\bibfield  {journal} {\bibinfo  {journal} {JHEP}\
  }\textbf {\bibinfo {volume} {08}},\ \bibinfo {pages} {050} (\bibinfo {year}
  {2015})},\ \Eprint {http://arxiv.org/abs/1412.1485} {arXiv:1412.1485
  [hep-ph]} \BibitemShut {NoStop}%
\bibitem [{\citenamefont {Orfanidis}\ and\ \citenamefont
  {Rittenberg}(1973)}]{Orfanidis:1973ix}%
  \BibitemOpen
  \bibfield  {author} {\bibinfo {author} {\bibfnamefont {S.~J.}\ \bibnamefont
  {Orfanidis}}\ and\ \bibinfo {author} {\bibfnamefont {V.}~\bibnamefont
  {Rittenberg}},\ }\href {\doibase 10.1016/0550-3213(73)90660-3} {\bibfield
  {journal} {\bibinfo  {journal} {Nucl. Phys.}\ }\textbf {\bibinfo {volume}
  {B59}},\ \bibinfo {pages} {570} (\bibinfo {year} {1973})}\BibitemShut
  {NoStop}%
\bibitem [{\citenamefont {Armstrong}\ \emph {et~al.}(1987)\citenamefont
  {Armstrong} \emph {et~al.}}]{Armstrong:1987nu}%
  \BibitemOpen
  \bibfield  {author} {\bibinfo {author} {\bibfnamefont {T.}~\bibnamefont
  {Armstrong}} \emph {et~al.} (\bibinfo {collaboration}
  {BROOKHAVEN-HOUSTON-PENNSYLVANIA STATE-RICE}),\ }\href {\doibase
  10.1103/PhysRevD.36.659} {\bibfield  {journal} {\bibinfo  {journal} {Phys.
  Rev.}\ }\textbf {\bibinfo {volume} {D36}},\ \bibinfo {pages} {659} (\bibinfo
  {year} {1987})}\BibitemShut {NoStop}%
\bibitem [{\citenamefont {Bertin}\ \emph {et~al.}(1997)\citenamefont {Bertin}
  \emph {et~al.}}]{Bertin:1997gn}%
  \BibitemOpen
  \bibfield  {author} {\bibinfo {author} {\bibfnamefont {A.}~\bibnamefont
  {Bertin}} \emph {et~al.} (\bibinfo {collaboration} {OBELIX}),\ }\bibfield
  {booktitle} {\emph {\bibinfo {booktitle} {{Low-energy anti-proton physics.
  Proceedings, 4th Biennial Conference, LEAP'96, Dinkelsbuehl, Germany, August
  27-31, 1996}}},\ }\href {\doibase 10.1016/S0920-5632(97)00280-6} {\bibfield
  {journal} {\bibinfo  {journal} {Nucl. Phys. Proc. Suppl.}\ }\textbf {\bibinfo
  {volume} {56}},\ \bibinfo {pages} {227} (\bibinfo {year} {1997})},\ \bibinfo
  {note} {[,227(1997)]}\BibitemShut {NoStop}%
\bibitem [{\citenamefont {Huo}\ \emph {et~al.}(2016)\citenamefont {Huo},
  \citenamefont {Matsumoto}, \citenamefont {Sming~Tsai},\ and\ \citenamefont
  {Yanagida}}]{Huo:2015nwa}%
  \BibitemOpen
  \bibfield  {author} {\bibinfo {author} {\bibfnamefont {R.}~\bibnamefont
  {Huo}}, \bibinfo {author} {\bibfnamefont {S.}~\bibnamefont {Matsumoto}},
  \bibinfo {author} {\bibfnamefont {Y.-L.}\ \bibnamefont {Sming~Tsai}}, \ and\
  \bibinfo {author} {\bibfnamefont {T.~T.}\ \bibnamefont {Yanagida}},\ }\href
  {\doibase 10.1007/JHEP09(2016)162} {\bibfield  {journal} {\bibinfo  {journal}
  {JHEP}\ }\textbf {\bibinfo {volume} {09}},\ \bibinfo {pages} {162} (\bibinfo
  {year} {2016})},\ \Eprint {http://arxiv.org/abs/1506.06929} {arXiv:1506.06929
  [hep-ph]} \BibitemShut {NoStop}%
\bibitem [{\citenamefont {Lee}\ and\ \citenamefont {Wong}(2016)}]{Lee:2015hma}%
  \BibitemOpen
  \bibfield  {author} {\bibinfo {author} {\bibfnamefont {T.-G.}\ \bibnamefont
  {Lee}}\ and\ \bibinfo {author} {\bibfnamefont {C.-Y.}\ \bibnamefont {Wong}},\
  }\href {\doibase 10.1103/PhysRevC.95.029901, 10.1103/PhysRevC.93.014616}
  {\bibfield  {journal} {\bibinfo  {journal} {Phys. Rev.}\ }\textbf {\bibinfo
  {volume} {C93}},\ \bibinfo {pages} {014616} (\bibinfo {year} {2016})},\
  \bibinfo {note} {[Erratum: Phys. Rev.C95,no.2,029901(2017)]},\ \Eprint
  {http://arxiv.org/abs/1509.06031} {arXiv:1509.06031 [nucl-th]} \BibitemShut
  {NoStop}%
\bibitem [{\citenamefont {Hayashi}\ \emph {et~al.}(2016)\citenamefont
  {Hayashi}, \citenamefont {Ichikawa}, \citenamefont {Matsumoto}, \citenamefont
  {Ibe}, \citenamefont {Ishigaki},\ and\ \citenamefont
  {Sugai}}]{Hayashi:2016kcy}%
  \BibitemOpen
  \bibfield  {author} {\bibinfo {author} {\bibfnamefont {K.}~\bibnamefont
  {Hayashi}}, \bibinfo {author} {\bibfnamefont {K.}~\bibnamefont {Ichikawa}},
  \bibinfo {author} {\bibfnamefont {S.}~\bibnamefont {Matsumoto}}, \bibinfo
  {author} {\bibfnamefont {M.}~\bibnamefont {Ibe}}, \bibinfo {author}
  {\bibfnamefont {M.~N.}\ \bibnamefont {Ishigaki}}, \ and\ \bibinfo {author}
  {\bibfnamefont {H.}~\bibnamefont {Sugai}},\ }\href {\doibase
  10.1093/mnras/stw1457} {\bibfield  {journal} {\bibinfo  {journal} {Mon. Not.
  Roy. Astron. Soc.}\ }\textbf {\bibinfo {volume} {461}},\ \bibinfo {pages}
  {2914} (\bibinfo {year} {2016})},\ \Eprint {http://arxiv.org/abs/1603.08046}
  {arXiv:1603.08046 [astro-ph.GA]} \BibitemShut {NoStop}%
\bibitem [{\citenamefont {Geringer-Sameth}\ \emph {et~al.}(2015)\citenamefont
  {Geringer-Sameth}, \citenamefont {Koushiappas},\ and\ \citenamefont
  {Walker}}]{Geringer-Sameth:2014yza}%
  \BibitemOpen
  \bibfield  {author} {\bibinfo {author} {\bibfnamefont {A.}~\bibnamefont
  {Geringer-Sameth}}, \bibinfo {author} {\bibfnamefont {S.~M.}\ \bibnamefont
  {Koushiappas}}, \ and\ \bibinfo {author} {\bibfnamefont {M.}~\bibnamefont
  {Walker}},\ }\href {\doibase 10.1088/0004-637X/801/2/74} {\bibfield
  {journal} {\bibinfo  {journal} {Astrophys. J.}\ }\textbf {\bibinfo {volume}
  {801}},\ \bibinfo {pages} {74} (\bibinfo {year} {2015})},\ \Eprint
  {http://arxiv.org/abs/1408.0002} {arXiv:1408.0002 [astro-ph.CO]} \BibitemShut
  {NoStop}%
\bibitem [{\citenamefont {Ackermann}\ \emph {et~al.}(2015)\citenamefont
  {Ackermann} \emph {et~al.}}]{Ackermann:2015zua}%
  \BibitemOpen
  \bibfield  {author} {\bibinfo {author} {\bibfnamefont {M.}~\bibnamefont
  {Ackermann}} \emph {et~al.} (\bibinfo {collaboration} {Fermi-LAT}),\ }\href
  {\doibase 10.1103/PhysRevLett.115.231301} {\bibfield  {journal} {\bibinfo
  {journal} {Phys. Rev. Lett.}\ }\textbf {\bibinfo {volume} {115}},\ \bibinfo
  {pages} {231301} (\bibinfo {year} {2015})},\ \Eprint
  {http://arxiv.org/abs/1503.02641} {arXiv:1503.02641 [astro-ph.HE]}
  \BibitemShut {NoStop}%
\bibitem [{\citenamefont {Fisk}(1976)}]{Fisk:1976aw}%
  \BibitemOpen
  \bibfield  {author} {\bibinfo {author} {\bibfnamefont {L.~A.}\ \bibnamefont
  {Fisk}},\ }\href {\doibase 10.1086/154387} {\bibfield  {journal} {\bibinfo
  {journal} {Astrophys. J.}\ }\textbf {\bibinfo {volume} {206}},\ \bibinfo
  {pages} {333} (\bibinfo {year} {1976})}\BibitemShut {NoStop}%
\bibitem [{\citenamefont {Stone}\ \emph {et~al.}(2013)\citenamefont {Stone},
  \citenamefont {Cummings}, \citenamefont {McDonald}, \citenamefont {Heikkila},
  \citenamefont {Lal},\ and\ \citenamefont {Webber}}]{Stone2013}%
  \BibitemOpen
  \bibfield  {author} {\bibinfo {author} {\bibfnamefont {E.~C.}\ \bibnamefont
  {Stone}}, \bibinfo {author} {\bibfnamefont {A.~C.}\ \bibnamefont {Cummings}},
  \bibinfo {author} {\bibfnamefont {F.~B.}\ \bibnamefont {McDonald}}, \bibinfo
  {author} {\bibfnamefont {B.~C.}\ \bibnamefont {Heikkila}}, \bibinfo {author}
  {\bibfnamefont {N.}~\bibnamefont {Lal}}, \ and\ \bibinfo {author}
  {\bibfnamefont {W.~R.}\ \bibnamefont {Webber}},\ }\href {\doibase
  10.1126/science.1236408} {\bibfield  {journal} {\bibinfo  {journal}
  {Science}\ }\textbf {\bibinfo {volume} {341}},\ \bibinfo {pages} {150}
  (\bibinfo {year} {2013})}\BibitemShut {NoStop}%
\bibitem [{\citenamefont {Boudaud}\ \emph {et~al.}(2017)\citenamefont
  {Boudaud}, \citenamefont {Lavalle},\ and\ \citenamefont
  {Salati}}]{Boudaud:2016mos}%
  \BibitemOpen
  \bibfield  {author} {\bibinfo {author} {\bibfnamefont {M.}~\bibnamefont
  {Boudaud}}, \bibinfo {author} {\bibfnamefont {J.}~\bibnamefont {Lavalle}}, \
  and\ \bibinfo {author} {\bibfnamefont {P.}~\bibnamefont {Salati}},\ }\href
  {\doibase 10.1103/PhysRevLett.119.021103} {\bibfield  {journal} {\bibinfo
  {journal} {Phys. Rev. Lett.}\ }\textbf {\bibinfo {volume} {119}},\ \bibinfo
  {pages} {021103} (\bibinfo {year} {2017})},\ \Eprint
  {http://arxiv.org/abs/1612.07698} {arXiv:1612.07698 [astro-ph.HE]}
  \BibitemShut {NoStop}%
\bibitem [{\citenamefont {Cirelli}\ \emph {et~al.}(2011)\citenamefont
  {Cirelli}, \citenamefont {Corcella}, \citenamefont {Hektor}, \citenamefont
  {Hutsi}, \citenamefont {Kadastik}, \citenamefont {Panci}, \citenamefont
  {Raidal}, \citenamefont {Sala},\ and\ \citenamefont
  {Strumia}}]{Cirelli:2010xx}%
  \BibitemOpen
  \bibfield  {author} {\bibinfo {author} {\bibfnamefont {M.}~\bibnamefont
  {Cirelli}}, \bibinfo {author} {\bibfnamefont {G.}~\bibnamefont {Corcella}},
  \bibinfo {author} {\bibfnamefont {A.}~\bibnamefont {Hektor}}, \bibinfo
  {author} {\bibfnamefont {G.}~\bibnamefont {Hutsi}}, \bibinfo {author}
  {\bibfnamefont {M.}~\bibnamefont {Kadastik}}, \bibinfo {author}
  {\bibfnamefont {P.}~\bibnamefont {Panci}}, \bibinfo {author} {\bibfnamefont
  {M.}~\bibnamefont {Raidal}}, \bibinfo {author} {\bibfnamefont
  {F.}~\bibnamefont {Sala}}, \ and\ \bibinfo {author} {\bibfnamefont
  {A.}~\bibnamefont {Strumia}},\ }\href {\doibase
  10.1088/1475-7516/2012/10/E01, 10.1088/1475-7516/2011/03/051} {\bibfield
  {journal} {\bibinfo  {journal} {JCAP}\ }\textbf {\bibinfo {volume} {1103}},\
  \bibinfo {pages} {051} (\bibinfo {year} {2011})},\ \bibinfo {note} {[Erratum:
  JCAP1210,E01(2012)]},\ \Eprint {http://arxiv.org/abs/1012.4515}
  {arXiv:1012.4515 [hep-ph]} \BibitemShut {NoStop}%
\bibitem [{\citenamefont {Buch}\ \emph {et~al.}(2015)\citenamefont {Buch},
  \citenamefont {Cirelli}, \citenamefont {Giesen},\ and\ \citenamefont
  {Taoso}}]{Buch:2015iya}%
  \BibitemOpen
  \bibfield  {author} {\bibinfo {author} {\bibfnamefont {J.}~\bibnamefont
  {Buch}}, \bibinfo {author} {\bibfnamefont {M.}~\bibnamefont {Cirelli}},
  \bibinfo {author} {\bibfnamefont {G.}~\bibnamefont {Giesen}}, \ and\ \bibinfo
  {author} {\bibfnamefont {M.}~\bibnamefont {Taoso}},\ }\href {\doibase
  10.1088/1475-7516/2015/9/037, 10.1088/1475-7516/2015/09/037} {\bibfield
  {journal} {\bibinfo  {journal} {JCAP}\ }\textbf {\bibinfo {volume} {1509}},\
  \bibinfo {pages} {037} (\bibinfo {year} {2015})},\ \Eprint
  {http://arxiv.org/abs/1505.01049} {arXiv:1505.01049 [hep-ph]} \BibitemShut
  {NoStop}%
\bibitem [{\citenamefont {Donato}\ \emph {et~al.}(2004)\citenamefont {Donato},
  \citenamefont {Fornengo}, \citenamefont {Maurin},\ and\ \citenamefont
  {Salati}}]{Donato:2003xg}%
  \BibitemOpen
  \bibfield  {author} {\bibinfo {author} {\bibfnamefont {F.}~\bibnamefont
  {Donato}}, \bibinfo {author} {\bibfnamefont {N.}~\bibnamefont {Fornengo}},
  \bibinfo {author} {\bibfnamefont {D.}~\bibnamefont {Maurin}}, \ and\ \bibinfo
  {author} {\bibfnamefont {P.}~\bibnamefont {Salati}},\ }\href {\doibase
  10.1103/PhysRevD.69.063501} {\bibfield  {journal} {\bibinfo  {journal} {Phys.
  Rev.}\ }\textbf {\bibinfo {volume} {D69}},\ \bibinfo {pages} {063501}
  (\bibinfo {year} {2004})},\ \Eprint {http://arxiv.org/abs/astro-ph/0306207}
  {arXiv:astro-ph/0306207 [astro-ph]} \BibitemShut {NoStop}%
\bibitem [{\citenamefont {Navarro}\ \emph {et~al.}(1997)\citenamefont
  {Navarro}, \citenamefont {Frenk},\ and\ \citenamefont
  {White}}]{Navarro:1996gj}%
  \BibitemOpen
  \bibfield  {author} {\bibinfo {author} {\bibfnamefont {J.~F.}\ \bibnamefont
  {Navarro}}, \bibinfo {author} {\bibfnamefont {C.~S.}\ \bibnamefont {Frenk}},
  \ and\ \bibinfo {author} {\bibfnamefont {S.~D.~M.}\ \bibnamefont {White}},\
  }\href {\doibase 10.1086/304888} {\bibfield  {journal} {\bibinfo  {journal}
  {Astrophys. J.}\ }\textbf {\bibinfo {volume} {490}},\ \bibinfo {pages} {493}
  (\bibinfo {year} {1997})},\ \Eprint {http://arxiv.org/abs/astro-ph/9611107}
  {arXiv:astro-ph/9611107 [astro-ph]} \BibitemShut {NoStop}%
\bibitem [{\citenamefont {Burkert}(1996)}]{Burkert:1995yz}%
  \BibitemOpen
  \bibfield  {author} {\bibinfo {author} {\bibfnamefont {A.}~\bibnamefont
  {Burkert}},\ }\bibfield  {booktitle} {\emph {\bibinfo {booktitle} {{IAU
  Symposium 171: New Light on Galaxy Evolution Heidelberg, Germany, June 26-30,
  1995}}},\ }\href {\doibase 10.1086/309560} {\bibfield  {journal} {\bibinfo
  {journal} {IAU Symp.}\ }\textbf {\bibinfo {volume} {171}},\ \bibinfo {pages}
  {175} (\bibinfo {year} {1996})},\ \bibinfo {note} {[Astrophys.
  J.447,L25(1995)]},\ \Eprint {http://arxiv.org/abs/astro-ph/9504041}
  {arXiv:astro-ph/9504041 [astro-ph]} \BibitemShut {NoStop}%
\bibitem [{\citenamefont {Maurin}\ \emph {et~al.}(2014)\citenamefont {Maurin},
  \citenamefont {Melot},\ and\ \citenamefont {Taillet}}]{Maurin:2013lwa}%
  \BibitemOpen
  \bibfield  {author} {\bibinfo {author} {\bibfnamefont {D.}~\bibnamefont
  {Maurin}}, \bibinfo {author} {\bibfnamefont {F.}~\bibnamefont {Melot}}, \
  and\ \bibinfo {author} {\bibfnamefont {R.}~\bibnamefont {Taillet}},\ }\href
  {\doibase 10.1051/0004-6361/201321344} {\bibfield  {journal} {\bibinfo
  {journal} {Astron. Astrophys.}\ }\textbf {\bibinfo {volume} {569}},\ \bibinfo
  {pages} {A32} (\bibinfo {year} {2014})},\ \Eprint
  {http://arxiv.org/abs/1302.5525} {arXiv:1302.5525 [astro-ph.HE]} \BibitemShut
  {NoStop}%
\bibitem [{\citenamefont {Albert}\ \emph {et~al.}(2017)\citenamefont {Albert}
  \emph {et~al.}}]{Fermi-LAT:2016uux}%
  \BibitemOpen
  \bibfield  {author} {\bibinfo {author} {\bibfnamefont {A.}~\bibnamefont
  {Albert}} \emph {et~al.} (\bibinfo {collaboration} {Fermi-LAT, DES}),\ }\href
  {\doibase 10.3847/1538-4357/834/2/110} {\bibfield  {journal} {\bibinfo
  {journal} {Astrophys. J.}\ }\textbf {\bibinfo {volume} {834}},\ \bibinfo
  {pages} {110} (\bibinfo {year} {2017})},\ \Eprint
  {http://arxiv.org/abs/1611.03184} {arXiv:1611.03184 [astro-ph.HE]}
  \BibitemShut {NoStop}%
\bibitem [{\citenamefont {Tavani}\ \emph {et~al.}(2018)\citenamefont {Tavani}
  \emph {et~al.}}]{DeAngelis:2017gra}%
  \BibitemOpen
  \bibfield  {author} {\bibinfo {author} {\bibfnamefont {M.}~\bibnamefont
  {Tavani}} \emph {et~al.} (\bibinfo {collaboration} {e-ASTROGAM}),\ }\href
  {\doibase 10.1016/j.jheap.2018.07.001} {\bibfield  {journal} {\bibinfo
  {journal} {JHEAp}\ }\textbf {\bibinfo {volume} {19}},\ \bibinfo {pages} {1}
  (\bibinfo {year} {2018})},\ \Eprint {http://arxiv.org/abs/1711.01265}
  {arXiv:1711.01265 [astro-ph.HE]} \BibitemShut {NoStop}%
\bibitem [{\citenamefont {Rando}\ \emph {et~al.}(2019)\citenamefont {Rando},
  \citenamefont {De~Angelis},\ and\ \citenamefont {Mallamaci}}]{Rando:2019fzq}%
  \BibitemOpen
  \bibfield  {author} {\bibinfo {author} {\bibfnamefont {R.}~\bibnamefont
  {Rando}}, \bibinfo {author} {\bibfnamefont {A.}~\bibnamefont {De~Angelis}}, \
  and\ \bibinfo {author} {\bibfnamefont {M.}~\bibnamefont {Mallamaci}}
  (\bibinfo {collaboration} {thee-ASTROGAM}),\ }\bibfield  {booktitle} {\emph
  {\bibinfo {booktitle} {{Proceedings, 26th Extended European Cosmic Ray
  Symposium and 35th Russian Cosmic Ray Conference (RCRC 2018): Barnaul,
  Russia, July 6-10, 2018}}},\ }\href {\doibase
  10.1088/1742-6596/1181/1/012044} {\bibfield  {journal} {\bibinfo  {journal}
  {J. Phys. Conf. Ser.}\ }\textbf {\bibinfo {volume} {1181}},\ \bibinfo {pages}
  {012044} (\bibinfo {year} {2019})}\BibitemShut {NoStop}%
\bibitem [{\citenamefont {Sawano}\ \emph {et~al.}()\citenamefont {Sawano},
  \citenamefont {Hattori},\ and\ \citenamefont {Higashi}}]{smile}%
  \BibitemOpen
  \bibfield  {author} {\bibinfo {author} {\bibfnamefont {T.}~\bibnamefont
  {Sawano}}, \bibinfo {author} {\bibfnamefont {K.}~\bibnamefont {Hattori}}, \
  and\ \bibinfo {author} {\bibfnamefont {N.}~\bibnamefont {Higashi}},\ }in\
  \href {\doibase 10.7529/ICRC2011/V09/1120} {\emph {\bibinfo {booktitle}
  {{Proceedings, 32nd International Cosmic Ray Conference (ICRC 2011): Beijing,
  China, August 11-18, 2011}}}},\ Vol.~\bibinfo {volume} {9},\ p.\ \bibinfo
  {pages} {183}\BibitemShut {NoStop}%
\bibitem [{\citenamefont {Aoki}\ \emph {et~al.}(2012)\citenamefont {Aoki} \emph
  {et~al.}}]{Aoki:2012nn}%
  \BibitemOpen
  \bibfield  {author} {\bibinfo {author} {\bibfnamefont {S.}~\bibnamefont
  {Aoki}} \emph {et~al.},\ }\href@noop {} {\  (\bibinfo {year} {2012})},\
  \Eprint {http://arxiv.org/abs/1202.2529} {arXiv:1202.2529 [astro-ph.IM]}
  \BibitemShut {NoStop}%
\bibitem [{\citenamefont {Aramaki}\ \emph {et~al.}(2019)\citenamefont
  {Aramaki}, \citenamefont {Hansson~Adrian}, \citenamefont {Karagiorgi},\ and\
  \citenamefont {Odaka}}]{Aramaki:2019bpi}%
  \BibitemOpen
  \bibfield  {author} {\bibinfo {author} {\bibfnamefont {T.}~\bibnamefont
  {Aramaki}}, \bibinfo {author} {\bibfnamefont {P.}~\bibnamefont
  {Hansson~Adrian}}, \bibinfo {author} {\bibfnamefont {G.}~\bibnamefont
  {Karagiorgi}}, \ and\ \bibinfo {author} {\bibfnamefont {H.}~\bibnamefont
  {Odaka}},\ }\href@noop {} {\  (\bibinfo {year} {2019})},\ \Eprint
  {http://arxiv.org/abs/1901.03430} {arXiv:1901.03430 [astro-ph.HE]}
  \BibitemShut {NoStop}%
\bibitem [{\citenamefont {Ellis}\ \emph {et~al.}(2014)\citenamefont {Ellis}
  \emph {et~al.}}]{Ellis:2012rn}%
  \BibitemOpen
  \bibfield  {author} {\bibinfo {author} {\bibfnamefont {R.}~\bibnamefont
  {Ellis}} \emph {et~al.} (\bibinfo {collaboration} {PFS Team}),\ }\href
  {\doibase 10.1093/pasj/pst019} {\bibfield  {journal} {\bibinfo  {journal}
  {Publ. Astron. Soc. Jap.}\ }\textbf {\bibinfo {volume} {66}},\ \bibinfo
  {pages} {R1} (\bibinfo {year} {2014})},\ \Eprint
  {http://arxiv.org/abs/1206.0737} {arXiv:1206.0737 [astro-ph.CO]} \BibitemShut
  {NoStop}%
\bibitem [{\citenamefont {Bedaque}\ \emph {et~al.}(2009)\citenamefont
  {Bedaque}, \citenamefont {Buchoff},\ and\ \citenamefont
  {Mishra}}]{Bedaque:2009ri}%
  \BibitemOpen
  \bibfield  {author} {\bibinfo {author} {\bibfnamefont {P.~F.}\ \bibnamefont
  {Bedaque}}, \bibinfo {author} {\bibfnamefont {M.~I.}\ \bibnamefont
  {Buchoff}}, \ and\ \bibinfo {author} {\bibfnamefont {R.~K.}\ \bibnamefont
  {Mishra}},\ }\href {\doibase 10.1088/1126-6708/2009/11/046} {\bibfield
  {journal} {\bibinfo  {journal} {JHEP}\ }\textbf {\bibinfo {volume} {11}},\
  \bibinfo {pages} {046} (\bibinfo {year} {2009})},\ \Eprint
  {http://arxiv.org/abs/0907.0235} {arXiv:0907.0235 [hep-ph]} \BibitemShut
  {NoStop}%
\bibitem [{\citenamefont {Liu}\ \emph {et~al.}(2013)\citenamefont {Liu},
  \citenamefont {Wu},\ and\ \citenamefont {Zhou}}]{Liu:2013vha}%
  \BibitemOpen
  \bibfield  {author} {\bibinfo {author} {\bibfnamefont {Z.-P.}\ \bibnamefont
  {Liu}}, \bibinfo {author} {\bibfnamefont {Y.-L.}\ \bibnamefont {Wu}}, \ and\
  \bibinfo {author} {\bibfnamefont {Y.-F.}\ \bibnamefont {Zhou}},\ }\href
  {\doibase 10.1103/PhysRevD.88.096008} {\bibfield  {journal} {\bibinfo
  {journal} {Phys. Rev.}\ }\textbf {\bibinfo {volume} {D88}},\ \bibinfo {pages}
  {096008} (\bibinfo {year} {2013})},\ \Eprint {http://arxiv.org/abs/1305.5438}
  {arXiv:1305.5438 [hep-ph]} \BibitemShut {NoStop}%
\bibitem [{\citenamefont {Bellazzini}\ \emph {et~al.}(2013)\citenamefont
  {Bellazzini}, \citenamefont {Cliche},\ and\ \citenamefont
  {Tanedo}}]{Bellazzini:2013foa}%
  \BibitemOpen
  \bibfield  {author} {\bibinfo {author} {\bibfnamefont {B.}~\bibnamefont
  {Bellazzini}}, \bibinfo {author} {\bibfnamefont {M.}~\bibnamefont {Cliche}},
  \ and\ \bibinfo {author} {\bibfnamefont {P.}~\bibnamefont {Tanedo}},\ }\href
  {\doibase 10.1103/PhysRevD.88.083506} {\bibfield  {journal} {\bibinfo
  {journal} {Phys. Rev.}\ }\textbf {\bibinfo {volume} {D88}},\ \bibinfo {pages}
  {083506} (\bibinfo {year} {2013})},\ \Eprint {http://arxiv.org/abs/1307.1129}
  {arXiv:1307.1129 [hep-ph]} \BibitemShut {NoStop}%
\bibitem [{\citenamefont {Sommerfeld}(1931)}]{Sommerfeld}%
  \BibitemOpen
  \bibfield  {author} {\bibinfo {author} {\bibfnamefont {A.}~\bibnamefont
  {Sommerfeld}},\ }\href {\doibase 10.1002/andp.19314030302} {\bibfield
  {journal} {\bibinfo  {journal} {Ann. d. Phy}\ }\textbf {\bibinfo {volume}
  {11}},\ \bibinfo {pages} {257} (\bibinfo {year} {1931})}\BibitemShut
  {NoStop}%
\bibitem [{\citenamefont {Hisano}\ \emph {et~al.}(2003)\citenamefont {Hisano},
  \citenamefont {Matsumoto},\ and\ \citenamefont {Nojiri}}]{Hisano:2002fk}%
  \BibitemOpen
  \bibfield  {author} {\bibinfo {author} {\bibfnamefont {J.}~\bibnamefont
  {Hisano}}, \bibinfo {author} {\bibfnamefont {S.}~\bibnamefont {Matsumoto}}, \
  and\ \bibinfo {author} {\bibfnamefont {M.~M.}\ \bibnamefont {Nojiri}},\
  }\href {\doibase 10.1103/PhysRevD.67.075014} {\bibfield  {journal} {\bibinfo
  {journal} {Phys. Rev.}\ }\textbf {\bibinfo {volume} {D67}},\ \bibinfo {pages}
  {075014} (\bibinfo {year} {2003})},\ \Eprint
  {http://arxiv.org/abs/hep-ph/0212022} {arXiv:hep-ph/0212022 [hep-ph]}
  \BibitemShut {NoStop}%
\bibitem [{\citenamefont {Hisano}\ \emph {et~al.}(2004)\citenamefont {Hisano},
  \citenamefont {Matsumoto},\ and\ \citenamefont {Nojiri}}]{Hisano:2003ec}%
  \BibitemOpen
  \bibfield  {author} {\bibinfo {author} {\bibfnamefont {J.}~\bibnamefont
  {Hisano}}, \bibinfo {author} {\bibfnamefont {S.}~\bibnamefont {Matsumoto}}, \
  and\ \bibinfo {author} {\bibfnamefont {M.~M.}\ \bibnamefont {Nojiri}},\
  }\href {\doibase 10.1103/PhysRevLett.92.031303} {\bibfield  {journal}
  {\bibinfo  {journal} {Phys. Rev. Lett.}\ }\textbf {\bibinfo {volume} {92}},\
  \bibinfo {pages} {031303} (\bibinfo {year} {2004})},\ \Eprint
  {http://arxiv.org/abs/hep-ph/0307216} {arXiv:hep-ph/0307216 [hep-ph]}
  \BibitemShut {NoStop}%
\bibitem [{\citenamefont {Hisano}\ \emph {et~al.}(2005)\citenamefont {Hisano},
  \citenamefont {Matsumoto}, \citenamefont {Nojiri},\ and\ \citenamefont
  {Saito}}]{Hisano:2004ds}%
  \BibitemOpen
  \bibfield  {author} {\bibinfo {author} {\bibfnamefont {J.}~\bibnamefont
  {Hisano}}, \bibinfo {author} {\bibfnamefont {S.}~\bibnamefont {Matsumoto}},
  \bibinfo {author} {\bibfnamefont {M.~M.}\ \bibnamefont {Nojiri}}, \ and\
  \bibinfo {author} {\bibfnamefont {O.}~\bibnamefont {Saito}},\ }\href
  {\doibase 10.1103/PhysRevD.71.063528} {\bibfield  {journal} {\bibinfo
  {journal} {Phys. Rev.}\ }\textbf {\bibinfo {volume} {D71}},\ \bibinfo {pages}
  {063528} (\bibinfo {year} {2005})},\ \Eprint
  {http://arxiv.org/abs/hep-ph/0412403} {arXiv:hep-ph/0412403 [hep-ph]}
  \BibitemShut {NoStop}%
\bibitem [{\citenamefont {Blum}\ \emph {et~al.}(2016)\citenamefont {Blum},
  \citenamefont {Sato},\ and\ \citenamefont {Slatyer}}]{Blum:2016nrz}%
  \BibitemOpen
  \bibfield  {author} {\bibinfo {author} {\bibfnamefont {K.}~\bibnamefont
  {Blum}}, \bibinfo {author} {\bibfnamefont {R.}~\bibnamefont {Sato}}, \ and\
  \bibinfo {author} {\bibfnamefont {T.~R.}\ \bibnamefont {Slatyer}},\ }\href
  {\doibase 10.1088/1475-7516/2016/06/021} {\bibfield  {journal} {\bibinfo
  {journal} {JCAP}\ }\textbf {\bibinfo {volume} {1606}},\ \bibinfo {pages}
  {021} (\bibinfo {year} {2016})},\ \Eprint {http://arxiv.org/abs/1603.01383}
  {arXiv:1603.01383 [hep-ph]} \BibitemShut {NoStop}%
\bibitem [{\citenamefont {Bellazzini}\ \emph {et~al.}(2016)\citenamefont
  {Bellazzini}, \citenamefont {Franceschini}, \citenamefont {Sala},\ and\
  \citenamefont {Serra}}]{Bellazzini:2015nxw}%
  \BibitemOpen
  \bibfield  {author} {\bibinfo {author} {\bibfnamefont {B.}~\bibnamefont
  {Bellazzini}}, \bibinfo {author} {\bibfnamefont {R.}~\bibnamefont
  {Franceschini}}, \bibinfo {author} {\bibfnamefont {F.}~\bibnamefont {Sala}},
  \ and\ \bibinfo {author} {\bibfnamefont {J.}~\bibnamefont {Serra}},\ }\href
  {\doibase 10.1007/JHEP04(2016)072} {\bibfield  {journal} {\bibinfo  {journal}
  {JHEP}\ }\textbf {\bibinfo {volume} {04}},\ \bibinfo {pages} {072} (\bibinfo
  {year} {2016})},\ \Eprint {http://arxiv.org/abs/1512.05330} {arXiv:1512.05330
  [hep-ph]} \BibitemShut {NoStop}%
\end{thebibliography}%
\end{document}